\documentclass[lettersize,journal]{IEEEtran}
\usepackage{amsmath,amsfonts}
\usepackage{algorithmic}
\usepackage{array}
\usepackage{textcomp}
\usepackage{stfloats}
\usepackage{url}
\usepackage{verbatim}
\usepackage{graphicx}
\def\BibTeX{{\rm B\kern-.05em{\sc i\kern-.025em b}\kern-.08em
	T\kern-.1667em\lower.7ex\hbox{E}\kern-.125emX}}
\usepackage{balance}
\usepackage{xcolor}
\usepackage[hidelinks]{hyperref}
\makeatletter
\def\endthebibliography{
\def\@noitemerr{\@latex@warning{Empty `thebibliography' environment}}
\endlist
}
\makeatother
\usepackage{cite}
\usepackage{tikz}
\usetikzlibrary{shapes}
\usetikzlibrary{arrows}

\usepackage{tikzpeople}
\tikzset{
Speaker/.pic={
	\filldraw[fill=gray!40,pic actions] 
	(-15pt,0) -- 
	coordinate[midway] (-front) 
	(15pt,0) -- 
	++([shift={(-6pt,8pt)}]0pt,0pt) coordinate (aux1) -- 
	++(-18pt,0) coordinate (aux2) 
	-- cycle 
	(aux1) -- ++(0,6pt) -- coordinate[midway] (-back) ++(-18pt,0) -- (aux2);
	\draw (-1mm,-0.4mm) arc (-180:0:1mm);
	\draw (-2mm,-0.6mm) arc (-180:0:2mm);
	\draw (-3mm,-0.8mm) arc (-180:0:3mm);
}}
\usepackage[usestackEOL]{stackengine}
\definecolor{mycolor1}{rgb}{0.04310,0.12160,0.56080}
\definecolor{mycolor2}{rgb}{0.10980,0.70980,0.09800}
\definecolor{mycolor3}{rgb}{0,0,0}
\definecolor{mycolor4}{rgb}{1.00000,0.07450,0.65100}
\definecolor{mycolor5}{rgb}{0.87060,0.46670,0.11370}
\definecolor{mygray}{RGB}{90,90,90}
\usepackage{pgfplots}
\usepackage{fontawesome}
\usepackage{subfig}
\usepackage{placeins}
\renewcommand{\Vec}{\mathbf}
\newcommand\norm[1]{\left\lVert#1\right\rVert}
\DeclareMathOperator*{\argmin}{argmin}
\usepackage{siunitx}
\newcolumntype{L}{>{\scriptstyle}l}
\newcolumntype{C}{>{\scriptstyle}c}
\newcolumntype{R}{>{\scriptstyle}r}

\usepackage{cancel}
\usepackage{romannum}
\usepackage{romannum}
\usepackage{amsthm}
\newtheorem{theorem}{Theorem}[section]
\usepackage{multirow}
\usepackage{booktabs}
\theoremstyle{definition}

\newtheorem{definition}{Condition}
\allowdisplaybreaks

\begin{document}
\title{Integrated Minimum Mean Squared Error Algorithms for Combined Acoustic Echo Cancellation and Noise Reduction} 
\author{\thanks{This research was carried out at the ESAT Laboratory of KU Leuven, in the frame of Research Council KU Leuven C14-21-0075 "A holistic approach to the design of integrated and distributed digital signal processing algorithms for audio and speech communication devices", and Aspirant Grant 11PDH24N (for A. Roebben) from the Research Foundation - Flanders (FWO). The scientific responsibility is assumed by its authors.}\thanks{© 2026 IEEE.  Personal use of this material is permitted.  Permission from IEEE must be obtained for all other uses, in any current or future media, including reprinting/republishing this material for advertising or promotional purposes, creating new collective works, for resale or redistribution to servers or lists, or reuse of any copyrighted component of this work in other works.}Arnout Roebben\thanks{Arnout Roebben, Toon van Waterschoot and Marc Moonen are with the Department of Electrical Engineering (ESAT), STADIUS Center for Dynamical Systems, Signal Processing and Data Analytics, KU Leuven, B-3001 Leuven, Belgium (e-mail: arnout.roebben@esat.kuleuven.be; toon.vanwaterschoot@esat.kuleuven.be; marc.moonen@esat.kuleuven.be).}, \IEEEmembership{Graduate Student Member, IEEE}, Toon van Waterschoot, \IEEEmembership{Member, IEEE}, Jan Wouters,\thanks{Jan Wouters is with the Department of Neurosciences, Research Group ExpORL, KU Leuven, B-3000 Leuven, Belgium (e-mail: jan.wouters@med.kuleuven.be).} and Marc Moonen, \IEEEmembership{Fellow, IEEE}}
\markboth{}
{Integrated minimum mean squared error algorithmsfor combined acoustic echo cancellation and noise reduction}

\maketitle

\begin{abstract}
    In many speech recording applications, noise and acoustic echo corrupt the desired speech. Consequently, combined noise reduction (NR) and acoustic echo cancellation (AEC) is required. Generally, a cascade approach is followed, i.e., the AEC and NR are designed in isolation by selecting a separate signal model, separate cost function, and separate solution strategy. The AEC and NR are then cascaded one after the other, not accounting for their interaction. In this paper, an integrated approach is proposed to consider this interaction in a general multi-microphone/multi-loudspeaker setup. Therefore, a single signal model of either the microphone signal vector or the extended signal vector, obtained by stacking microphone and loudspeaker signals, is selected, a single mean squared error cost function is formulated, and a common solution strategy is used. Using this microphone signal model, a multi-channel Wiener filter (MWF) is derived. Using the extended signal model, it is shown that an extended MWF (MWF\textsubscript{ext}) can be derived, and several equivalent expressions can be found, which are nevertheless shown to be interpretable as cascade algorithms. Specifically, the MWF\textsubscript{ext} is shown to be equivalent to algorithms where the AEC precedes the NR (AEC-NR), the NR precedes the AEC (NR-AEC), and the extended NR (NR\textsubscript{ext}) precedes the AEC and post-filter (PF) (NR\textsubscript{ext}-AEC-PF). Under rank-deficiency conditions the MWF\textsubscript{ext} is non-unique. Equivalence then amounts to the expressions being specific, not necessarily minimum-norm solutions, for this MWF\textsubscript{ext}. The practical performances differ due to non-stationarities and imperfect correlation matrix estimation, with the AEC-NR and NR\textsubscript{ext}-AEC-PF attaining best overall performance.
\end{abstract}


\section{Introduction} \label{section:introduction}
\IEEEPARstart{I}{n} many speech recording applications, e.g., hands-free telephony in a car or telecommunication using hearing instruments, noise and acoustic echo corrupt the desired speech as illustrated in Fig. \ref{fig:signal_model} \cite{hanslerTopicsAcousticEcho2006}. The noise originates from within the room (the so-called near-end room), while the echo originates from loudspeakers playing signals recorded in another room (the so-called far-end room). To suppress the near-end room noise and echo, noise reduction (NR) and acoustic echo cancellation (AEC) algorithms are required. 
\begin{figure}
	\centering
	
\scalebox{0.55}{
%
%
\begin{tikzpicture}
	\node[businessman,female,minimum width=1cm,skin=blue!30!white] (bm) {};
	\draw (5mm,4mm) arc (90:-90:1mm);
	\draw (5mm,5mm) arc (90:-90:2mm);
	\draw (5mm,6mm) arc (90:-90:3mm);
	
	\node[draw=none,below of=bm,node distance=6.5cm] (t1) {};
	\node[draw=none,right of=t1,node distance=-3.3cm] (t2) {};
	\pic [rotate=90,local bounding box=sp1,right of=t2,node distance=2cm] {Speaker};
	
	\node[draw=none,below of=sp1,node distance=-1cm] (t4) {};
	\node[draw=none,right of=t4,node distance=-0.25cm] { \huge$\vdots$};
	
	\pic [rotate=90,right of=t2,node distance=3.8cm,local bounding box=sp2] {Speaker};

	\node[draw=none, minimum width=0cm,below of=t4,node distance=5cm] (t5) {};
	\node[bob,minimum width=1cm,right of=t5,node distance=1.5cm,skin=blue!30!white,hair=yellow!50!black] (b) {};
	\node[draw=none, minimum width=0cm,below of=t4,node distance=3.7cm] (t6) {};
	\node[overlay,ellipse callout,callout relative pointer={(-0.2cm,-0.4cm)},right of=t6,node distance=2.3cm,aspect=2.5,fill=gray!40,minimum size=1cm] (tb1) {$\dots$};
	\node[charlie,female,minimum width=1cm,right of=b,node distance=2.2cm,mirrored,skin=blue!30!white,hair=black] (c) {};
	\node[overlay,ellipse callout,callout relative pointer={(0.2cm,-0.4cm)},%
	aspect=2.5,fill=gray!40,minimum size=1cm
	,right of=tb1,node distance=0.7cm] {$\dots$};

	\node[draw=none,minimum size=0cm,right of=sp1,node distance=7cm] (t7) {};
	\node[draw=none,minimum size=0cm,below of=t7,node distance=-1.8cm] (temp) {};
	\node[draw=black,fill=gray!40,circle,minimum size=0.6cm,right of=temp,node distance=-0.28cm] (mic1) {};
	\draw ([yshift=8pt]mic1.west) -- ([yshift=-8pt]mic1.west);
	
	\node[draw=none,below of=mic1,node distance=0.8cm] (t8) {};
	\node[draw=none,right of=t8,node distance=0cm] (vd1) {\huge $\vdots$};
	
	\node[draw=black,fill=gray!40,circle,minimum size=0.6cm,below of=t8,node distance=1cm] (mic2) {};
	\draw ([yshift=8pt]mic2.west) -- ([yshift=-8pt]mic2.west);
	
	\node[draw=none,below of=t8,minimum size=0cm,node distance=0.79cm] (t9) {};
	\node[draw=black,right of=t9,minimum width=2.3cm,node distance=2cm,minimum height=3.7cm] (b) {\Large Algorithm};
	
	\node[draw=none,right of=t9,minimum size=0cm,node distance=3.8cm] (out) {\Large $\hat{s}_r$};
	\draw [->] (b) -- (out);		
	
	\node[draw=none,right of=mic1,node distance=0.4cm] (t10) {};
	\draw [->] (t10) -- (t10-|b.west);
	\node[draw=none,right of=mic2,node distance=0.4cm] (t11) {};
	\draw [->] (t11) -- (t11-|b.west);			
	\node[draw=none,below of=t8,node distance=0.12cm] (t12) {};
	\node[draw=none,right of=t12,node distance=0.4cm] (t13) {};
	\draw [->] (t13) -- (t13-|b.west);	
	
	\node[draw=black,above right=-0.92cm and 1.5cm of sp1,minimum width=1.3cm,minimum height=2.6cm] (F) {\Large $F(.)$};
	
	\node[draw=none,right of=sp1,node distance=0.4cm] (t14) {};
	\draw [->] (t14) -- (t14-|F.west);	
	\node[draw=none,right of=sp2,node distance=0.4cm] (t15) {};
	\draw [->] (t15) -- (t15-|F.west);		
	
	\node[draw=none,right of=t14,node distance=0.7cm,minimum size=0cm] (t19a) {};
	\node[draw=none,above of=t19a,node distance=0.19cm,minimum size=0cm] (t19) {};
	\node[draw=none,below of=t19,node distance=0.93cm,minimum size=0cm] (t20a) {};	
	\node[draw=none,below of=t19,node distance=0.8cm,minimum size=0cm] (t20b) {};	
	\node[draw=none,right of=t20b,node distance=-0.12cm,minimum size=0cm] (t20) {};
	\draw [dashed,*-] (t19) -- (t20a);	
	\draw [->,dashed] (t20) -- (t20-|b.west);	
	\node[draw=none,right of=t15,node distance=0.4cm,minimum size=0cm] (t21a) {};
	\node[draw=none,above of=t21a,node distance=0.2cm,minimum size=0cm] (t21) {};
	\node[draw=none,below of=t21,node distance=3.53cm,minimum size=0cm] (t22a) {};	
	\node[draw=none,below of=t21,node distance=3.4cm,minimum size=0cm] (t22b) {};	
	\node[draw=none,right of=t22b,node distance=-0.12cm,minimum size=0cm] (t22) {};
	\draw [dashed,*-] (t21) -- (t22a);	
	\draw [->,dashed] (t22) -- (t22-|b.west);		
	
	\node[draw=none,below of=vd1,node distance=1.91cm] {\huge $\vdots$};
	\node[draw=none,below of=t8,node distance=2.01cm] (t23) {};
	\node[draw=none,right of=t23,node distance=0.4cm] (t24) {};
	\draw [->,dashed] (t24) -- (t24-|b.west);		
	
	\node[draw=none,left of=t8,node distance=0.1cm] (t16) {};
	\node[draw=none,right of=F,node distance=4.09cm] (t17) {};
	\node[draw=none,below of=bm,node distance=-0.3cm] (t25) {};
	\node[draw=none,right of=t25,node distance=0.9cm] (t26) {};
	\node[draw=none,below of=c,node distance=-1.4cm] (t27) {};
	\node[draw=none,right of=t27,node distance=-0.3cm] (t28) {};	
	\draw [->] (t26) -- (t17);	
	\draw [->] (t28) -- (t17);	
	\draw [->] (F) -- (t17);	
	
	\node[draw=none,right of=bm,node distance=2.2cm] (t29) {};
	\node[draw=none,below of=t29,node distance=1.8cm,anchor=east] (ds) {\Large Desired speech $\Vec{s}$};		
	\node[draw=none,right of=sp1,node distance=4.7cm] (t32) {};
	\node[draw=none,below of=t29,node distance=4.7cm,anchor=east] (ds) {\Large Noise $\Vec{n}$};	
	\node[draw=none,right of=sp1,node distance=1.2cm] (t33) {};
	\node[draw=none,below of=t33,node distance=-0.32cm] (ds) {\Large ${l}_L$};	
	\node[draw=none,above of=t33,node distance=2.15cm] (ds) {\Large ${l}_1$};	
	\node[draw=none,below of=t29,node distance=3.35cm,anchor=east] (ds) {\Large Echo $\Vec{e}$};	
	
	\node[draw=none,right of=bm,node distance=3.35cm] (t34) {};
	\node[draw=none,below of=t34,node distance=2cm] (ds) {\Large $\Vec{m}$};	
\end{tikzpicture}
} 

	\caption{Algorithms for combined acoustic echo cancellation (AEC) and noise reduction (NR) aim at providing an estimate of the desired speech $\hat{s}_r$ in reference microphone $r$ by suppressing the near-end room noise signal vector $\Vec{n}$ and echo signal vector $\Vec{e}$, originating from the loudspeaker signals $l_j$, $j\in\{1,\cdots,L\}$, by means of the echo path map $F(.)$. To this end, the microphone signal vector $\Vec{m}$ and possibly the loudspeaker signals $l_j$ are utilised. The figurines have been generated using \cite{tikzpeople}.}
	\label{fig:signal_model} 
\end{figure}

AEC aims at estimating the desired speech by suppressing the echo \cite{hanslerTopicsAcousticEcho2006}. AEC algorithms traditionally exploit the availability of loudspeaker signals to compute an estimate of this echo, which can then be subtracted from the microphone signals \cite{hanslerTopicsAcousticEcho2006,Albu_FastBlockExactGaussSeidel2004}. Other approaches exist as well, such as approaches based on single-channel gains \cite{yangStereophonicAcousticEcho2012}, data-driven end-to-end neural networks (NNs) \cite{zhangNeuralMultiChannelMultiMicrophone2023}, and hybrid combinations of model-based and data-driven approaches \cite{fazelCADAECContextAwareDeep2020,seidel_neural_2024,seidel_convergence_2024}.

NR aims at estimating the desired speech by suppressing the near-end room noise \cite{benestyStudyWienerFilter2005,docloMultichannelSignalEnhancement2015c}. NR algorithms traditionally exploit the availability of microphone signals. Examples of NR algorithms include beamformers, such as the multi-channel Wiener filter (MWF) and generalised sidelobe canceller (GSC) \cite{benestyStudyWienerFilter2005,docloMultichannelSignalEnhancement2015c}, Kalman filters \cite{gannotSpeechProcessingUtilizing2012}, single-channel gains \cite{hendriksDFTDomainBasedSingleMicrophone2013b}, data-driven end-to-end NNs \cite{jiangNovelSkipConnection2023} and hybrid combinations of model-based and data-driven approaches \cite{zhangADLMVDRAllDeep2021}. 

Algorithms for combined AEC and NR consequently aim at estimating the desired speech by jointly suppressing the echo and near-end room noise \cite{hanslerTopicsAcousticEcho2006}. Combined AEC and NR algorithms exploit the availability of microphone and loudspeaker signals as illustrated in Fig. \ref{fig:signal_model}. To achieve this combined AEC and NR, generally, a cascade approach is adhered to, i.e., the AEC and NR are designed in isolation by selecting a separate signal model, formulating a separate cost function, and using a separate solution strategy. The AEC and NR are then cascaded one after the other, leading to algorithms where the AEC precedes the NR (AEC-NR) \cite{gustafssonCombinedAcousticEcho1998a,cohenJointBeamformingEcho2018,luisvaleroLowComplexityMultiMicrophoneAcoustic2019a,Albu_CombinedEchoNoise2004}, the NR precedes the AEC (NR-AEC) \cite{martinAcousticEchoCancellation1997,docloCombinedAcousticEcho2000,schrammenChangePredictionLow2019,roebbenCascadedNoiseReduction2024}, or variations thereof \cite{jeannesCombinedNoiseEcho2001,herbordtAcousticHumanMachineFrontEnd2003,burtonNewStructureCombining2007}.

Despite the AEC being less perturbed by near-end room noise in the NR-AEC, the AEC-NR tends to outperform the NR-AEC in terms of echo suppression. Indeed, the AEC in the NR-AEC needs to track (the adaptivity of) the NR, which is not the case for the AEC in the AEC-NR \cite{reuvenJointAcousticEcho2004,docloCombinedAcousticEcho2000,luisvaleroLowComplexityMultiMicrophoneAcoustic2019a}. Further, although the AEC needs to process the noisy microphone signal vectors, the AEC-NR can use the NR to suppress residual echo \cite{gustafssonPostfilterEchoNoise1999}. The NR-AEC, on the other hand, is advantageous computational-complexity-wise as only one echo path estimation per loudspeaker is required, opposed to one echo path estimation for each loudspeaker-microphone pair for the AEC-NR \cite{reuvenJointAcousticEcho2004,docloCombinedAcousticEcho2000,luisvaleroLowComplexityMultiMicrophoneAcoustic2019a}.

In this paper, the cascade approach is contrasted to an integrated approach where a single algorithm for joint AEC and NR is designed by selecting a single signal model, formulating a single cost function, and using a common solution strategy. Whereas the cascade approach, thus, designs two separate algorithms with distinct parameters, the integrated approach designs a single algorithm with shared parameters.

In \cite{konforti_multichannel_2023} and \cite{herbordtJointOptimizationLCMV2004,maruoOptimalSolutionsBeamformer2011b}, this integrated approach has been applied with a linearly constrained minimum variance (LCMV) cost function, together with a microphone signal model \cite{konforti_multichannel_2023} or extended signal model of stacked microphone and loudspeaker signal vectors \cite{herbordtJointOptimizationLCMV2004,maruoOptimalSolutionsBeamformer2011b}. In \cite{nathwaniJointAcousticEcho2018}, a combined state-space model for the echo path and the autoregressive desired speech has been considered, using a cost function and solution strategy based on Kalman filtering and expectation maximisation (EM). Integrated single-channel gains have been studied in \cite{habetsMMSELogspectralAmplitude2006,parkIntegratedAcousticEcho2012,jayakumarIntegratedAcousticEcho2016}. 

Data-driven NNs generally do not assume any particular signal model, but do optimise a single cost function to perform both the AEC and NR task \cite{zhangDeepLearningJoint2019a,indenbom_deepvqe_2023,seidel_efficient_2023,braun_task_2022}. There can either be one NN jointly performing both tasks \cite{seidel_efficient_2023,indenbom_deepvqe_2023}, or dedicated subnetworks can separately perform one single task each \cite{braun_task_2022}. These NNs can be preceded by a model-based AEC \cite{seidel_efficient_2024,shetu_hybrid_2024,shetu_align-ulcnet_2024,franzen_deep_2022}, hence leading to hybrid methods. This preceding AEC itself can also be of hybrid nature \cite{haubner_synergistic_2021}. 
	
For specific choices of the signal model and cost function, the algorithms resulting from the cascade and integrated approaches can nevertheless be shown to be equivalent. In \cite{gustafssonCombinedAcousticEcho1998a,jeannesCombinedNoiseEcho2001,docloMultimicrophoneNoiseReduction2003, romboutsIntegratedApproachAcoustic2005a,ruizDistributedCombinedAcoustic2022}, a mean squared error (MSE) cost function has been used together with an extended signal model. While it is stressed that this integrated approach leads to a single algorithm with shared parameters for the AEC and NR, it has been shown that the resulting algorithm can be interpreted as the AEC-NR cascade algorithm in the $1$-microphone/$1$-loudspeaker setup \cite{gustafssonCombinedAcousticEcho1998a,jeannesCombinedNoiseEcho2001}, in the $2$-microphone/$1$-loudspeaker setup with uncorrelated near-end room noise in both microphones \cite{jeannesCombinedNoiseEcho2001}, in the multi-microphone/$1$-loudspeaker setup when using a linear signal model and invertibility assumptions \cite{schwartz_efficient_2024}, and in the multi-microphone/multi-loudspeaker setup when using a specific generalised eigenvalue decomposition (GEVD) implementation and (multiframe) linear signal model for the echo paths \cite{ruizDistributedCombinedAcoustic2022}. Similarly, in the multi-microphone/$1$-loudspeaker setup it has been shown that the resulting algorithm can be interpreted as an NR-AEC cascade algorithm \cite{docloMultimicrophoneNoiseReduction2003}, and in the multi-microphone/multi-loudspeaker as an extended MWF (MWF\textsubscript{ext}), i.e., an MWF on an extended signal model when using the specific GEVD implementation and (multiframe) linear signal model \cite{ruizDistributedCombinedAcoustic2022}. Nevertheless, while both the AEC-NR, NR-AEC and MWF\textsubscript{ext} are theoretically equivalent under these specific conditions, their practical performances differ due to parameter estimation differences, e.g., as described supra for the AEC-NR and NR-AEC. 

The MSE cost has thus only been partially explored: only for specific choices of the signal model and cost function, and only showing equivalence for the AEC-NR, NR-AEC and MWF\textsubscript{ext}. In this paper, the aim is to apply the integrated approach to the most general multi-microphone/multi-loudspeaker setup, where the loudspeaker and microphone signals are possibly linearly related, allowing to generalise previous work, to propose new algorithms, and to study practical performance differences across algorithms. Therefore, the main research questions, tackled in this paper are as follows:\vspace{-0.2em}
\begin{enumerate}
	\item What algorithms are retrieved by applying the integrated approach to a combined AEC and NR problem in the general multi-microphone/multi-loudspeaker setup, possibly with linear dependencies between the loudspeaker and microphone signals, and using an MSE cost?
	\item How do non-stationarities and imperfect correlation matrix estimation influence the practical performance? 
\end{enumerate}
\vspace{-0.2em}

As for the first research question, a single signal model of either the microphone signal vector, or the extended signal vector obtained by stacking microphone and loudspeaker signals will be selected, a single MSE cost function will be formulated, and a common solution strategy will be used. Selecting the microphone signal model, an MWF will be derived, whereas selecting the extended signal model, an extended MWF (MWF\textsubscript{ext}) will be derived. This MWF\textsubscript{ext} will be shown to be theoretically equivalent to several expressions, which turn out to be interpretable as specific cascade algorithms. Specifically, the MWF\textsubscript{ext} will be shown to be equivalent to the AEC-NR and NR-AEC, thus generalising \cite{gustafssonCombinedAcousticEcho1998a,jeannesCombinedNoiseEcho2001,docloMultimicrophoneNoiseReduction2003, romboutsIntegratedApproachAcoustic2005a,ruizDistributedCombinedAcoustic2022}. Further, the MWF\textsubscript{ext} will be shown to be equivalent to an algorithm where an extended NR (NR\textsubscript{ext}) precedes the AEC and a post-filter (PF) (NR\textsubscript{ext}-AEC-PF) under the assumption of the echo paths being additive maps, i.e., preserving the addition operation. This algorithm without PF, NR\textsubscript{ext}-AEC, has originally been proposed as a cascade algorithm for combined acoustic feedback cancellation (AFC) and NR \cite{ruizCascadeMultiChannelNoise2022c}, and as a cascade algorithm for combined AEC and NR \cite{roebbenCascadedNoiseReduction2024}. In this paper, the algorithm will be adapted to the general AEC and NR problem, and the PF will be shown to be necessary for MSE optimality. Additionally, under rank-deficiency conditions the MWF\textsubscript{ext} is non-unique, such that it will be shown that under these rank-deficiency conditions the theoretical equivalence of the AEC-NR, NR-AEC and NR\textsubscript{ext}-AEC-PF expressions to the MWF\textsubscript{ext} amounts to the expressions being specific, not necessarily minimum-norm solutions, for this MWF\textsubscript{ext}.

As for the second research question, non-stationarities and imperfect correlation matrix estimation effects will be analysed theoretically, and validated experimentally. The AEC-NR and NR\textsubscript{ext}-AEC-PF attain best overall performance.

In Section \ref{section:cascaded_approach} the cascade approach is reviewed first, discussing the AEC and NR design in isolation. This is contrasted to the integrated approach in Section \ref{section:integrated_approach}. The practical performance differences due to non-stationarities and imperfect correlation matrix estimation are analysed theoretically in Section \ref{section:practial_considerations}. A computational complexity comparison is provided in Section \ref{section:numerical_complexity}. After introducing the simulation setup in Section \ref{section:experimental_procedures}, the practical performances are also experimentally analysed in Section \ref{section:results_and_discussion}. Finally, Section \ref{section:conclusion} draws the conclusions. MATLAB code is available in \cite{roebbenGithubRepositoryIntegrated2024}. 

\FloatBarrier
\section{Cascade approach} \label{section:cascaded_approach}
Section \ref{section:cascaded_approach-AEC} and Section \ref{section:cascaded_approach-NR} describe the design of the isolated AEC and NR algorithms, optimal in the MSE sense for their isolated cost functions. The near-end room noise presence is thus neglected in the design of the AEC and the echo presence in the design of the NR. Section \ref{section:cascaded_approach-AEC_NR} considers the cascade approach for combined AEC and NR.

The signal model is presented in the $z$-domain to accommodate the duality between frequency domain and time domain. The conversion from $z$- to frequency domain is realised by replacing index $z$ with frequency-bin index $f$, and possibly frame index $k$. The conversion from $z$- to time domain is realised by replacing the $z$-domain variables with time-lagged vectors, possibly multiplied with Toeplitz matrices, and replacing index $z$ with time index $t$.

\subsection{Acoustic echo cancellation (AEC)} \label{section:cascaded_approach-AEC}
AEC aims at estimating the desired speech by suppressing the echo originating from the loudspeakers. To this end, the echo signals are estimated from the loudspeaker signals and subtracted from the microphone signals \cite{hanslerTopicsAcousticEcho2006}\cite[Chapter 5]{benestyAdvancesNetworkAcoustic2001}.
\ \\
\subsubsection{Signal model} \label{section:cascaded_approach-AEC-signal_model}
Considering an $M$-microphone/$L$-loudspeaker setup, the microphone signals $m_i(z)$, $i\in\{1,\cdots, M\}$, can be stacked into the microphone signal vector $\Vec{m}(z)\in\mathbb{C}^{M\times 1}$ as
\begin{equation} \label{eq:signal_model_def_x_z_domain}
	\Vec{m}(z) = \begin{bmatrix}
		m_1(z)&
		m_2(z)&
		\cdots&
		m_M(z)
	\end{bmatrix}^\top_{\textstyle \raisebox{2pt}{.}}
\end{equation}
This microphone signal vector can be decomposed into a desired speech signal vector $\Vec{s}(z)\in\mathbb{C}^{M\times 1}$ and an echo signal vector $\Vec{e}(z)\in\mathbb{C}^{M\times 1}$ as 
\begin{equation} \label{eq:signal_model_AEC_s+u}
	\Vec{m}(z) =\Vec{s}(z) + \Vec{e}(z)_{\textstyle \raisebox{2pt}{,}}
\end{equation}
where $\Vec{e}(z)$ originates from the loudspeaker signal vector $\Vec{l}(z)$ 
\begin{equation}
	\Vec{l}(z) = \begin{bmatrix}
		l_1(z)&
		l_2(z)&
		\cdots&
		l_L(z)
	\end{bmatrix}^\top_{\textstyle \raisebox{2pt}{}}
\end{equation}
by means of a map $F(.): \mathbb{C}^{L\times 1}\rightarrow\mathbb{C}^{M\times 1}$, i.e., $\Vec{e}(z)=F(\Vec{l}(z))$. While $F(.)$ can be a general map, it can also be restricted by, e.g., assuming $F(.)$ to be an additive or linear map. If the loudspeaker signal vector $\Vec{l}(z)$ consists of a far-end room speech component $\Vec{l}^s(z)$ and a far-end room noise component $\Vec{l}^n(z)$, an additive map preserves this speech-noise relation as \cite{roebbenCascadedNoiseReduction2024}
\begin{subequations} \label{eq:additive_map}
	\begin{align}
	\Vec{e}(z) &= F(\Vec{l}^s(z) + \Vec{l}^n(z))= F(\Vec{l}^s(z)) + F(\Vec{l}^n(z))\\ & = \Vec{e}^s(z) + \Vec{e}^n(z)_{\textstyle \raisebox{2pt}{,}}
	\end{align}
\end{subequations}
with $\Vec{e}^s(z)$ and $\Vec{e}^n(z)$ the far-end room speech and noise components in the echo respectively. In a linear map,
\begin{subequations}
	\begin{align} \Vec{e}(z)&=F\left(\Vec{l}(z)\right)\\ &=F_{\text{lin}}(z)\Vec{l}(z)_{\textstyle \raisebox{2pt}{,}}
		\end{align}
\end{subequations}
with 
\begin{equation}
	F_{\text{lin}}(z) = \begin{bmatrix}
		f_1^1(z) & \cdots & f_1^L(z)\\
		\vdots & \ddots & \vdots\\
		f_M^1(z) & \cdots & f_M^L(z)
	\end{bmatrix}_{\textstyle \raisebox{0pt}{,}}
\end{equation}
and $f_i^j(z)$ the transfer function of the echo path between the $j$th loudspeaker and the $i$th microphone. A continuous-time linear echo path combined with multi-rate digital signal processing (with aliasing), e.g., corresponds to an additive map rather than a linear map.
\ \\
\subsubsection{Cost function} \label{section:cascaded_approach-AEC-cost_function}
The goal of the AEC is to minimise the MSE between the echo signal for a chosen reference microphone $r\in\{1,\cdots,M\}$ and the filtered loudspeaker signal vector $\Vec{w}_{\text{AEC}}(z)^H\Vec{l}(z)$, with $\Vec{w}_{\text{AEC}}(z)\in\mathbb{C}^{L\times1}$ \cite[Chapter 5]{benestyAdvancesNetworkAcoustic2001}:
\begin{equation} \label{eq:cost_function_AEC}
	\Vec{w}_{\text{AEC}}(z) = \argmin_{\Vec{w}(z)} \mathbb{E}\left\{\norm{e_r(z)-\Vec{w}(z)^H\Vec{l}(z)}^2_2\right\}_{\textstyle \raisebox{2pt}{.}}
\end{equation}

\subsubsection{Solution strategy} \label{section:cascaded_approach-AEC-solution_strategy}
Defining $R_{ll}(z)=\mathbb{E}\left\{\Vec{l}(z)\Vec{l}(z)^H\right\}\in\mathbb{C}^{L\times L}$ and $R_{le}(z)=\mathbb{E}\left\{\Vec{l}(z)\Vec{e}(z)^H\right\}\in\mathbb{C}^{L\times M}$ as the loudspeaker correlation matrix and the loudspeaker-echo cross-correlation matrix respectively, the solution to (\ref{eq:cost_function_AEC}) is given as the solution to the Wiener-Hopf equations \cite[Chapter 5]{benestyAdvancesNetworkAcoustic2001}
\begin{equation} \label{eq:wiener_hopf_AEC}
	R_{ll}(z)\Vec{w}_{\text{AEC}}(z)=R_{{le}}(z)\Vec{t}_{r_{\textstyle \raisebox{2pt}{,}}}
\end{equation}
with $\Vec{t}_r\in\mathbb{C}^{M\times 1}$ a unit
vector with $1$ at position $r$ and $0$ elsewhere.
The solution to (\ref{eq:wiener_hopf_AEC}) is given as
\begin{equation} \label{eq:solution_AEC}
	\Vec{w}_{\text{AEC}}(z) = R_{ll}(z)^g R_{{le}}(z)\Vec{t}_{r_{\textstyle \raisebox{2pt}{,}}}
\end{equation}
with $.^g$ denoting the generalised inverse\footnote{A generalised inverse $.^g$ only needs to satisfy the property $R_{ll}(z)R_{ll}(z)^gR_{ll}(z)=R_{ll}(z)$ \cite{jamesGeneralisedInverse1978}. A pseudo-inverse additionally needs to satisfy the properties $R_{ll}(z)^gR_{ll}(z)R_{ll}(z)^g=R_{ll}(z)^g$, $\left(R_{ll}(z)R_{ll}(z)^g\right)^H=R_{ll}(z)R_{ll}(z)^g$, and $\left(R_{ll}(z)^gR_{ll}(z)\right)^H=R_{ll}(z)^gR_{ll}(z)$ \cite{penroseGeneralizedInverseMatrices1955}.}$^,$\footnote{This generalised inverse is atypical in the literature as commonly the inverse or pseudo-inverse is used, e.g., \cite[Chapter 1]{benestyAdvancesNetworkAcoustic2001}. Nevertheless, the generalised inverse is already introduced for consistency with Section \ref{section:integrated_approach}.} \cite{jamesGeneralisedInverse1978}. This generalised inverse is non-unique if $R_{ll}(z)$ is rank-deficient, although specific choices can be made such as the pseudo-inverse $.^\dagger$ which is uniquely defined, and with which (\ref{eq:solution_AEC}) corresponds to the minimum-norm solution \cite{penroseGeneralizedInverseMatrices1955}. If $R_{ll}(z)$ is full-rank, i.e., the loudspeaker signals are linearly independent, then $.^g$ can be replaced with the inverse $.^{-1}$.

The desired speech estimate $\hat{s}_{r,\text{AEC}}\in\mathbb{C}$ in microphone $r$ is given as \cite[Chapter 5]{benestyAdvancesNetworkAcoustic2001}
\begin{equation} \label{eq:desired_estimate_AEC}
	\hat{s}_{r,\text{AEC}}(z) = \Vec{t}_{r}^H\Vec{m}(z)-\Vec{w}_{{AEC}}(z)^H\Vec{l}(z)_{\textstyle \raisebox{2pt}{.}}
\end{equation}
Depending on possible post-processing of the signals after applying the AEC, this procedure can be repeated for multiple choices of reference microphone.
\subsection{Noise reduction (NR)} \label{section:cascaded_approach-NR}
NR aims at estimating the desired speech by suppressing the near-end room noise through filtering the microphone signals \cite{benestyStudyWienerFilter2005}.
\ \\
\subsubsection{Signal model} \label{section:cascaded_approach-NR-signal_model}
Considering an $M$-microphone setup, the microphone signal vector $\Vec{m}(z)\in\mathbb{C}^{M\times 1}$ consists of a desired speech signal vector $\Vec{s}(z)\in\mathbb{C}^{M\times 1}$ and a near-end room noise signal vector $\Vec{n}(z)\in\mathbb{C}^{M\times 1}$
\begin{equation} \label{eq:signal_model_NR}
	\Vec{m}(z) = \Vec{s}(z) + \Vec{n}(z)_{\textstyle \raisebox{2pt}{.}}
\end{equation}
\subsubsection{Cost function} \label{section:cascaded_approach-NR-cost_function}
The goal of the NR is to minimise the MSE between the desired speech signal for a chosen reference microphone $r\in\{1,\cdots,M\}$ and the filtered microphone signal vector $\Vec{w}_{\text{NR}}(z)^H\Vec{m}(z)$, with $\Vec{w}_{\text{NR}}(z)\in\mathbb{C}^{M\times1}$ \cite{benestyStudyWienerFilter2005,serizelLowrankApproximationBased2014}:
\begin{equation} \label{eq:cost_function_NR}
	\Vec{w}_{\text{NR}}(z) = \argmin_{\Vec{w}(z)} \mathbb{E}\left\{\norm{s_r(z)-\Vec{w}(z)^H\Vec{m}(z)}_2^2\right\}_{\textstyle \raisebox{2pt}{.}}
\end{equation}
\subsubsection{Solution strategy} \label{section:cascaded_approach-NR-solution_strategy}
Defining $R_{mm}(z) = \mathbb{E}\left\{\Vec{m}(z)\Vec{m}(z)^H\right\}\in\mathbb{C}^{M\times M}$ and $R_{ss}(z) = \mathbb{E}\left\{\Vec{s}(z)\Vec{s}(z)^H\right\}\in\mathbb{C}^{M\times M}$ as the microphone correlation matrix and the desired speech correlation matrix respectively, and assuming $\Vec{s}(z)$ and $\Vec{n}(z)$ to be uncorrelated, the solution to (\ref{eq:cost_function_NR}) is given as \cite{benestyStudyWienerFilter2005}
\begin{equation} \label{eq:solution_NR}
	\Vec{w}_{\text{NR}}(z) = R_{mm}(z)^{g}R_{ss}(z)\Vec{t}_{r_{\textstyle \raisebox{2pt}{.}}}
\end{equation}
If $R_{mm}(z)$ is rank-deficient, then $.^g$ can again be chosen as $.^\dagger$, such that (\ref{eq:solution_NR}) corresponds to the minimum-norm solution. Indeed, $R_{mm}(z)$ can be rank-deficient, e.g., modelling $<M$ localised noise sources. If $R_{mm}(z)$ is full-rank, then $.^g$ can be replaced with $.^{-1}$. The desired speech estimate $\hat{s}_{r,\text{NR}}(z)\in\mathbb{C}$ in microphone $r$ is given as \cite{benestyStudyWienerFilter2005}
\begin{equation} \label{eq:desired_estimate_NR}
	\hat{s}_{r,\text{NR}}(z) = \Vec{w}_{\text{NR}}(z)^H\Vec{m}(z)_{\textstyle \raisebox{2pt}{.}}
\end{equation}
Depending on possible post-processing of the signals after applying the NR, this procedure can be repeated for multiple choices of reference microphone.

\subsection{Combined AEC and NR} \label{section:cascaded_approach-AEC_NR}
To resolve the combined AEC and NR problem, both algorithms are cascaded one after the other, where the AEC precedes the NR (AEC-NR) \cite{gustafssonCombinedAcousticEcho1998a,cohenJointBeamformingEcho2018,luisvaleroLowComplexityMultiMicrophoneAcoustic2019a,Albu_CombinedEchoNoise2004}, the NR precedes the AEC (NR-AEC) \cite{martinAcousticEchoCancellation1997,docloCombinedAcousticEcho2000,schrammenChangePredictionLow2019,roebbenCascadedNoiseReduction2024} or variations thereof are considered \cite{jeannesCombinedNoiseEcho2001,herbordtAcousticHumanMachineFrontEnd2003,burtonNewStructureCombining2007}. As such, the cascade approach may suffer from algorithmic conflicts as the interaction between the isolated algorithms is not taken into account. Indeed, (residual) echo and near-end room noise may leak between the different stages of the cascade algorithms, thereby reducing performance.

\section{Integrated approach} \label{section:integrated_approach}
In this section, the integrated approach is applied by selecting a signal model, including both the near-end room noise and echo, of either the microphone signal vector only or of an extended signal vector obtained by stacking microphone and loudspeaker signals (Section \ref{section:integrated_approach-signal_model}), formulating an MSE cost function (Section \ref{section:integrated_approach-cost_function}), and using a common solution strategy (Section \ref{section:integrated_approach-solution_strategy}). Using the microphone signal model, an MWF will be obtained, whereas using the extended signal model an MWF\textsubscript{ext} will be obtained. Several theoretically equivalent expressions to the MWF\textsubscript{ext} are derived, which turn out to be nevertheless interpretable as specific cascade algorithms, as illustrated in Fig. \ref{fig:schematics_overview}. Specifically, the MWF\textsubscript{ext} is shown to be theoretically equivalent to algorithms where the AEC precedes the NR (AEC-NR), the NR precedes the AEC (NR-AEC), and the extended NR (NR\textsubscript{ext}) precedes the AEC and a post-filter (PF) (NR\textsubscript{ext}-AEC-PF). Under rank-deficiency conditions the MWF\textsubscript{ext} is non-unique, such that this theoretical equivalence amounts to the expressions being specific, not necessarily minimum-norm solutions, for this MWF\textsubscript{ext}. For conciseness, the $z$-indices will be omitted from now on.

\begin{figure}
	\centering
	\subfloat[MWF]{
		
\scalebox{0.38}{
%
%
\begin{tikzpicture}
\node[draw=none] (t2) {};
\node[draw=none,below of=t2,node distance=5cm] (t3) {};
\node [draw=none,rotate=90,right of=t3,node distance=0.4cm,draw=none] (sp1) {};

\node[draw=none,below of=sp1,node distance=-1cm] (t4) {};

\node [draw=none,rotate=90,right of=sp1,node distance=1.8cm,draw=none] (sp2) {};

\node[draw=none,minimum size=0cm,above of=sp1,node distance=4.8cm,draw=none] (temp) {};
\node[draw=black,fill=gray!40,circle,minimum size=0.6cm,right of=temp,node distance=-0.18cm] (mic1) {};
\draw ([yshift=8pt]mic1.west) -- ([yshift=-8pt]mic1.west);		

\node[draw=none,below of=mic1,node distance=0.8cm,draw=none] (t1) {};
\node[draw=none,right of=t1,node distance=0cm] {\huge$\vdots$};

\node[draw=black,fill=gray!40,circle,minimum size=0.6cm,below of=t1,node distance=1cm] (mic2) {};
\draw ([yshift=8pt]mic2.west) -- ([yshift=-8pt]mic2.west);

\node[draw=none,below of=t4,minimum size=0cm,node distance=-1.3cm] (t5) {};
\node[draw=none,right of=t5,minimum width=2.3cm,node distance=2.1cm,minimum height=5.6cm] (b11) {};
	\node[draw,above of=b11,minimum width=2.7cm,node distance=1.65cm,minimum height=2.3cm] (b1) {\huge MWF};

\node[draw=none,right of=b1,minimum size=0cm,node distance=2cm] (out) {\huge $\hat{s}_r$};
\draw [->,-{Stealth[length=1mm, width=1mm]}] (b1) -- (out);

\node[draw=none,right of=mic1,node distance=0.49cm] (t6) {};
\draw [->,-{Stealth[length=1mm, width=1mm]}] (t6) -- (t6-|b1.west);
\node[draw=none,right of=mic2,node distance=0.49cm] (t7) {};
\draw [->,-{Stealth[length=1mm, width=1mm]}] (t7) -- (t7-|b1.west);			
\node[draw=none,below of=t1,node distance=0.12cm] (t8) {};
\node[draw=none,right of=t8,node distance=0.49cm] (t9) {};
\draw [->,-{Stealth[length=1mm, width=1mm]}] (t9) -- (t9-|b1.west);	

\node[draw=none,right of=sp1,node distance=0.4cm] (t10) {};
\node[draw=none,right of=sp2,node distance=0.4cm] (t11) {};
\node[draw=none,below of=t4,node distance=0.1cm] (t12) {};
\node[draw=none,right of=t12,node distance=0.37cm] (t13) {};

\draw[decorate,decoration={brace,amplitude=5pt}]
(-0.7,-2) -- (-0.7,0.6);
\node[draw=none,left of=mic1,node distance=1.1cm] (t14) {};
\node[draw=none,below of=t14,node distance=0.9cm] (x1) {\huge $\Vec{m}$};%

\draw[decorate,decoration={brace,amplitude=5pt},draw=none]
(-1.3,-5.3) -- (-1.3,0.7);
\node[draw=none,left of=mic1,node distance=1.5cm] (t15) {};
\end{tikzpicture}
} 

		\label{fig:MWF_overview}
	}\vspace{-0.3em}
	\subfloat[MWF\textsubscript{ext}]{
		
\scalebox{0.38}{
%
%
\begin{tikzpicture}
	\node[draw=none] (t2) {};
	\node[draw=none,below of=t2,node distance=5cm] (t3) {};
	\node[draw=none,right of=t3,node distance=-0.3cm] (t6) {};
	\pic [rotate=270,local bounding box=sp1,right of=t6,node distance=0.4cm] {Speaker};
	
	\node[draw=none,below of=sp1,node distance=-1cm] (t4) {};
	\node[draw=none,right of=t4,node distance=0cm] {\huge$\vdots$};
	
	\pic [rotate=270,right of=sp1,node distance=-2cm,local bounding box=sp2] {Speaker};
	
	\node[draw=none,minimum size=0cm,above of=sp1,node distance=4.8cm] (t8) {};
	\node[draw=none,minimum size=0cm,left of=t8,node distance=-0.3cm] (temp) {};
	\node[draw=black,fill=gray!40,circle,minimum size=0.6cm,right of=temp,node distance=-0.18cm] (mic1) {};
	\draw ([yshift=8pt]mic1.west) -- ([yshift=-8pt]mic1.west);		
	
	\node[draw=none,below of=mic1,node distance=0.8cm] (t1) {};
	\node[draw=none,right of=t1,node distance=0cm] {\huge$\vdots$};
	
	\node[draw=black,fill=gray!40,circle,minimum size=0.6cm,below of=t1,node distance=1cm] (mic2) {};
	\draw ([yshift=8pt]mic2.west) -- ([yshift=-8pt]mic2.west);
	
	\node[draw=none,below of=t4,minimum size=0cm,node distance=-1.3cm] (t5) {};
	\node[draw=black,right of=t5,minimum width=2.7cm,node distance=2.1cm,minimum height=5.6cm] (b1) {\huge MWF\textsubscript{ext}};
	
	\node[draw=none,right of=b1,minimum size=0cm,node distance=2cm] (out) {\huge $\hat{s}_r$};
	\draw [->,-{Stealth[length=1mm, width=1mm]}] (b1) -- (out);
	
	\node[draw=none,right of=mic1,node distance=0.30cm] (t6) {};
	\draw [->,-{Stealth[length=1mm, width=1mm]}] (t6) -- (t6-|b1.west);
	\node[draw=none,right of=mic2,node distance=0.30cm] (t7) {};
	\draw [->,-{Stealth[length=1mm, width=1mm]}] (t7) -- (t7-|b1.west);			
	\node[draw=none,below of=t1,node distance=0.12cm] (t8) {};
	\node[draw=none,right of=t8,node distance=0.30cm] (t9) {};
	\draw [->,-{Stealth[length=1mm, width=1mm]}] (t9) -- (t9-|b1.west);	
	
	\node[draw=none,right of=sp1,node distance=0.4cm] (t10) {};
	\draw [->,-{Stealth[length=1mm, width=1mm]}] (t10) -- (t10-|b1.west);	
	\node[draw=none,right of=sp2,node distance=0.4cm] (t11) {};
	\draw [->,-{Stealth[length=1mm, width=1mm]}] (t11) -- (t11-|b1.west);	
	\node[draw=none,below of=t4,node distance=0.1cm] (t12) {};
	\node[draw=none,right of=t12,node distance=0.37cm] (t13) {};
	\draw [->,-{Stealth[length=1mm, width=1mm]}] (t13) -- (t13-|b1.west);	
	
	\draw[decorate,decoration={brace,amplitude=5pt}]
	(-0.7,-2.8) -- (-0.7,-0.2);
	\node[draw=none,left of=mic1,node distance=1.1cm] (t14) {};
	\node[draw=none,below of=t14,node distance=0.9cm] (x1) {\huge $\Vec{m}$};%
	
	\draw[decorate,decoration={brace,amplitude=5pt}]
	(-0.7,-5.9) -- (-0.7,-2.9);
	\node[draw=none,left of=mic1,node distance=1cm] (t14-2) {};
	\node[draw=none,below of=t14-2,node distance=3.8cm]{\huge $\Vec{l}$};%
	
	\draw[decorate,decoration={brace,amplitude=5pt}]
	(-1.5,-6) -- (-1.5,0);
	\node[draw=none,left of=mic1,node distance=1.9cm] (t15) {};
	\node[draw=none,below of=t15,node distance=2.3cm] {\huge $\tilde{\Vec{m}}$};%
\end{tikzpicture}
} 

		\label{fig:MWFext_overview}
	}\vspace{-0.3em}
	\subfloat[AEC-NR]{
		
\scalebox{0.38}{
%
	\begin{tikzpicture}
		\centering
	\node[draw=none] (t2) {};
	\node[draw=none,below of=t2,node distance=5cm] (t3) {};
	\node[draw=none,right of=t3,node distance=-0.3cm] (t6) {};
	\pic [rotate=270,local bounding box=sp1,right of=t6,node distance=0.4cm] {Speaker};
	
	\node[draw=none,below of=sp1,node distance=-1cm] (t4) {};
	\node[draw=none,right of=t4,node distance=0cm] {\huge$\vdots$};
	
	\pic [rotate=270,right of=sp1,node distance=-2cm,local bounding box=sp2] {Speaker};
	
	\node[draw=none,minimum size=0cm,above of=sp1,node distance=4.8cm] (t8) {};
	\node[draw=none,minimum size=0cm,left of=t8,node distance=-0.3cm] (temp) {};
	\node[draw=black,fill=gray!40,circle,minimum size=0.6cm,right of=temp,node distance=-0.18cm] (mic1) {};
	\draw ([yshift=8pt]mic1.west) -- ([yshift=-8pt]mic1.west);		
	
	\node[draw=none,below of=mic1,node distance=0.8cm] (t1) {};
	\node[draw=none,right of=t1,node distance=0cm] {\huge$\vdots$};
	
	\node[draw=black,fill=gray!40,circle,minimum size=0.6cm,below of=t1,node distance=1cm] (mic2) {};
	\draw ([yshift=8pt]mic2.west) -- ([yshift=-8pt]mic2.west);
	
	\node[draw=none,below of=t4,minimum size=0cm,node distance=-1.3cm] (t5) {};
	\node[draw=none,right of=t5,minimum width=2.3cm,node distance=2.1cm,minimum height=5.6cm] (b11) {};
	\node[draw,above of=b11,minimum width=2.7cm,node distance=0cm,minimum height=5.6cm] (b1) {\huge AEC};
	
	\node[draw=none,below of=t4,minimum size=0cm,node distance=-2.95cm] (r5) {};
	\node[draw=none,right of=r5,minimum width=0cm,node distance=2.1cm,minimum height=0cm] (rb11) {};	
	\node[draw=black,right of=rb11,minimum width=2.7cm,node distance=3.1cm,minimum height=2.3cm] (b2) {\huge NR};

	\node[draw=none,right of=mic1,node distance=0.30cm] (t6) {};
	\draw [->,-{Stealth[length=1mm, width=1mm]}] (t6) -- (t6-|b1.west);
	\node[draw=none,right of=mic2,node distance=0.30cm] (t7) {};
	\draw [->,-{Stealth[length=1mm, width=1mm]}] (t7) -- (t7-|b1.west);			
	\node[draw=none,below of=t1,node distance=0.12cm] (t8) {};
	\node[draw=none,right of=t8,node distance=0.30cm] (t9) {};
	\draw [->,-{Stealth[length=1mm, width=1mm]}] (t9) -- (t9-|b1.west);	
	
	\node[draw=none,right of=sp1,node distance=0.4cm] (t10) {};
	\draw [->,-{Stealth[length=1mm, width=1mm]}] (t10) -- (t10-|b1.west);	
	\node[draw=none,right of=sp2,node distance=0.4cm] (t11) {};
	\draw [->,-{Stealth[length=1mm, width=1mm]}] (t11) -- (t11-|b1.west);	
	\node[draw=none,below of=t4,node distance=0.1cm] (t12) {};
	\node[draw=none,right of=t12,node distance=0.37cm] (t13) {};
	\draw [->,-{Stealth[length=1mm, width=1mm]}] (t13) -- (t13-|b1.west);	
	
	\node[draw=none,right of=mic1,node distance=3.2cm] (t14) {};
	\draw [->,-{Stealth[length=1mm, width=1mm]}] (t14) -- (t14-|b2.west);
	\node[draw=none,right of=mic2,node distance=3.2cm] (t15) {};
	\draw [->,-{Stealth[length=1mm, width=1mm]}] (t15) -- (t15-|b2.west);
	\node[draw=none,right of=t1,node distance=3.2cm] (t18) {};
	\draw [->,-{Stealth[length=1mm, width=1mm]}] (t18) -- (t18-|b2.west);
	
	
	\node[draw=none,right of=b2,minimum size=0cm,node distance=2cm] (out) {\huge $\hat{s}_r$};
	\draw [->,-{Stealth[length=1mm, width=1mm]}] (b2) -- (out);
\end{tikzpicture}
} 

		\label{fig:AEC-NR_overview}
	}\vspace{-0.3em}
	\subfloat[NR-AEC]{
		
\scalebox{0.38}{
%
%
\begin{tikzpicture}
	\centering
	\node[draw=none] (t2) {};
	\node[draw=none,below of=t2,node distance=5cm] (t3) {};
	\node[draw=none,right of=t3,node distance=-0.3cm] (t6) {};
	\pic [rotate=270,local bounding box=sp1,right of=t6,node distance=0.4cm] {Speaker};
	
	\node[draw=none,below of=sp1,node distance=-1cm] (t4) {};
	\node[draw=none,right of=t4,node distance=0cm] {\huge$\vdots$};
	
	\pic [rotate=270,right of=sp1,node distance=-2cm,local bounding box=sp2] {Speaker};
	
	\node[draw=none,minimum size=0cm,above of=sp1,node distance=4.8cm] (t8) {};
	\node[draw=none,minimum size=0cm,left of=t8,node distance=-0.3cm] (temp) {};
	\node[draw=black,fill=gray!40,circle,minimum size=0.6cm,right of=temp,node distance=-0.18cm] (mic1) {};
	\draw ([yshift=8pt]mic1.west) -- ([yshift=-8pt]mic1.west);		
	
	\node[draw=none,below of=mic1,node distance=0.8cm] (t1) {};
	\node[draw=none,right of=t1,node distance=0cm] {\huge$\vdots$};
	
	\node[draw=black,fill=gray!40,circle,minimum size=0.6cm,below of=t1,node distance=1cm] (mic2) {};
	\draw ([yshift=8pt]mic2.west) -- ([yshift=-8pt]mic2.west);
	
	\node[draw=none,below of=t4,minimum size=0cm,node distance=-1.3cm] (t5) {};
	\node[draw=none,right of=t5,minimum width=2.3cm,node distance=2.1cm,minimum height=5.6cm] (b11) {};
	\node[draw,above of=b11,minimum width=2.7cm,node distance=1.65cm,minimum height=2.3cm] (b1) {\huge NR};
	
	\node[draw=black,right of=b11,minimum width=2.7cm,node distance=3.1cm,minimum height=5.6cm] (b2) {\huge AEC};
	
	\node[draw=none,right of=mic1,node distance=0.30cm] (t6) {};
	\draw [->,-{Stealth[length=1mm, width=1mm]}] (t6) -- (t6-|b1.west);
	\node[draw=none,right of=mic2,node distance=0.30cm] (t7) {};
	\draw [->,-{Stealth[length=1mm, width=1mm]}] (t7) -- (t7-|b1.west);			
	\node[draw=none,below of=t1,node distance=0.12cm] (t8) {};
	\node[draw=none,right of=t8,node distance=0.30cm] (t9) {};
	\draw [->,-{Stealth[length=1mm, width=1mm]}] (t9) -- (t9-|b1.west);	
	
	\node[draw=none,right of=sp1,node distance=0.4cm] (t10) {};
	\draw [->,-{Stealth[length=1mm, width=1mm]}] (t10) -- (t10-|b2.west);	
	\node[draw=none,right of=sp2,node distance=0.4cm] (t11) {};
	\draw [->,-{Stealth[length=1mm, width=1mm]}] (t11) -- (t11-|b2.west);	
	\node[draw=none,below of=t4,node distance=0.1cm] (t12) {};
	\node[draw=none,right of=t12,node distance=0.37cm] (t13) {};
	\draw [->,-{Stealth[length=1mm, width=1mm]}] (t13) -- (t13-|b2.west);	
	
	\node[draw=none,right of=mic1,node distance=3.2cm] (t14) {};
	\draw [->,-{Stealth[length=1mm, width=1mm]}] (t14) -- (t14-|b2.west);
	\node[draw=none,right of=mic2,node distance=3.2cm] (t15) {};
	\draw [->,-{Stealth[length=1mm, width=1mm]}] (t15) -- (t15-|b2.west);
	\node[draw=none,right of=t1,node distance=3.2cm] (t18) {};
	\draw [->,-{Stealth[length=1mm, width=1mm]}] (t18) -- (t18-|b2.west);
	
	
	\node[draw=none,right of=b2,minimum size=0cm,node distance=2cm] (out) {\huge $\hat{s}_r$};
	\draw [->,-{Stealth[length=1mm, width=1mm]}] (b2) -- (out);
	
\end{tikzpicture}
} 

		\label{fig:NR-AEC_overview}
	}\vspace{-0.3em}
	\subfloat[NR\textsubscript{ext}-AEC-PF]{
		
\scalebox{0.38}{
%
%
\begin{tikzpicture}
	\centering
	\node[draw=none] (t2) {};
	\node[draw=none,below of=t2,node distance=5cm] (t3) {};
	\node[draw=none,right of=t3,node distance=-0.3cm] (t6) {};
	\pic [rotate=270,local bounding box=sp1,right of=t6,node distance=0.4cm] {Speaker};
	
	\node[draw=none,below of=sp1,node distance=-1cm] (t4) {};
	\node[draw=none,right of=t4,node distance=0cm] {\huge$\vdots$};
	
	\pic [rotate=270,right of=sp1,node distance=-2cm,local bounding box=sp2] {Speaker};
	
	\node[draw=none,minimum size=0cm,above of=sp1,node distance=4.8cm] (t8) {};
	\node[draw=none,minimum size=0cm,left of=t8,node distance=-0.3cm] (temp) {};
	\node[draw=black,fill=gray!40,circle,minimum size=0.6cm,right of=temp,node distance=-0.18cm] (mic1) {};
	\draw ([yshift=8pt]mic1.west) -- ([yshift=-8pt]mic1.west);		
	
	\node[draw=none,below of=mic1,node distance=0.8cm] (t1) {};
	\node[draw=none,right of=t1,node distance=0cm] {\huge$\vdots$};
	
	\node[draw=black,fill=gray!40,circle,minimum size=0.6cm,below of=t1,node distance=1cm] (mic2) {};
	\draw ([yshift=8pt]mic2.west) -- ([yshift=-8pt]mic2.west);
	
	\node[draw=none,below of=t4,minimum size=0cm,node distance=-1.3cm] (t5) {};
	\node[draw=black,right of=t5,minimum width=2.7cm,node distance=2.1cm,minimum height=5.6cm] (b1) {\huge NR\textsubscript{ext}};
	
	\node[draw=black,right of=b1,minimum width=2.7cm,node distance=3.1cm,minimum height=5.6cm] (b2) {\huge AEC};
	
	\node[draw=none,right of=b2,minimum width=2.6cm,node distance=3.1cm,minimum height=5.6cm] (b33) {};
	\node[draw,above of=b33,minimum width=2.7cm,node distance=1.65cm,minimum height=2.3cm] (b3) {\huge PF};
	
	\node[draw=none,right of=mic1,node distance=0.30cm] (t6) {};
	\draw [->,-{Stealth[length=1mm, width=1mm]}] (t6) -- (t6-|b1.west);
	\node[draw=none,right of=mic2,node distance=0.30cm] (t7) {};
	\draw [->,-{Stealth[length=1mm, width=1mm]}] (t7) -- (t7-|b1.west);			
	\node[draw=none,below of=t1,node distance=0.12cm] (t8) {};
	\node[draw=none,right of=t8,node distance=0.30cm] (t9) {};
	\draw [->,-{Stealth[length=1mm, width=1mm]}] (t9) -- (t9-|b1.west);	
	
	\node[draw=none,right of=sp1,node distance=0.4cm] (t10) {};
	\draw [->,-{Stealth[length=1mm, width=1mm]}] (t10) -- (t10-|b1.west);	
	\node[draw=none,right of=sp2,node distance=0.4cm] (t11) {};
	\draw [->,-{Stealth[length=1mm, width=1mm]}] (t11) -- (t11-|b1.west);	
	\node[draw=none,below of=t4,node distance=0.1cm] (t12) {};
	\node[draw=none,right of=t12,node distance=0.37cm] (t13) {};
	\draw [->,-{Stealth[length=1mm, width=1mm]}] (t13) -- (t13-|b1.west);	
	
	\node[draw=none,right of=mic1,node distance=3.21cm] (t14) {};
	\draw [->,-{Stealth[length=1mm, width=1mm]}] (t14) -- (t14-|b2.west);
	\node[draw=none,right of=mic2,node distance=3.21cm] (t15) {};
	\draw [->,-{Stealth[length=1mm, width=1mm]}] (t15) -- (t15-|b2.west);
	\node[draw=none,right of=sp1,node distance=3.33cm] (t16) {};
	\draw [->,-{Stealth[length=1mm, width=1mm]}] (t16) -- (t16-|b2.west);
	\node[draw=none,right of=sp2,node distance=3.27cm] (t17) {};
	\draw [->,-{Stealth[length=1mm, width=1mm]}] (t17) -- (t17-|b2.west);
	\node[draw=none,right of=t1,node distance=3.21cm] (t18) {};
	\draw [->,-{Stealth[length=1mm, width=1mm]}] (t18) -- (t18-|b2.west);
	\node[draw=none,right of=t4,node distance=3.33cm] (t19) {};
	\draw [->,-{Stealth[length=1mm, width=1mm]}] (t19) -- (t19-|b2.west);
	
	\node[draw=none,right of=mic1,node distance=6.3cm] (t20) {};
	\draw [->,-{Stealth[length=1mm, width=1mm]}] (t20) -- (t20-|b3.west);
	\node[draw=none,right of=mic2,node distance=6.3cm] (t21) {};
	\draw [->,-{Stealth[length=1mm, width=1mm]}] (t21) -- (t21-|b3.west);
	\node[draw=none,right of=t1,node distance=6.3cm] (t22) {};
	\draw [->,-{Stealth[length=1mm, width=1mm]}] (t22) -- (t22-|b3.west);
	
	\node[draw=none,right of=b3,minimum size=0cm,node distance=2cm] (out) {\huge $\hat{s}_r$};
	\draw [->,-{Stealth[length=1mm, width=1mm]}] (b3) -- (out);		
	
%
%
	
\end{tikzpicture}
} 

		\label{fig:NRext-AEC_overview}
	}
	\caption{Cascade interpretation of the integrated algorithms. (a) The MWF is obtained using the microphone signal model. (b) The MWF\textsubscript{ext} is obtained using the extended signal model, which can be shown to be theoretically equivalent to (c) the AEC-NR, (d) the NR-AEC, and (e) the NR\textsubscript{ext}-AEC-PF.}
	\label{fig:schematics_overview}
\end{figure}

\subsection{Signal model} \label{section:integrated_approach-signal_model}
The microphone signal vector $\Vec{m}\in\mathbb{C}^{M\times 1}$ is defined as a mixture of a desired speech signal vector $\Vec{s}\in\mathbb{C}^{M\times1}$, an echo signal vector $\Vec{e}\in\mathbb{C}^{M\times1}$ and a near-end room noise signal vector $\Vec{n}\in\mathbb{C}^{M\times1}$: 
\begin{equation} \label{eq:signal_model_def_x}
	\begin{IEEEeqnarraybox}{rll}
		\Vec{m} &= \Vec{s} + \Vec{e} + \Vec{n}_{\textstyle \raisebox{2pt}{.}}
	\end{IEEEeqnarraybox}
\end{equation}
This signal model (\ref{eq:signal_model_def_x}) will be used to derive the MWF in Section \ref{section:integrated_approach-solution_strategy-MWF}.

An extended signal model can be constructed as well, stacking the microphone signal vector $\Vec{m}$ and loudspeaker signal vector $\Vec{l}$ into $\tilde{\Vec{m}}\in\mathbb{C}^{(M+L)\times 1}$ as
\begin{subequations} \label{eq:signal_model_def_x_tilde}
	\begin{align}
	\tilde{\Vec{m}} &= \begin{bmatrix} \Vec{m}^\top & \Vec{l}^\top \end{bmatrix}^\top\\
	&= \tilde{\Vec{s}} + \tilde{\Vec{e}}+\tilde{\Vec{n}}\\
	&= \begin{bmatrix} \Vec{s} \\ \Vec{0}_{L\times1} \end{bmatrix}+\begin{bmatrix} \Vec{e} \\ \Vec{l} \end{bmatrix}+\begin{bmatrix} \Vec{n} \\ \Vec{0}_{L\times1} \end{bmatrix}_{\textstyle \raisebox{2pt}{,}}
\end{align}
\end{subequations}	
Herein, $\Vec{0}_{L\times 1}\in\mathbb{C}^{L\times 1}$ corresponds to a vector containing only zero elements. This signal model (\ref{eq:signal_model_def_x_tilde}) will be used to derive the MWF and MWF\textsubscript{ext} in Section \ref{section:integrated_approach-solution_strategy-MWF} and Section \ref{section:integrated_approach-solution_strategy-MWFext} respectively, and the AEC-NR and NR-AEC in Section \ref{section:integrated_approach-solution_strategy-AEC_NR} and Section \ref{section:integrated_approach-solution_strategy-NR_AEC} respectively.

If the echo path is an additive map as well, $\Vec{e}$ can be decomposed into the sum of far-end room speech and noise components, i.e., $\Vec{e}^s$ and $\Vec{e}^n$, originating from $\Vec{l}^s$ and $\Vec{l}^n$ respectively, such that (\ref{eq:signal_model_def_x}) and (\ref{eq:signal_model_def_x_tilde}) can be decomposed into
\begin{equation} \label{eq:signal_model_def_x_us+un}
	\Vec{m} =\Vec{s} + \Vec{e}^s + \Vec{e}^n + \Vec{n}_{\textstyle \raisebox{2pt}{,}}
\end{equation}
and
\begin{subequations} \label{eq:signal_model_def_x_map}
	\begin{align} 
		\tilde{\Vec{m}} 
		&= \tilde{\Vec{s}} + \tilde{\Vec{e}}^s + \tilde{\Vec{e}}^n + \tilde{\Vec{n}}\\&=\begin{bmatrix} \Vec{s} \\ \Vec{0}_{L\times1} \end{bmatrix}+\begin{bmatrix} \Vec{e}^s \\ \Vec{l}^s \end{bmatrix}+\begin{bmatrix} \Vec{e}^n \\ \Vec{l}^n \end{bmatrix}+\begin{bmatrix} \Vec{n} \\ \Vec{0}_{L\times1} \end{bmatrix}_{\textstyle \raisebox{2pt}{.}}
		\end{align}
\end{subequations}	
The signal model (\ref{eq:signal_model_def_x_tilde}) will be used to derive the NR\textsubscript{ext}-AEC-PF in Section \ref{section:integrated_approach-solution_strategy-NRext_AEC}.

As $\Vec{s}$ and $\Vec{e}^s$ represent speech signal vectors, they can be assumed to attain an on-off behaviour. To detect these on-off periods, voice activity detectors (VADs) are assumed to be available, i.e., $\text{VAD}_{s}$ and $\text{VAD}_{e^s}$ differentiate between on-off periods in $\Vec{s}$ and $\Vec{e}^s$ respectively. As $\Vec{n}$ and $\Vec{e}^n$, on the other hand, represent noise signal vectors, they can be assumed to be always-on. Thus, the following regimes can be defined:
\begin{itemize}
	\item $\text{VAD}_{s}=1$, $\text{VAD}_{e^s}=1$: $\Vec{m}=\Vec{s}+\Vec{e}+\Vec{n}$ recorded,
	\item $\text{VAD}_{s}=1$, $\text{VAD}_{e^s}=0$: $\Vec{s}+\Vec{e}^n+\Vec{n}$ recorded,
	\item $\text{VAD}_{s}=0$, $\text{VAD}_{e^s}=1$: $\Vec{e}+\Vec{n}$ recorded,
	\item $\text{VAD}_{s}=0$, $\text{VAD}_{e^s}=0$: $\Vec{e}^n + \Vec{n}$ recorded.
\end{itemize}
For general non-additive echo path maps, $\text{VAD}_{e^s}$ then does not detect on-off periods of $\Vec{e}^s$ itself as possibly only non-linear combinations of the loudspeaker signals can be observed, but $\text{VAD}_{e^s}$ then still discriminates between activity of all components in the echo signal and activity of only partial components. In the literature (e.g., \cite{romboutsIntegratedApproachAcoustic2005a,maruoOptimalSolutionsBeamformer2011b}), the echo is often assumed to only contain a stationary always-on signal, which corresponds to only keeping $\Vec{e}^n$ and discarding $\text{VAD}_{e^s}$. 

Further, the following assumptions are made:
\begin{itemize}
	\item $\Vec{s}$, $\Vec{n}$, $\Vec{e}^s$ and $\Vec{e}^n$ are uncorrelated, $\Vec{l}^s$ and $\Vec{l}^n$ are uncorrelated, and $\Vec{l}^s$ and $\Vec{l}^n$ are uncorrelated with $\Vec{s}$ and $\Vec{n}$.
	\item For the theoretical derivation all (speech) signals will be assumed stationary during their on periods. However, in Section \ref{section:practial_considerations}, this assumption will be relaxed by considering the influence of non-stationary signal vectors. Indeed, in practice, $\Vec{s}$ (and $\Vec{e}^s$) are non-stationary, whereas $\Vec{n}$ (and $\Vec{e}^n$) are nevertheless considered to be stationary.
\end{itemize}

The general $M$-microphone/$L$-loudspeaker setup is considered, and no assumptions are made regarding the rank of $R_{ss}$, $R_{nn}$, $R_{mm}$ or $R_{ll}$, thus possibly modelling localised desired speech and near-end room noise sources and linearly related loudspeaker and microphone signals. 

\subsection{Cost function} \label{section:integrated_approach-cost_function}
The goal of the integrated AEC and NR algorithm is to minimise the MSE between the desired speech signal ${s}_r\in\mathbb{C}$ and a filtered signal vector, simultaneously suppressing the echo signal vector $\Vec{e}$ and near-end room noise signal vector $\Vec{n}$, by either considering the microphone signal model (\ref{eq:signal_model_def_x}), i.e.,
\begin{equation} \label{eq:MWF_cost}
	\Vec{w}_{\text{int}} = \argmin_{\Vec{w}} \mathbb{E}\left\{\norm{s_r-\Vec{w}^H\Vec{m}}_2^2\right\}_{\textstyle \raisebox{2pt}{,}}
\end{equation}
to design $\Vec{w}_{\text{int}}\in\mathbb{C}^{M\times1}$, or the extended signal model (\ref{eq:signal_model_def_x_tilde}), i.e.,
\begin{equation} \label{eq:MWFext_cost}
	\begin{IEEEeqnarraybox}{rll}
		\tilde{\Vec{w}}_{\text{int}} &= \argmin_{\tilde{\Vec{w}}} \mathbb{E}\left\{\norm{s_r-\tilde{\Vec{w}}^H\tilde{\Vec{m}}}_2^2\right\}_{\textstyle \raisebox{2pt}{,}}
	\end{IEEEeqnarraybox}
\end{equation}
to design $\tilde{\Vec{w}}_{\text{int}}\in\mathbb{C}^{\left(M+L\right)\times1}$. As illustrated in Fig. \ref{fig:signal_model}, the loudspeaker signals are only filtered by $\tilde{\Vec{w}}_{\text{int}}$ when estimating $s_r$, thus without affecting the playback in the near-end room. 

\subsection{Solution strategy} \label{section:integrated_approach-solution_strategy}
\subsubsection{Multi-channel Wiener filter (MWF)} \label{section:integrated_approach-solution_strategy-MWF}
Using the microphone signal model (\ref{eq:signal_model_def_x}), the solution to (\ref{eq:MWF_cost}) is obtained as 
\begin{equation} \label{eq:MWF}
	\boxed{\begin{IEEEeqnarraybox}{rll}
			\Vec{w}_{\text{int}} &= R_{{m}{m}}^{g}R_{s{s}}{{\Vec{t}}_r}_{\textstyle \raisebox{2pt}{,}}
	\end{IEEEeqnarraybox}}
\end{equation}
hence corresponding to the multi-channel Wiener filter (MWF) (Fig. \ref{fig:MWF_overview}). $R_{mm}=R_{ss}+R_{ee}+R_{nn}$ can be computed during periods of desired speech and echo activity ($\text{VAD}_s=1 \land \text{VAD}_{e^s}=1$). $R_{ss}$ can be computed by constructing $R_{nn}+R_{ee}$ during periods of simultaneous desired speech inactivity and echo activity ($\text{VAD}_s=0 \land \text{VAD}_{e^s}=1$), and subtracting the result from $R_{mm}$.
\ \\
\subsubsection{Extended multi-channel Wiener filter (MWF\textsubscript{ext})} \label{section:integrated_approach-solution_strategy-MWFext}
Similar to the MWF, using the extended signal model (\ref{eq:signal_model_def_x_tilde}), the solution to (\ref{eq:MWFext_cost}) is obtained as an MWF with extended correlation matrices $R_{\tilde{m}\tilde{m}} = \mathbb{E}\left\{\tilde{\Vec{m}}\tilde{\Vec{m}}^H\right\}\in\mathbb{C}^{\left(M+L\right)\times\left(M+L\right)}$ and $R_{\tilde{s}\tilde{s}} = \mathbb{E}\left\{\tilde{\Vec{s}}\tilde{\Vec{s}}^H\right\}\in\mathbb{C}^{\left(M+L\right)\times\left(M+L\right)}$
\begin{equation} \label{eq:MWFext}
	\boxed{\begin{IEEEeqnarraybox}{rll}
			\tilde{\Vec{w}}_{\text{int}} &= R_{\tilde{m}\tilde{m}}^{g}R_{\tilde{s}\tilde{s}}\tilde{\Vec{t}}_{r_{\textstyle \raisebox{2pt}{,}}}
	\end{IEEEeqnarraybox}}
\end{equation}
hence, named the extended MWF (MWF\textsubscript{ext}) (Fig. \ref{fig:MWFext_overview}). $R_{\tilde{m}\tilde{m}}$ can be computed when $\text{VAD}_s=1 \land \text{VAD}_{e^s}=1$, while $R_{\tilde{s}\tilde{s}}$ can be computed by constructing $R_{\tilde{e}\tilde{e}}+R_{\tilde{n}\tilde{n}}$ when $\text{VAD}_s=0 \land \text{VAD}_{e^s}=1$ and subtracting the result from $R_{\tilde{m}\tilde{m}}$.
\ \\
\subsubsection{AEC precedes NR (AEC-NR)} \label{section:integrated_approach-solution_strategy-AEC_NR}
A closed-form expression for a valid generalised matrix inverse in (\ref{eq:MWFext}) is as follows (cfr. Supplementary material Section \Romannum{9})
\begin{equation} \label{eq:AECNR_derivation_inverse_Rxx}
	\begin{IEEEeqnarraybox}{rll}
		&R_{\tilde{m}\tilde{m}}^{g,g} = \begin{bmatrix}
			\Sigma_{mm}^\dagger & -\Sigma_{mm}^\dagger R_{el}R_{ll}^\dagger\\ -R_{ll}^\dagger R_{le}\Sigma_{mm}^\dagger & R_{ll}^\dagger + R_{ll}^\dagger R_{le}\Sigma_{mm}^\dagger R_{el}R_{ll}^\dagger
		\end{bmatrix}_{\textstyle \raisebox{2pt}{,}}
	\end{IEEEeqnarraybox}
\end{equation}
with $\Sigma_{mm}=R_{mm}-R_{el}R_{ll}^\dagger R_{le}$. Herein, $R_{\tilde{m}\tilde{m}}^{g,g}\in\mathbb{C}^{(M+L)\times(M+L)}$ is thus one possible valid expression for all of the generally non-unique generalised inverses $R_{\tilde{m}\tilde{m}}^{g}$ (cfr. Supplementary material Section \Romannum{9}). If $\Sigma_{mm}$ is full-rank, i.e., $\text{rank}\!\left(\Sigma_{mm}\right)=M$, then (\ref{eq:AECNR_derivation_inverse_Rxx}) corresponds to the pseudo-inverse. If additionally $\text{rank}\!\left(R_{ll}\right)=L$, then (\ref{eq:AECNR_derivation_inverse_Rxx}) corresponds to the inverse.

Plugging (\ref{eq:AECNR_derivation_inverse_Rxx}) into (\ref{eq:MWFext}), the following expression is obtained, with $\mathbb{I}_{M\times M}\in\mathbb{C}^{M\times M}$ the identity matrix:

\begin{equation}\label{eq:AECNR}
	\boxed{\begin{IEEEeqnarraybox}{rll}
	&\tilde{\Vec{w}}_{\text{int}} = \underbrace{\begin{bmatrix}
		\mathbb{I}_{M\times M} & 0_{M\times L}\\ -R_{ll}^\dagger R_{le} & 0_{L\times L}
	\end{bmatrix}}_{\text{AEC}}\underbrace{\begin{bmatrix}
		\Sigma_{mm}^{\dagger}R_{ss}\\ {0}_{L\times M}
	\end{bmatrix}}_{\text{NR}}{{\Vec{t}}_{r_{\textstyle \raisebox{15pt}{,}}}}
\end{IEEEeqnarraybox}}
\end{equation}
which can be interpreted as an AEC preceding an NR (Fig. \ref{fig:AEC-NR_overview}). Indeed, $R_{ll}^\dagger R_{le}$ corresponds to the minimum-norm estimate of the echo paths as described in Section \ref{section:cascaded_approach-AEC}. Furthermore, $\Sigma_{mm}$ can be interpreted as the microphone correlation matrix after applying the AEC, such that $\Sigma_{mm}^\dagger R_{ss}$ corresponds to an MWF aimed at suppressing near-end room noise and residual echo after the AEC, and thus acts as a multi-channel post-filter. It will nevertheless be referred to as NR for consistency with the cascaded design definitions in Section \ref{section:cascaded_approach}, and for consistency with the literature \cite{jeannesCombinedNoiseEcho2001,ruizDistributedCombinedAcoustic2022,romboutsIntegratedApproachAcoustic2005a,kuo_integrated_2005}. To show this equivalence, the microphone correlation matrix after applying the AEC is defined as
\begin{equation} \label{eq:AECNR_corr}
	\begin{IEEEeqnarraybox}{rll}
		&\mathbb{E}\left\{\left(\Vec{m}-R_{el}R_{ll}^\dagger \Vec{l}\right)\left(\Vec{m}-R_{el}R_{ll}^\dagger \Vec{l}\right)^H\right\}\\
		&= R_{mm}-2R_{el}R_{ll}^\dagger R_{le} + R_{el}R_{ll}^\dagger R_{ll} R_{ll}^\dagger R_{le_{\textstyle \raisebox{2pt}{,}}}
	\end{IEEEeqnarraybox}
\end{equation}
wherein, $R_{ll}^\dagger R_{ll}R_{ll}^\dagger=R_{ll}^\dagger$ as per definition of the pseudo-inverse \cite{penroseGeneralizedInverseMatrices1955}, such that (\ref{eq:AECNR_corr}) is indeed equal to $\Sigma_{mm}$. Consequently, if $\text{rank}\!\left(\Sigma_{mm}\right)=M$, then the MWF\textsubscript{ext} (\ref{eq:MWFext}) is uniquely defined and equivalent to the AEC-NR (\ref{eq:AECNR}) as (\ref{eq:AECNR_derivation_inverse_Rxx}) then corresponds to the uniquely-defined pseudo-inverse. If $\text{rank}\!\left(\Sigma_{mm}\right)< M$, (\ref{eq:MWFext}) is not uniquely defined as $R_{\tilde{m}\tilde{m}}^g$ is not unique. Nevertheless, (\ref{eq:AECNR}) then still corresponds to one of the solutions of the Wiener-Hopf equations $R_{\tilde{m}\tilde{m}}\tilde{\Vec{w}}=R_{\tilde{s}\tilde{s}}\tilde{\Vec{t}}_r$, be it not necessarily to the minimum-norm solution.

To compute the correlation matrices in (\ref{eq:AECNR}), $R_{ll}$ and $R_{le}$ can be readily constructed when $\text{VAD}_{e^s}=1$ as $\Vec{s}$, $\Vec{n}$ and $\Vec{e}$ are uncorrelated, although in practice $R_{le}$ is computed when $\text{VAD}_s=0 \land \text{VAD}_{e^s}=1$ to reduce excess error from $\Vec{s}$ \cite[Chapter 8]{benestyAdvancesNetworkAcoustic2001}. $\Sigma_{mm}$ can then be constructed by computing the microphone correlation matrix after applying the AEC when $\text{VAD}_{s}=1 \land \text{VAD}_{e^s}=1$, whereas $R_{ss}$ can be computed by subtracting the microphone correlation matrix after applying the AEC when $\text{VAD}_s=0 \land \text{VAD}_{e^s}=1$, i.e., subtracting $R_{ee}+R_{nn}-R_{el}R_{ll}^\dagger R_{le}$, from $\Sigma_{mm}$.

If the echo path is linear and the AEC filters are chosen sufficiently long to model this echo path, i.e., $\Vec{e}=F_{\text{lin}}\Vec{l}$ and $R_{le}=R_{ll}F_{\text{lin}}^H$, $\tilde{\Vec{w}}_{\text{int}}$ (\ref{eq:AECNR}) is equivalent to $\tilde{\Vec{w}}_{\text{int,lin}}\in\mathbb{C}^{\left(M+L\right)\times 1}$:
\begin{equation} \label{eq:AECNR_lin}
	\boxed{\tilde{\Vec{w}}_{\text{int,lin}} = \underbrace{\begin{bmatrix}
				\mathbb{I}_{M\times M} & 0_{M\times L}\\ -R_{ll}^\dagger R_{ll} F_{\text{lin}}^H & 0_{L\times L}
		\end{bmatrix}}_{\text{AEC}}\underbrace{\begin{bmatrix}\left(R_{ss}+R_{nn}\right)^{-1}R_{ss}\\ {0}_{L\times M}\end{bmatrix}}_{\text{NR}}{{\Vec{t}}_{r_{\textstyle \raisebox{15pt}{,}}}}}
\end{equation}
where the NR does not need to suppress residual echo as the AEC already fully cancels the echo. If $R_{ll}$ is full-rank, then $R_{ll}^\dagger R_{ll}F_{\text{lin}}=F_{\text{lin}}$, corresponding to straightforward AEC. If $R_{ll}$ is rank-deficient, i.e., if the loudspeaker signals are linearly dependent, the AEC $R_{ll}^\dagger R_{ll} F^H$ does not correspond to the true echo path matrix $F_{\text{lin}}$, but to a weighted version thereof. This linear weighting results in a far-end room dependence of the AEC, as $R_{ll}$ depends on the far-end room scenario. The AEC, therefore, also needs to adapt to changes in the far-end room, such that (\ref{eq:AECNR_lin}) provides a generalised expression for the multi-channel AEC problem, e.g., extending \cite{shimauchiStereoProjectionEcho1995}.
\ \\
\subsubsection{NR precedes AEC (NR-AEC)} \label{section:integrated_approach-solution_strategy-NR_AEC}
By reordering the block matrices in (\ref{eq:AECNR}), the following expression is obtained
\begin{equation}\label{eq:NRAEC}
	{\begin{IEEEeqnarraybox}{rll}
			&\tilde{\Vec{w}}_{\text{int}} = \underbrace{\begin{bmatrix}
					\Sigma_{mm}^{\dagger}R_{ss} & 0_{M\times L}\\ 0_{L\times M} & \mathbb{I}_{L\times L}
			\end{bmatrix}}_{\text{NR}}\underbrace{\begin{bmatrix}
					\mathbb{I}_{M\times M}\\ -R_{ll}^\dagger R_{le}\left(\Sigma_{mm}^{\dagger}R_{ss}\right)
			\end{bmatrix}}_{\text{AEC}} {\Vec{t}}_{r_{\textstyle \raisebox{2pt}{.}}}
	\end{IEEEeqnarraybox}}
\end{equation}
Here, the NR aimed to suppress the noise and residual echo now precedes the AEC (Fig. \ref{fig:NR-AEC_overview}), as opposed to the AEC-NR (\ref{eq:AECNR}). As the echo signal vector after applying the NR is affected by the NR, the AEC not only has to model the echo paths but rather the combination of the echo paths and the NR, which leads to the additional factor $\Sigma_{mm}^\dagger R_{ss}$ in the AEC. 

The advantage performance-wise of the NR-AEC over the AEC-NR is that the AEC operates under reduced near-end room noise influence, thereby using a cascade implementation where the NR precedes the AEC. However, to do so, the loudspeaker correlation matrix and the loudspeaker-echo cross-correlation matrix are only computed after applying an NR, such that $\Sigma_{mm}$ is generally replaced by $R_{mm}$, resulting in an expression $\tilde{\Vec{w}}_{\text{mod}}\in\mathbb{C}^{\left(M+L\right)\times 1}$ with a modified NR and AEC, NR\textsubscript{mod} and AEC\textsubscript{mod} respectively \cite{martinAcousticEchoCancellation1997,docloCombinedAcousticEcho2000,schrammenChangePredictionLow2019,roebbenCascadedNoiseReduction2024}, 
\begin{equation}\label{eq:NRAEC_mod}
	\boxed{\begin{IEEEeqnarraybox}{rll}
			&\tilde{\Vec{w}}_{\text{mod}} = \underbrace{\begin{bmatrix}
					R_{mm}^{\dagger}R_{ss} & 0_{M\times L}\\ 0_{L\times M} & \mathbb{I}_{L\times L}
			\end{bmatrix}}_{\text{NR\textsubscript{mod}}}\underbrace{\begin{bmatrix}
					\mathbb{I}_{M\times M}\\ -R_{ll}^\dagger R_{le}\left(R_{mm}^{\dagger}R_{ss}\right)
			\end{bmatrix}}_{\text{AEC\textsubscript{mod}}} {\Vec{t}}_{r_{\textstyle \raisebox{2pt}{,}}}
	\end{IEEEeqnarraybox}}
\end{equation}
which then loses its MSE optimality. $R_{mm}$ is computed when $\text{VAD}_s=1 \land \text{VAD}_{e^s}=1$, and $R_{ss}$ is computed by subtracting $R_{ee}+R_{nn}$, collected when $\text{VAD}_s=0 \land \text{VAD}_{e^s}=1$, from $R_{mm}$. $R_{ll}$ and $R_{le}\left(R_{mm}^{\dagger}R_{ss}\right)$ are then computed as the loudspeaker correlation matrix and the loudspeaker-echo cross-correlation matrix after applying the NR\textsubscript{mod}.
\ \\
\subsubsection{Extended NR precedes AEC and PF (NR\textsubscript{ext}-AEC-PF)} \label{section:integrated_approach-solution_strategy-NRext_AEC}
The additional assumption is now made that the echo path is an additive map, hence using the signal model (\ref{eq:signal_model_def_x_map}). As detailed in Section \ref{section:cascaded_approach-AEC}, this assumption is more general than the assumption of a linear echo path as it only assumes that the plus operator is preserved by the map (Section \ref{section:cascaded_approach-AEC-signal_model}). This makes it possible to decouple the effect of $\Vec{l}^s$ and $\Vec{l}^n$ in the microphone into $\Vec{e}^s$ and $\Vec{e}^n$. Time-varying echo paths are also allowed, as time-variance does not affect this property. Furthermore, assume that $R_{ll}^\dagger R_{le}$ is a solution of the Wiener-Hopf equations $R_{l^sl^s}W=R_{l^se^s}$, i.e., $R_{l^sl^s}R_{ll}^\dagger R_{le}=R_{l^se^s}$. The latter assumption corresponds to the minimum-norm echo path estimate using the loudspeaker signals also being a solution of the Wiener-Hopf equations for the echo path estimate using only the far-end room speech component in the loudspeaker signals, thereby extending the assumption in \cite{roebbenCascadedNoiseReduction2024} to the case where $R_{ll}$ and $R_{l^sl^s}$ can be rank-deficient.

Under these assumptions, an NR\textsubscript{ext} filter $W_{\text{NR\textsubscript{ext}}}\in\mathbb{C}^{\left(M+L\right)\times \left(M+L\right)}$ can be defined as the MSE-optimal filter to suppress both $\tilde{\Vec{n}}$ and $\tilde{\Vec{e}}^n$ while preserving both $\tilde{\Vec{s}}$ and $\tilde{\Vec{e}}^s$:
\begin{equation} \label{eq:NRextAEC-WNRext}
	W_{\text{NR\textsubscript{ext}}}=R_{\tilde{m}\tilde{m}}^{g,g}\left(R_{\tilde{s}\tilde{s}}+R_{\tilde{e}^s\tilde{e}^s}\right)_{\textstyle \raisebox{2pt}{.}}
\end{equation} 
Applying $W_{\text{NR\textsubscript{ext}}}$ to $\tilde{\Vec{m}}$ leads to an extended signal vector,
\begin{equation}
	\tilde{\Vec{m}}^`=\begin{bmatrix}
		\Vec{m}^`\\\Vec{l}^`
	\end{bmatrix}=W_{\text{NR\textsubscript{ext}}}^H\tilde{\Vec{m}}_{\textstyle \raisebox{2pt}{.}}
\end{equation} 

Using (\ref{eq:NRextAEC-WNRext}) and the assumption that $R_{l^sl^s}R_{ll}^\dagger R_{le}=R_{l^se^s}$, the following filter can be shown to be equivalent to (\ref{eq:MWFext}) if $R_{\tilde{m}\tilde{m}}^{g,g}$ is chosen as defined in (\ref{eq:AECNR_derivation_inverse_Rxx}) for $R_{\tilde{m}\tilde{m}}^g$ both in (\ref{eq:MWFext}) and (\ref{eq:NRextAEC}) (cfr. Supplementary material Section \Romannum{10})
\begin{equation} \label{eq:NRextAEC}
	\boxed{\begin{IEEEeqnarraybox}{rll}
			\tilde{\Vec{w}}_{\text{int}} {=}&\underbrace{R_{\tilde{m}\tilde{m}}^{g,g}\left(R_{\tilde{s}\tilde{s}}+R_{\tilde{e}^s\tilde{e}^s}\right)}_{\text{NR\textsubscript{ext}}}\underbrace{\begin{bmatrix}
		\mathbb{I}_{M\times M} & 0_{M\times L}\\ -R_{l^`l^`}^{\dagger}R_{l^`m^`} & 0_{L\times L}
\end{bmatrix}}_{\text{AEC}}\cdot\\
&\underbrace{\begin{bmatrix}
	\Sigma_{m^`m^`}^\dagger R_{s^`s}\\ {0}_{L\times M}
\end{bmatrix}}_{\text{PF}}{\Vec{t}}_{r_{\textstyle \raisebox{2pt}{,}}}
	\end{IEEEeqnarraybox}}
\end{equation}
In the post-filter (PF), $\Sigma_{m^`m^`}=R_{m^`m^`}-R_{m^`l^`}R_{l^`l^`}^\dagger R_{l^`m^`}$ corresponds to the microphone correlation matrix after applying the NR\textsubscript{ext} and the AEC. Consequently, (\ref{eq:NRextAEC}) can be interpreted as an NR\textsubscript{ext} preceding an AEC and a PF (Fig. \ref{fig:NRext-AEC_overview}), where the NR\textsubscript{ext} is aimed at suppressing the near-end room noise and the far-end room noise component in the echo, and the AEC is aimed at removing the far-end room speech and residual noise components in the echo. The PF is aimed at suppressing residual noise and echo, while preserving the desired speech. $R_{\tilde{m}\tilde{m}}$ can be collected when both the desired speech and the far-end room speech component in the echo are active, i.e., when $\text{VAD}_s=1 \land \text{VAD}_{e^s}=1$. $R_{\tilde{s}\tilde{s}}+R_{\tilde{e}^s\tilde{e}^s}$ can be computed by collecting $R_{\tilde{e}^n\tilde{e}^n}+R_{\tilde{n}\tilde{n}}$ when $\text{VAD}_s=0 \land \text{VAD}_{e^s}=0$, and subtracting the result from $R_{\tilde{m}\tilde{m}}$. $R_{l`l`}$ and $R_{l^`m^`}$ can be readily collected after applying the NR\textsubscript{ext} when $\text{VAD}_{e^s}=1$ as $\Vec{s}^`$, $\Vec{e}^`$ and $\Vec{n}^`$ are uncorrelated. In practice, excess error due to $\Vec{s}^`$ can nevertheless be avoided by computing $R_{l`m`}$ when $\text{VAD}_s=0 \land \text{VAD}_{e^s}=1$. For the PF, $\Sigma_{m^`m^`}$ can be collected when $\text{VAD}_{s}=1\land \text{VAD}_{e^s}=1$ after applying the NR\textsubscript{ext} and AEC. $R_{s^`s}$ can be computed by subtracting $R_{e^`e}-R_{m^`l^`}R_{l^`l^`}^\dagger R_{l^`e}+R_{n^`n}$, collected when $\text{VAD}_s=0 \land \text{VAD}_{e^s}=1$, from $\Sigma_{m^`m}$, collected when $\text{VAD}_s=1 \land \text{VAD}_{e^s}=1$. Alternatively, when $\begin{bmatrix}
	\mathbb{I}_{M\times M} & 0_{M\times L}
\end{bmatrix}W_{\text{NR\textsubscript{ext}}}\begin{bmatrix}
	\mathbb{I}_{M\times M} \\ 0_{L\times M}
\end{bmatrix}$ is of full-rank, $R_{s^`s}$ can be computed by subtracting $R_{e^`e^`}-R_{m^`l^`}R_{l^`l^`}^\dagger R_{l^`m^`}+R_{n^`n^`}$ from $\Sigma_{m^`m^`}$ and post-multiplying with $\left(\begin{bmatrix}
\mathbb{I}_{M\times M} & 0_{M\times L}
\end{bmatrix}W_{\text{NR\textsubscript{ext}}}\begin{bmatrix}
\mathbb{I}_{M\times M} \\ 0_{L\times M}
\end{bmatrix}\right)^{-1}$.

If $\text{rank}\!\left(\Sigma_{mm}\right)=M$, then the MWF\textsubscript{ext} is uniquely defined and equivalent to the NR\textsubscript{ext}-AEC-PF (\ref{eq:NRextAEC}) as (\ref{eq:AECNR_derivation_inverse_Rxx}) then corresponds to the uniquely-defined pseudo-inverse. If $\text{rank}\!\left(\Sigma_{mm}\right)<M$, then (\ref{eq:AECNR_derivation_inverse_Rxx}) is not uniquely defined as $R_{\tilde{m}\tilde{m}}^g$ is not unique. Nevertheless, (\ref{eq:NRextAEC}) then still corresponds to one of the solutions of the Wiener-Hopf equations $R_{\tilde{m}\tilde{m}}\tilde{\Vec{w}}=R_{\tilde{s}{s}}{\Vec{t}}_r$, be it not necessarily the minimum-norm solution. 

The difference between the NR-AEC (\ref{eq:NRAEC}) and the first two filters (NR\textsubscript{ext} and AEC) in the NR\textsubscript{ext}-AEC-PF (\ref{eq:NRextAEC}) can be seen as follows \cite{roebbenCascadedNoiseReduction2024}: Whereas the NR in (\ref{eq:NRAEC}) operates solely on the microphones, aimed at suppressing the near-end room noise, the NR\textsubscript{ext} in (\ref{eq:NRextAEC}) operates on both the microphones and the loudspeakers (without affecting the playback), aimed at suppressing both the near-end room noise and the far-end room noise component in the echo. Consequently, the AEC in (\ref{eq:NRAEC}) aims at removing the entire echo signal vector, while the AEC in (\ref{eq:NRextAEC}) mostly aims at suppressing the far-end room speech component in the echo (and the residual far-end room noise component in the echo). Further, whereas the AEC in (\ref{eq:NRAEC}) needs to adapt to the preceding NR, the AEC in (\ref{eq:NRextAEC}) actually does not need to adapt to the preceding NR\textsubscript{ext}, because (cfr. Supplementary material Section \Romannum{10}),
\begin{equation} \label{eq:NRextAEC-AEC}
	R_{l^`l^`}^\dagger R_{l^`m^`}=R_{l^sl^s}^\dagger R_{{l^s e^s}_{\textstyle \raisebox{2pt}{,}}}
\end{equation}
which corresponds to the MSE-optimal AEC based on the far-end room speech component in the echo, which is thus independent of the NR\textsubscript{ext}. Supplementary material Section \Romannum{10} thus generalises \cite{roebbenCascadedNoiseReduction2024}, where this independence of NR\textsubscript{ext} and AEC was shown when considering a full-rank $R_{mm}$ and $R_{ll}$.

The NR\textsubscript{ext}-AEC-PF can also be interpreted as the AEC-NR to which a preceding NR\textsubscript{ext} is added. This NR\textsubscript{ext} then already aims at partially reducing the near-end room noise and far-end room noise component in the echo before applying the AEC-NR. The PF in the NR\textsubscript{ext}-AEC-PF has the same functionality as the NR in the AEC-NR, but will be referred to as PF to avoid confusion with this NR\textsubscript{ext}. 

If the echo path is linear and the AEC filters are chosen sufficiently long to model this echo path, i.e., $\Vec{e}=F_{\text{lin}}\Vec{l}$ and $R_{le}=R_{ll}F_{\text{lin}}^H$, (\ref{eq:NRextAEC}) simplifies to $\tilde{\Vec{w}}_{\text{int,lin}}$ (cfr. Supplementary material Section \Romannum{11})
\begin{equation} \label{eq:NRextAEC-lin1}
	\boxed{\begin{IEEEeqnarraybox}{rll}
		\tilde{\Vec{w}}_{\text{int,lin}} &{=}{\underbrace{R_{\tilde{m}\tilde{m}}^{g,g}\left(R_{\tilde{s}\tilde{s}}+R_{\tilde{e}^s\tilde{e}^s}\right)}_{\text{NR\textsubscript{ext}}}\underbrace{\begin{bmatrix}
				\mathbb{I}_{M\times M}\\ -R_{l^sl^s}^{\dagger}R_{l^sl^s}F_{\text{lin}}^H
		\end{bmatrix}}_{\text{AEC}}{\Vec{t}}_{r_{\textstyle \raisebox{2pt}{,}}}}
	\end{IEEEeqnarraybox}}
\end{equation}
where the PF is reduced to $\begin{bmatrix}\mathbb{I}_{M\times M}\\0_{L\times M} \end{bmatrix}$ and hence omitted.

\section{Practical considerations} \label{section:practial_considerations}

Although theoretical equivalence exists between the expressions in Section \ref{section:integrated_approach}, the practical performances of their cascade algorithm implementations differ due to non-stationarities and imperfect correlation matrix estimation. In practice the expressions of Section \ref{section:integrated_approach} are either implemented in the frequency domain, replacing index $z$ with frequency-bin index $f$ and possibly frame index $k$, or in the time domain, replacing the $z$-domain variables with time-lagged vectors and replacing index $z$ with time index $t$. To illustrate the implications of the practical considerations, the frequency-domain expressions will be given next, although a similar reasoning can be made for the time-domain expressions. 

Regarding the non-stationarities, (short-term) stationarity of the spatial position of the desired speech, near-end room noise and loudspeakers can be assumed \cite{romboutsIntegratedApproachAcoustic2005a}. Spectral (short-term) stationarity of the near-end room noise and the far-end room noise component in the loudspeakers and echo can also be assumed as they model, e.g., sensor and diffuse noise sources. Correspondingly, the effect of spectral non-stationarities in the desired speech and (the far-end room speech component in) the loudspeakers and echo is focused upon. These non-stationarities, e.g., lead to a different contribution of the loudspeaker correlation matrix depending on the regime, such that $R_{ll}^{\{s \! = \! 1,e^s \! = \! 1\}}(f)$ recorded when $\text{VAD}_s=1 \land \text{VAD}_{e^s}=1$ and $R_{ll}^{\{s \! = \! 0,e^s \! = \! 1\}}(f)$ recorded when $\text{VAD}_s=0 \land \text{VAD}_{e^s}=1$ differ, i.e., $R_{ll}^{\{s \! = \! 1,e^s \! = \! 1\}}(f)\neq R_{ll}^{\{s \! = \! 0,e^s \! = \! 1\}}(f)$. However, due to the assumed (short-term) stationarity of the near-end room noise, $R_{nn}^{\{\text{s} \! = \! 1, \text{e\textsuperscript{s}} \! = \! 1\}}(f)=R_{nn}^{\{\text{s} \! = \! 1, \text{e\textsuperscript{s}} \! = \! 0\}}(f)=R_{nn}^{\{\text{s} \! = \! 0, \text{e\textsuperscript{s}} \! = \! 1\}}(f)=R_{nn}^{\{\text{s} \! = \! 0, \text{e\textsuperscript{s}} \! = \! 0\}}(f)=R_{nn}(f)$, and a similar property holds for the far-end room noise component in the echo. 

Regarding the correlation matrix estimation, as there is only access to a finite number of samples and only during the regimes as specified by $\text{VAD}_s(k,f)$ and $\text{VAD}_{e^s}(k,f)$ (Section \ref{section:integrated_approach-signal_model}), the correlation matrices are estimated using time averaging during these regimes. To illustrate this procedure, the microphone correlation matrix $\hat{R}_{mm}^{\{\text{s} \! = \! 1, \text{e\textsuperscript{s}} \! = \! 1\}}(f)\in\mathbb{C}^{M\times M}$ in the MWF (\ref{eq:MWF}) can be estimated across the $K$ frames where $\text{VAD}_s(k,f)=1 \land\text{VAD}_{e^s}(k,f)=1 $ as
\begin{equation}\label{eq:time_averaging_correlation}
	\hat{R}_{mm}^{\{\text{s} \! = \! 1, \text{e\textsuperscript{s}} \! = \! 1\}}(f) = \frac{1}{K}\sum_{k=1}^{K}\Vec{m}(k,f)\Vec{m}(k,f)^H_{\textstyle \raisebox{2pt}{.}}
\end{equation}Herein, $\hat{.}$ is used to denote estimated variables. As during the regimes specified by $\text{VAD}_s(k,f)$ and $\text{VAD}_{e^s}(k,f)$ there is, e.g., no access to desired speech only, $\hat{R}_{ss}^{\{\text{s} \! = \! 1, \text{e\textsuperscript{s}} \! = \! 1\}}$ cannot be readily estimated by time averaging. Nevertheless, $\hat{R}_{ss}^{\{\text{s} \! = \! 1, \text{e\textsuperscript{s}} \! = \! 1\}}(f)$ can be estimated by subtracting the correlation matrices $\hat{R}_{ee}^{\{\text{s} \! = \! 0, \text{e\textsuperscript{s}} \! = \! 1\}}(f)+\hat{R}_{nn}^{\{\text{s} \! = \! 0, \text{e\textsuperscript{s}} \! = \! 1\}}(f)$ estimated when $\text{VAD}_s=0 \land \text{VAD}_{e^s}=1$ from $\hat{R}_{mm}^{\{\text{s} \! = \! 1, \text{e\textsuperscript{s}} \! = \! 1\}}(f)$. 

This subtraction can also be performed using a generalised eigenvalue decomposition (GEVD) on this matrix pencil $\{\hat{R}_{mm}^{\{\text{s} \! = \! 1, \text{e\textsuperscript{s}} \! = \! 1\}}(f),\hat{R}_{ee}^{\{\text{s} \! = \! 0, \text{e\textsuperscript{s}} \! = \! 1\}}(f)+\hat{R}_{nn}^{\{\text{s} \! = \! 0, \text{e\textsuperscript{s}} \! = \! 1\}}(f)\}$ to enforce a rank constraint on $\hat{R}_{ss}^{\{\text{s} \! = \! 1, \text{e\textsuperscript{s}} \! = \! 1\}}$ \cite{serizelLowrankApproximationBased2014,baiTemplatesSolutionAlgebraic2000}. This rank constraint can be imposed as the ground truth $R_{ss}$ generally models a limited number of desired speech sources and consequently generally has rank $R_s<M$. To this end, $\hat{R}_{mm}^{\{\text{s} \! = \! 1, \text{e\textsuperscript{s}} \! = \! 1\}}(f)$ and $\hat{R}_{ee}^{\{\text{s} \! = \! 0, \text{e\textsuperscript{s}} \! = \! 1\}}(f)+\hat{R}_{nn}^{\{\text{s} \! = \! 0, \text{e\textsuperscript{s}} \! = \! 1\}}(f)$ are jointly diagonalised, and only the $R_s$ modes with highest signal-to-noise ratio (SNR) are retained when subtracting both. Collecting the generalised eigenvectors in the columns of $\hat{Q}(f)\in\mathbb{C}^{M\times M}$, denoting the generalised eigenvalues of $\hat{R}_{mm}^{\{\text{s} \! = \! 1, \text{e\textsuperscript{s}} \! = \! 1\}}(f)$ by $\hat{\lambda}_{m_i}(f)$, and denoting the generalised eigenvalues of $\hat{R}_{e}^{\{\text{s} \! = \! 0, \text{e\textsuperscript{s}} \! = \! 1\}}(f)+\hat{R}_{nn}^{\{\text{s} \! = \! 0, \text{e\textsuperscript{s}} \! = \! 1\}}(f)$ by $\hat{\lambda}_{(e+n)_i}(f)$, the GEVD adheres to the following equations \cite{serizelLowrankApproximationBased2014}
	\begin{subequations} \label{eq:MWF_GEVD}
			\begin{align}
			&\hat{R}_{mm}^{\{\text{s} \! = \! 1, \text{e\textsuperscript{s}} \! = \! 1\}}(f) = \hat{Q}(f)\hat{\Lambda}_m(f)\hat{Q}(f)^H\\& \ \ = \hat{Q}(f)\text{diag}\!\left(\hat{\lambda}_{m_1}(f),\cdots,\hat{\lambda}_{m_M}(f)\right)\hat{Q}(f)^H\\
			&\hat{R}_{ee}^{\{\text{s} \! = \! 0, \text{e\textsuperscript{s}} \! = \! 1\}}\!(f) \!+\! \hat{R}_{nn}^{\{\text{s} \! = \! 0, \text{e\textsuperscript{s}} \! = \! 1\}}\!(f) \! = \! \hat{Q}(f)\hat{\Lambda}_{(e+n)}(f)\hat{Q}(f)^H\\
					& \ \ = \hat{Q}(f)\text{diag}\!\left(\hat{\lambda}_{(e+n)_1}(f),\cdots,\hat{\lambda}_{(e+n)_M}(f)\right)\hat{Q}(f)^H_{\textstyle \raisebox{2pt}{.}}
				\end{align}
		\end{subequations}
Using this GEVD, the rank-$R_s$ GEVD approximation $\hat{R}_{ss}^{\{\text{s} \! = \! 1, \text{e\textsuperscript{s}} \! = \! 1\}}(f)$ can then be computed as \cite{serizelLowrankApproximationBased2014} 
\begin{equation} \label{eq:MWF_Rss_GEVD}
			\begin{IEEEeqnarraybox}{rll}
					&\hat{R}_{ss}^{\{\text{s} \! = \! 1, \text{e\textsuperscript{s}} \! = \! 1\}}= \hat{Q}(f)\cdot\text{diag}\bigl(\hat{\lambda}_{m_1}(f)-\hat{\lambda}_{(e+n)_1}(f),\cdots,\\ &\qquad \qquad \quad \ \hat{\lambda}_{m_{R_s}}(f)-\hat{\lambda}_{(e+n)_{R_s}(f)},0,\cdots,0\bigr)\hat{Q}(f)^H_{\textstyle \raisebox{2pt}{.}}
				\end{IEEEeqnarraybox}
\end{equation}
In what follows, the influence of non-stationarities and imperfect correlation matrix estimation is discussed in more detail for each of the algorithms. Afterwards, the connection is made with recent advances in data-driven approaches.
\ \\
\subsubsection{Multi-channel Wiener filter (MWF)} \label{section:practial_considerations-filter_estimation-MWF}
Regarding the (far-end speech component in the) loudspeakers and echo, non-stationary echo (together with imperfect correlation matrix estimation) leads to the same contribution of the noise correlation matrix but a different contribution of the echo correlation matrix in $\hat{R}_{mm}^{\{\text{s} \! = \! 1, \text{e\textsuperscript{s}} \! = \! 1\}}(f)$ and $\hat{R}_{ee}^{\{\text{s} \! = \! 0, \text{e\textsuperscript{s}} \! = \! 1\}}(f)+\hat{R}_{nn}^{\{\text{s} \! = \! 0, \text{e\textsuperscript{s}} \! = \! 1\}}(f)$, such that by subtracting both, next to the desired speech correlation matrix, the echo correlation matrix is partially retained. The MWF then aims at estimating the desired speech and partially the echo. A GEVD approximation does not resolve this issue. 

Regarding the desired speech, the non-stationarity effect is more contained than for the loudspeakers and echo. Indeed, while the echo correlation matrix is present on both sides of the matrix pencil $\{\hat{R}_{{m}{m}}^{\{\text{s} \! = \! 1, \text{e\textsuperscript{s}} \! = \! 1\}}(f),\hat{R}_{{e}{e}}^{\{\text{s} \! = \! 0, \text{e\textsuperscript{s}} \! = \! 1\}}(f)+\hat{R}_{{n}{n}}^{\{\text{s} \! = \! 0, \text{e\textsuperscript{s}} \! = \! 1\}}(f)\}$, the desired speech correlation matrix is only present on the left-hand side. Consequently, the effect of desired speech non-stationarities is effectively limited to the MWF only being optimal with respect to the desired speech's average temporal characteristics.
\ \\
\subsubsection{Extended multi-channel Wiener filter (MWF\textsubscript{ext})} \label{section:practial_considerations-filter_estimation-MWFext}
Regarding the (far-end speech component in the) loudspeakers and echo, in \cite{ruizDistributedCombinedAcoustic2022}, it is argued that plain subtraction of ${R}_{\tilde{m}\tilde{m}}^{\{\text{s} \! = \! 1, \text{e\textsuperscript{s}} \! = \! 1\}}(f)$ by ${R}_{\tilde{e}\tilde{e}}^{\{\text{s} \! = \! 0, \text{e\textsuperscript{s}} \! = \! 1\}}(f)+{R}_{\tilde{n}\tilde{n}}^{\{\text{s} \! = \! 0, \text{e\textsuperscript{s}} \! = \! 1\}}(f)$ leads to partial reconstruction of the echo due to non-stationary loudspeaker and echo signal vectors having a different contribution in the extended echo correlation matrix on both sides. Nevertheless, it is argued that the GEVD-implementation of the MWF\textsubscript{ext} using a linear signal model with sufficiently long filters to model the echo path is not affected by these non-stationarities \cite{ruizDistributedCombinedAcoustic2022}. Indeed, referring to the extended signal model (\ref{eq:signal_model_def_x_tilde}), a vector $\Vec{0}_{L\times 1}$ is stacked under $\Vec{s}(k,f)$ and $\Vec{n}(k,f)$, while $\Vec{l}(k,f)$ is stacked under $\Vec{e}(k,f)$. This difference in structure is reflected in the generalised eigenvectors, as the GEVD of $\{{R}_{\tilde{m}\tilde{m}}^{\{\text{s} \! = \! 1, \text{e\textsuperscript{s}} \! = \! 1\}}(f),{R}_{\tilde{e}\tilde{e}}^{\{\text{s} \! = \! 0, \text{e\textsuperscript{s}} \! = \! 1\}}(f) + {R}_{\tilde{n}\tilde{n}}^{\{\text{s} \! = \! 0, \text{e\textsuperscript{s}} \! = \! 1\}}(f)\}$ is defined as
\begin{subequations}
	\begin{align}
		&{R}_{\tilde{m}\tilde{m}}^{\{\text{s} \! = \! 1, \text{e\textsuperscript{s}} \! = \! 1\}}(f) = {\tilde{Q}}(f){\tilde{\Lambda}}_{\tilde{m}}(f){\tilde{Q}}(f)^H\\
		&{R}_{\tilde{e}\tilde{e}}^{\{\text{s} \! = \! 0, \text{e\textsuperscript{s}} \! = \! 1\}}\!(f) \! +\! {R}_{\tilde{n}\tilde{n}}^{\{\text{s} \! = \! 0, \text{e\textsuperscript{s}} \! = \! 1\}}\!(f) \! =\! {\tilde{Q}}(f){\tilde{\Lambda}}_{(\tilde{e}+\tilde{n})}(f){\tilde{Q}}(f)^H_{\textstyle \raisebox{2pt}{.}}
	\end{align}
\end{subequations}
with $\tilde{Q}(f),{\tilde{\Lambda}}_{\tilde{m}}(f),{\tilde{\Lambda}}_{(\tilde{e}+\tilde{n})}(f) \in\mathbb{C}^{(M+L)\times (M+L)}$ containing the generalised eigenvectors and eigenvalues respectively. As for a linear echo path with sufficiently long filters to model this echo path $\text{rank}(R_{\tilde{e}\tilde{e}})\leq L$, $\tilde{Q}(f)$ reflects the zero-structure as \cite{ruizDistributedCombinedAcoustic2022}
\begin{equation}
	\tilde{Q}(f) =\left[ \begin{array}{c|c@{}}
		Q_1(f) & \multirow{ 2}{*}{$\tilde{Q}_2(f)$}\\
		\cmidrule{1-1}
		0_{M\times M}&\\
	\end{array} \right]_{\textstyle \raisebox{2pt}{.}}
\end{equation}
Herein, $\begin{bmatrix} Q_{1}(f)^\top & 0_{M\times M}\end{bmatrix}^\top\in\mathbb{C}^{(M+L)\times M}$ is related to the desired speech and near-end room noise, and $\tilde{Q}_2(f)\in\mathbb{C}^{(M+L)\times L}$ to the echo, such that only retaining $\tilde{Q}_2(f)$ allows for suppressing the echo even given its non-stationarity \cite{ruizDistributedCombinedAcoustic2022}. 

While this property is true for a linear signal model with sufficiently long filters to model the echo paths, in practice correlation matrix estimation using time averaging likely leads to the extended echo correlation matrices being of full-rank, i.e., being of rank $M+L$, due to imperfect correlation matrix estimation or due to the application of a general echo path (on linearly independent loudspeaker signals). The contribution of this extended echo correlation matrix on each side of the matrix pencil $\{\hat{\tilde{R}}_{\tilde{m}\tilde{m}}^{\{\text{s} \! = \! 1, \text{e\textsuperscript{s}} \! = \! 1\}}(f),\hat{\tilde{R}}_{\tilde{e}\tilde{e}}^{\{\text{s} \! = \! 0, \text{e\textsuperscript{s}} \! = \! 1\}}(f)+\hat{\tilde{R}}_{\tilde{n}\tilde{n}}^{\{\text{s} \! = \! 0, \text{e\textsuperscript{s}} \! = \! 1\}}(f)\}$ will then differ, such that all generalised eigenvectors will be influenced by the loudspeaker and echo signal vectors, and the echo will be partially retained by the MWF\textsubscript{ext}.

As for the MWF (Section \ref{section:practial_considerations-filter_estimation-MWF}), the effect of desired speech non-stationarities is limited, leading to an MWF\textsubscript{ext} optimal to the desired speech's average temporal characteristics. 
\ \\
\subsubsection{AEC precedes NR (AEC-NR)} \label{section:practial_considerations-filter_estimation-AECNR}
As both $\hat{R}_{ll}^{\{\text{s} \! = \! 0, \text{e\textsuperscript{s}} \! = \! 1\}}(f)$ and $\hat{R}_{le}^{\{\text{s} \! = \! 0, \text{e\textsuperscript{s}} \! = \! 1\}}(f)$ are updated under the same conditions, i.e., $\text{VAD}_s=0 \land \text{VAD}_{e^s}=1$, both matrices are similarly affected by loudspeaker and echo non-stationarities. Consequently, while it is true that in general the estimated echo path may differ based on the loudspeaker and echo non-stationarities, i.e., $\hat{R}_{ll}^{\{\text{s} \! = \! 1, \text{e\textsuperscript{s}} \! = \! 1\}}(f)^\dagger\hat{R}_{le}^{\{\text{s} \! = \! 1, \text{e\textsuperscript{s}} \! = \! 1\}}(f)$ not necessarily equalling $\hat{R}_{ll}^{\{\text{s} \! = \! 0, \text{e\textsuperscript{s}} \! = \! 1\}}(f)^\dagger\hat{R}_{le}^{\{\text{s} \! = \! 0, \text{e\textsuperscript{s}} \! = \! 1\}}(f)$, the estimated echo path is nevertheless optimal with respect to the loudspeaker's and echo's average temporal characteristics during that regime. Additionally, under additional assumptions of the echo path, stronger statements can be made. For example, modelling a volume increase $\alpha$ jointly for all loudspeakers when $\text{VAD}_s=1$ (i.e., $\alpha\Vec{l}(k,f)$) with respect to $\text{VAD}_s=0$, leads to $\alpha^2\hat{R}_{ll}^{\{\text{s} \! = \! 1, \text{e\textsuperscript{s}} \! = \! 1\}}=\hat{R}_{ll}^{\{\text{s} \! = \! 0, \text{e\textsuperscript{s}} \! = \! 1\}}$. Under this condition, when the echo path is linear with possible undermodelling, it nevertheless holds that $\hat{R}_{ll}^{\{\text{s} \! = \! 1, \text{e\textsuperscript{s}} \! = \! 1\}}(f)^\dagger\hat{R}_{le}^{\{\text{s} \! = \! 1, \text{e\textsuperscript{s}} \! = \! 1\}}(f)=\hat{R}_{ll}^{\{\text{s} \! = \! 0, \text{e\textsuperscript{s}} \! = \! 1\}}(f)^\dagger\hat{R}_{le}^{\{\text{s} \! = \! 0, \text{e\textsuperscript{s}} \! = \! 1\}}(f)$. As the AEC is not updated when $\text{VAD}_s=1$, non-stationarities of the desired speech do not affect the AEC.

The non-stationarities of the loudspeaker and echo affect the NR similarly to the MWF as described in Section \ref{section:practial_considerations-filter_estimation-MWF}, although to a reduced extent as the AEC already partially reduces the echo. In the limit case where the echo path is linear, the filters are sufficiently long to model echo paths, and there is no noise presence, the AEC can already remove the entire echo, such that the NR is not affected by loudspeaker and echo non-stationarities. The effect of desired speech non-stationarities is the same as for the MWF (Section \ref{section:practial_considerations-filter_estimation-MWF}).
\ \\
\subsubsection{NR precedes AEC (NR-AEC)} \label{section:practial_considerations-filter_estimation-NRAEC}
In its modified form (\ref{eq:NRAEC_mod}), the NR corresponds to the MWF (\ref{eq:MWF}), such that the effect of non-stationary echo, loudspeaker and desired speech for the NR is the same as described in Section \ref{section:practial_considerations-filter_estimation-MWF}.

While the NR thus partially reconstructs the echo due to loudspeaker and echo non-stationarities, the AEC further reduces the echo using an estimated echo path optimal with respect to the loudspeaker's and echo's average temporal characteristics, such that the NR-AEC in its modified form can be interpreted as an MWF adjusted with an AEC to suppress residual echo. As described in Section \ref{section:practial_considerations-filter_estimation-AECNR}, the AEC is not affected by desired speech non-stationarities due to the AEC only being updated when $\text{VAD}_s(k,f)=0$.
\ \\
\subsubsection{NR\textsubscript{ext} precedes AEC and PF (NR\textsubscript{ext}-AEC-PF)} \label{section:practial_considerations-filter_estimation-NRextAEC}
The NR\textsubscript{ext} is less affected by non-stationarities in the far-end room speech component in the loudspeakers and echo than the MWF\textsubscript{ext} as these components only appear in $\hat{R}_{\tilde{m}\tilde{m}}^{\{\text{s} \! = \! 1, \text{e\textsuperscript{s}} \! = \! 1\}}(f)$ when computing $\hat{R}_{\tilde{s}\tilde{s}}^{\{\text{s} \! = \! 1, \text{e\textsuperscript{s}} \! = \! 1\}}(f)+\hat{R}_{\tilde{e}^s\tilde{e}^s}^{\{\text{s} \! = \! 1, \text{e\textsuperscript{s}} \! = \! 1\}}(f)$ by subtracting $\hat{R}_{\tilde{e}^n\tilde{e}^n}^{\{\text{s} \! = \! 0, \text{e\textsuperscript{s}} \! = \! 0\}}(f)+\hat{R}_{\tilde{n}\tilde{n}}^{\{\text{s} \! = \! 0, \text{e\textsuperscript{s}} \! = \! 0\}}(f)$ from $\hat{R}_{\tilde{m}\tilde{m}}^{\{\text{s} \! = \! 1, \text{e\textsuperscript{s}} \! = \! 1\}}(f)$, possibly using a GEVD. The effect of desired speech non-stationarities is thus limited as $\hat{R}_{\tilde{s}\tilde{s}}^{\{\text{s} \! = \! 1, \text{e\textsuperscript{s}} \! = \! 1\}}(f)$ only appears in $\hat{R}_{\tilde{m}\tilde{m}}^{\{\text{s} \! = \! 1, \text{e\textsuperscript{s}} \! = \! 1\}}(f)$.

The AEC is similarly affected by non-stationarities as in the AEC-NR (Section \ref{section:practial_considerations-filter_estimation-AECNR}) and the NR-AEC (Section \ref{section:practial_considerations-filter_estimation-NRAEC}).

The PF takes the form of an MWF (after applying the NR\textsubscript{ext} and the AEC), such that the PF is affected by echo, loudspeaker and desired speech non-stationarities similar to the MWF (Section \ref{section:practial_considerations-filter_estimation-MWF}). Nevertheless, as the NR\textsubscript{ext}-AEC already partially removes the near-end room noise and echo, the PF is less affected by the loudspeaker and echo non-stationarities than the MWF.\\

\subsubsection{Connection to data-driven approaches} \label{section:hybrid}
The expressions (\ref{eq:MWF}), (\ref{eq:MWFext}), (\ref{eq:AECNR}), (\ref{eq:NRAEC_mod}), and (\ref{eq:NRextAEC}) can also be aligned with advances in data-driven approaches, as these expressions provide model structures optimal with respect to the MSE cost function, of which the model-based parameter estimation in terms of correlation matrices can be replaced by data-driven parameter estimation, e.g., using NNs.

For example, the AEC can be equivalently implemented with a recursive least squares (RLS) adaptive filter \cite{sayed_state-space_1994}, which can be unified with Kalman filtering \cite{sayed_state-space_1994}. As there exist many variants of hybrid Kalman filters for AEC with NN-based parameter estimation \cite{seidel_convergence_2024,seidel_neural_2024}, the AEC-NR, NR-AEC and NR\textsubscript{ext}-AEC-PF can make use of these hybrid Kalman filters. Another example would be to supply the MWF, NR, or PF through a NN (a so-called mask), optimised through an MSE or other cost function, or to replace the MWF, NR, or PF with a NN. This leads to algorithms with linear/hybrid AEC filters followed by a NN as in \cite{seidel_efficient_2024,shetu_hybrid_2024,shetu_align-ulcnet_2024,franzen_deep_2022,haubner_synergistic_2021} with the AEC-NR of (\ref{eq:AECNR}).

\section{Computational complexity} \label{section:numerical_complexity}
Next to the different practical considerations discussed in Section \ref{section:practial_considerations}, the different algorithms also have a different computational complexity. To this end, a floating-point operation (FLOP) count is performed using big $\mathcal{O}$ notation as a function of the dimensions of the correlation matrices involved, for the cascade algorithm implementations of the expressions as specified in Section \ref{section:integrated_approach-solution_strategy}. More specifically, the expressions (\ref{eq:MWF}), (\ref{eq:MWFext}), (\ref{eq:AECNR}), (\ref{eq:NRAEC_mod}), and (\ref{eq:NRextAEC}) in function of correlation matrices and pseudo-inverses thereof are considered in their frequency-domain formulation (cfr. Section \ref{section:practial_considerations}). Nevertheless, similar expressions exist for the time-domain expressions, and more numerically efficient computational implementations do exist to compute these expressions as, e.g., detailed in \cite{hanslerTopicsAcousticEcho2006}. The complexity of the different stages making up the algorithms is detailed first, wherein stages with similar complexity have been grouped together, leading the an overall computational complexity comparison. The resulting computational complexity for the different algorithms is also given in Table \ref{table:numerical_complexity}.
\begin{table}
\begin{center}
\begin{tabular}{ l l  l  l }
\hline
\textbf{Algorithm} & \multicolumn{3}{l}{\textbf{Computational complexity}} \\
\cmidrule{2-4}
& \textbf{MWF\textsubscript{ext}}/\textbf{NR\textsubscript{ext}} & \textbf{AEC} & \textbf{MWF}/\textbf{NR}/\textbf{PF}\\
\hline
MWF & & & $\mathcal{O}\left(M^3\right)$\\
MWF\textsubscript{ext} & $\mathcal{O}\left(\left(M+L\right)^3\right)$ & & \\
AEC-NR & & $\mathcal{O}\left(L^3\right)$ & $\mathcal{O}\left(M^3\right)$ \\
NR-AEC & & $\mathcal{O}\left(L^3\right)$ & $\mathcal{O}\left(M^3\right)$ \\
NR\textsubscript{ext}-AEC-PF & $\mathcal{O}\left(\left(M+L\right)^3\right)$ & $\mathcal{O}\left(L^3\right)$ & $\mathcal{O}\left(M^3\right)$
\end{tabular}
\end{center}
\caption{The computational complexity in big $\mathcal{O}$ notation of (\ref{eq:MWF}), (\ref{eq:MWFext}), (\ref{eq:AECNR}), (\ref{eq:NRAEC_mod}), and (\ref{eq:NRextAEC}) as a function of the dimensions of the correlation matrices involved. The NR\textsubscript{ext}-AEC-PF has the highest complexity, while the MWF has the lowest.}
\label{table:numerical_complexity}
\end{table}

\subsubsection{MWF\textsubscript{ext}/NR\textsubscript{ext}} The MWF\textsubscript{ext} and NR\textsubscript{ext} consider both the microphone and loudspeaker signals. Both algorithms require the updating of the correlation matrices using outer products ($\mathcal{O}(\left(M+L\right)^2)$), the computation of $\hat{R}_{\tilde{s}\tilde{s}}$ (MWF\textsubscript{ext}) or $\hat{R}_{\tilde{s}\tilde{s}}+\hat{R}_{\tilde{e}^s\tilde{e}^s}$ (NR\textsubscript{ext}) either using plain matrix subtraction ($\mathcal{O}((M+L)^2)$) or subtraction guided by a GEVD ($\mathcal{O}((M+L)^3)$), the computation of a pseudo-inverse ($O((M+L)^3)$), and a matrix-matrix multiplication ($\mathcal{O}((M+L)^3)$) \cite{golub_matrix_2013}. The filter application requires an inner product ($\mathcal{O}(M+L)$) \cite{golub_matrix_2013}. Thus, the MWF\textsubscript{ext} and NR\textsubscript{ext} have a complexity of $\mathcal{O}((M+L)^3)$, although the explicit computation of the pseudo-inverse could be avoided using the Woodbury identity \cite{golub_matrix_2013}.
\subsubsection{AEC} The AEC requires the updating of the loudspeaker-loudspeaker and loudspeaker-echo correlation matrices using outer products ($\mathcal{O}(L^2)$ and $O(LM)$), the computation of the pseudo-inverse of this loudspeaker-loudspeaker correlation matrix ($\mathcal{O}(L^3)$), and a matrix-matrix multiplication of this pseudo-inverse with the loudspeaker-echo correlation matrix ($\mathcal{O}(L^2M)$). The filter application requires an inner product and vector subtraction ($\mathcal{O}(2L+M)$). Thus, a complexity of $\mathcal{O}(L^3)$ is obtained, although the explicit computation of the pseudo-inverse could again be avoided using the Woodbury identity, thereby resorting to RLS or other adaptive filter implementations \cite{golub_matrix_2013}.
\subsubsection{MWF/NR/PF} The MWF, NR, and PF require similar computations as the MWF\textsubscript{ext} and NR\textsubscript{ext} but only consider the microphone signals, resulting in a complexity of $\mathcal{O}(M^3)$.
\subsubsection{Comparison} As the MWF only involves the microphone signals, it is computationally the cheapest. The NR\textsubscript{ext}-AEC-PF also involves loudspeaker signals, and requires three stages, such that it is computationally the most expensive. The computational complexity of the MWF\textsubscript{ext}, AEC-NR, and NR-AEC is intermediate, e.g., the NR\textsubscript{ext}-AFC-Pf can be interpreted as the AEC-NR with prior NR\textsubscript{ext} stage. When using efficient adaptive filtering implementations to compute the AEC, and when there is only one reference microphone, the NR-AEC is more computationally efficient than the AEC-NR as only one AEC filter per loudspeaker is required rather than one AEC filter for each loudspeaker-microphone pair \cite{reuvenJointAcousticEcho2004,docloCombinedAcousticEcho2000,luisvaleroLowComplexityMultiMicrophoneAcoustic2019a}.
\section{Experiment design} \label{section:experimental_procedures}
Although the MWF\textsubscript{ext}, AEC-NR, NR-AEC and NR\textsubscript{ext}-AEC-PF are theoretically equivalent, they differ in their practical estimation as described in Section \ref{section:practial_considerations}, such that their practical performances will differ. To experimentally validate these performance differences, the algorithms are compared to one another in the acoustic scenarios described in Section \ref{section:experimental_procedures-acoustic_scenarios}, using the algorithm settings described in Section \ref{section:experimental_procedures-algorithmic_settings}, and the performance measures described in Section \ref{section:experimental_procedures-performance_measures}. As the main focus of this paper is on algorithms derived from signal models, i.e., (\ref{eq:signal_model_def_x}) and (\ref{eq:signal_model_def_x_tilde}), the comparison will focus on these model-based algorithms. The reader is referred to \cite{seidel_convergence_2024}, for a comparison of the benefits and downsides of model-based, hybrid and data-driven algorithms. A relative comparison between these algorithms, and the impact of variations in signal-to-noise ratio (SNR) and signal-to-echo ratio (SER) is focussed upon, rather than aiming at maximal performance.

\subsection{Acoustic scenarios} \label{section:experimental_procedures-acoustic_scenarios}
\subsubsection{Setup-1} Five scenarios with varying desired speech source, near-end room noise source and loudspeaker positions are considered in a $\SI{5}{\meter}\times \SI{5}{\meter}\times \SI{3}{\meter}$ room with a wall reflection coefficient of $0.15$ ($T_{60}=\SI{0.11}{\second}$), corresponding to the scenarios examined in \cite{roebbenCascadedNoiseReduction2024}. To this end, the source-to-microphone and loudspeaker-to-microphone impulse responses of length $128$ samples are created using the randomised image method (RIM) \cite{desenaModelingRectangularGeometries2015} with a $16\cdot10^{3}$ $\SI{}{\hertz}$ sampling rate and randomised distances of $\SI{0.13}{\meter}$. $M = 2$ microphones are positioned at $\begin{bmatrix} 2 & 1.9 & 1\end{bmatrix}$$\SI{}{\meter}$ and $\begin{bmatrix} 2 & 1.8 & 1\end{bmatrix}$$\SI{}{\meter}$, while the position of the one desired speech source, one near-end room noise source and $L = 2$ loudspeakers are varied by placing these sources at congruent angles in a circle with a $\SI{0.2}{\meter}$ radius around the mean microphone position. The desired speech source in the loudspeakers consists of sentences of the hearing in noise test (HINT) database, concatenated with $\SI{5}{\second}$ of silence \cite{nilssonDevelopmentHearingNoise1994}, and the near-end room noise source consists of babble noise to model competing speakers \cite{auditecAuditoryTestsRevised1997}. Additionally, the far-end room speech component in the loudspeakers consists of HINT sentences, and the far-end room noise component in the loudspeakers consists of white noise, e.g., to model sensor and far-end room noise, of which the power ratio is set to $\SI{0}{\decibel}$. The relative power ratio between the echo signals in the microphones is set to $\SI{0}{\decibel}$. All signals are $\SI{30}{\second}$ long. 
The power ratio between the far-end room speech and noise components in the loudspeakers, and the power ratio between the echo signals both equal $\SI{0}{\decibel}$. The input SNR (SNR\textsuperscript{in}) and SER (SER\textsuperscript{in}) ratio in the reference microphone are varied between $\SI{-15}{\decibel}$ and $\SI{15}{\decibel}$.
\subsubsection{Setup-2} As in Setup-1, five scenarios with varying desired speech source, near-end room noise source and loudspeaker positions are considered. However, rather than using simulated impulse responses, measured impulse responses from the meeting room in the Aachen impulse response (AIR) database are selected \cite{jeub_binaural_2009}. As desired speech source and speech component in the loudspeakers $\SI{30}{\second}$ long sequences of HINT sentences are used \cite{nilssonDevelopmentHearingNoise1994}, and as noise source and noise component in the loudspeakers office noise from the diverse environments multi-channel acoustic noise database (DEMAND) is used \cite{thiemann2013diverse}. One noise source and one loudspeaker are considered, for which the SNR\textsuperscript{in} and SER\textsuperscript{in} are set to $\SI{5}{\decibel}$ and $\SI{0}{\decibel}$ respectively. The power ratio between the far-end room speech and noise components in the loudspeakers is set to $\SI{0}{\decibel}$ before feeding them through a (mild) non-linearity $\frac{\text{arctan}\!\left(\alpha(\Vec{l}^s+\Vec{l}^n)\right)}{\alpha}$ with $\alpha=1$ to model loudspeaker non-linearities as per \cite{seidel_convergence_2024}. Two microphones are considered.

\subsection{Algorithm settings} \label{section:experimental_procedures-algorithmic_settings}
In order to avoid the correlation matrix initialisation influencing the performance of the algorithms, and to focus on the steady-state performance, the correlation matrices are estimated across the entire $\SI{30}{\second}$ of data. 

Furthermore, the filters are calculated in the short-time Fourier transform (STFT) domain, the algorithms can also be implemented in the time domain, thereby replacing the index $z$ with frequency-bin index $f$ and frame index $k$, although as noted in Section \ref{section:cascaded_approach} and Section \ref{section:practial_considerations}, the algorithms can be equivalently implemented in the time domain. A squared root Hann window, window size of $512$ samples, window shift of $256$ samples and a sampling rate of $16\cdot10^{3}$ $\SI{}{\hertz}$ are used in Setup-1. To accommodate for the larger impulse response length in the AIR database, a window size of $2048$ samples and window shift of $1024$ samples is used in Setup-2.

As one desired speech source is assumed, the rank of $\hat{R}_{ss}^{\{\text{s} \! = \! 1, \text{e\textsuperscript{s}} \! = \! 1\}}(f)$ in the MWF\textsubscript{ext}, in the NR of the AEC-NR and the NR-AEC, and in the PF of the NR\textsubscript{ext}-AEC-PF is enforced to be equal to one by using a GEVD \cite{serizelLowrankApproximationBased2014}. Similarly, as one desired speech source and two independent loudspeaker sources are considered, the rank of $\hat{R}_{\tilde{s}\tilde{s}}^{\{\text{s} \! = \! 1, \text{e\textsuperscript{s}} \! = \! 1\}}(f)+\hat{R}_{\tilde{e}^s\tilde{e}^s}^{\{\text{s} \! = \! 1, \text{e\textsuperscript{s}} \! = \! 1\}}(f)$ in the NR\textsubscript{ext} of the NR\textsubscript{ext}-AEC-PF is enforced to equal three by using a GEVD. Indeed, in \cite{serizelLowrankApproximationBased2014}, it is shown that the rank of these estimated correlation matrices should be enforced to equal the number of sources using GEVD approximations. As by the assumption $\hat{R}_{l^sl^s}(f)\hat{R}_{ll}(f)^\dagger \hat{R}_{le}(f)=\hat{R}_{l^se^s}(f)$, $\begin{bmatrix}
	\mathbb{I}_{M\times M} & 0_{M\times L}
\end{bmatrix}\hat{W}_{\text{NR\textsubscript{ext}}}(f)\begin{bmatrix}
	{0}_{M\times L}\\\mathbb{I}_{L\times L}
\end{bmatrix}=0_{M\times M}$ (Section \ref{section:integrated_approach-solution_strategy-NRext_AEC}), this zero-structure is enforced. Corresponding to the literature, e.g., \cite{gustafssonCombinedAcousticEcho1998a,docloCombinedAcousticEcho2000,ruizDistributedCombinedAcoustic2022}, the algorithms are implemented in their cascade configuration, feeding the processed signals from one stage in the cascade to the next. The NR-AEC is thus implemented in its modified form (\ref{eq:NRAEC_mod}).

Ideal VADs for the desired speech and the far-end room speech component in the echo are assumed in Setup-1 to restrain the influence of VAD errors. In Setup-2, ideal VADs are compared to a NN-based VAD from \cite{ravanelli_speechbrain_2021}. Loudspeaker signals are fed to the NN to obtain the VAD flags related to $\text{VAD}_{e^s}$, and microphone signals are fed to the NN to obtain the VAD flags related to $\text{VAD}_s$.

\subsection{Performance measures} \label{section:experimental_procedures-performance_measures}
Noise reduction performance is calculated using the intelligibility-weighted SNR improvement $\left(\Delta \text{SNR\textsuperscript{I}}\right)$, echo reduction performance using the intelligibility-weighted SER improvement $\left(\Delta \text{SER\textsuperscript{I}}\right)$ and speech distortion (SD) performance using the intelligibility-weighted speech distortion $\left(\text{SD\textsuperscript{I}}\right)$ \cite{greenbergIntelligibilityweightedMeasuresSpeechtointerference1993,sprietAdaptiveFilteringTechniques2004a}, which are calculated as
{\allowdisplaybreaks\begin{subequations} \label{eq:metrics_delta}
	\begin{align}
		\Delta \text{SNR\textsuperscript{I}} &= \sum_{n_o=1}^{N_o}I_{n_o} \left(\text{SNR\textsubscript{$n_o$}\textsuperscript{\hspace{-1.3em}out}}-\text{SNR\textsubscript{$n_o$}\textsuperscript{\hspace{-1.3em}in}}\;\right)\\
		\Delta \text{SER\textsuperscript{I}} &= \sum_{n_o=1}^{N_o}I_{n_o} \left(\text{SER\textsubscript{$n_o$}\textsuperscript{\hspace{-1.3em}out}}-\text{SER\textsubscript{$n_o$}\textsuperscript{\hspace{-1.3em}in}}\;\right)\\
		\text{SD\textsuperscript{I}} &= \sum_{n_o=1}^{N_o}I_{n_o}\text{SD\textsubscript{$n_o$}}_{\textstyle \raisebox{2pt}{,}}
	\end{align}
\end{subequations}}
with,
\begin{equation} \label{eq:metrics_definitions}
	\begin{IEEEeqnarraybox}{rll} 
		\text{SNR\textsubscript{$n_o$}} &= 10\text{log}_{10}\left(\frac{P_{n_o}^{s}}{P_{n_o}^n}\right)\text{, }
		\text{SER\textsubscript{$n_o$}} = 10\text{log}_{10}\left(\frac{P_{n_o}^{s}}{P_{n_o}^e}\right)\\
		&\text{and } \text{SD\textsubscript{$n_o$}} = 10\text{log}_{10}\left(\frac{P_{n_o}^{s,\text{out}}}{P_{n_o}^{s,\text{in}}}\right)_{\textstyle \raisebox{2pt}{.}}
	\end{IEEEeqnarraybox}
\end{equation}
Herein, $P^s_{n_0}$, $P^n_{n_0}$ and $P^e_{n_0}\in\mathbb{R}$ refer to the desired speech, near-end room noise and echo signal powers in the $n_0$th one-third octave band $n_0\in\{1,\cdots,N_0\}$. The SNR, SER and SD measures are intelligibility-weighted using the band importance $I_{n_o}\in\mathbb{R}$ of \cite[Table 3]{acousticalsocietyofamericaANSIS3519971997}. A higher $\Delta \text{SER\textsuperscript{I}}$, $\Delta \text{SNR\textsuperscript{I}}$, and an $\text{SD\textsuperscript{I}}$ closer to zero indicates higher performance. 

The extended short-time objective intelligibility (ESTOI) improvement $\left(\Delta\text{ESTOI}=\text{ESTOI\textsuperscript{out}-ESTOI\textsuperscript{in}}\right)$ is applied as it predicts intelligibility when desired speech is corrupted with modulated interferers, such as echo and noise \cite{jensenAlgorithmPredictingIntelligibility2016}. Speech quality is evaluated with the perceptual evaluation of speech quality (PESQ) \cite{itu_pesq}. To support these results using a different model, intelligibility and quality are also evaluated with the hearing-aid speech perception index (HASPI) version 2 improvement $\left(\Delta\text{HASPI}=\text{HASPI\textsuperscript{out}-HASPI\textsuperscript{in}}\right)$ and hearing-aid speech quality index (HASQI) version 2 improvement $\left(\Delta\text{HASQI}=\text{HASQI\textsuperscript{out}-HASQI\textsuperscript{in}}\right)$ \cite{katesHearingAidSpeechPerception2021,katesHearingAidSpeechQuality2014}. For HASPI and HASQI no hearing loss is assumed. Higher $\Delta$ESTOI, $\Delta$PESQ, $\Delta$HASPI, and $\Delta$HASQI indicates higher performance. 

Finally, in Setup-2, the NN-based acoustic echo cancellation mean opinion score (AECMOS) is also evaluated \cite{aecmos_2022}. To this end, $\Delta$MOS\textsubscript{echo} and $\Delta$MOS\textsubscript{other} measure improvement in AEC performance similar to $\Delta$SER\textsuperscript{I} and improvement in degradation due to other causes such as noise and speech distortion, thereby bundling $\Delta$SNR\textsuperscript{I} and SD\textsuperscript{I}. Higher $\Delta$MOS\textsubscript{echo} and $\Delta$MOS\textsubscript{other} indicates higher performance.

\section{Results and discussion} \label{section:results_and_discussion}
As the NR\textsubscript{ext}-AEC-PF constitutes a new algorithm, Section \ref{section:results_and_discussion-NRext} first separately studies the performance of this algorithm, detailing the performance implications of each filter in the cascade. Thereafter, the performance of the integrated algorithms is compared to one another in Section \ref{section:results_and_discussion-comparison}.

\subsection{NR\textsubscript{ext} precedes AEC and PF (NR\textsubscript{ext}-AEC-PF)} \label{section:results_and_discussion-NRext}
\begin{figure}
	\centering
	\includegraphics[width=0.9\linewidth]{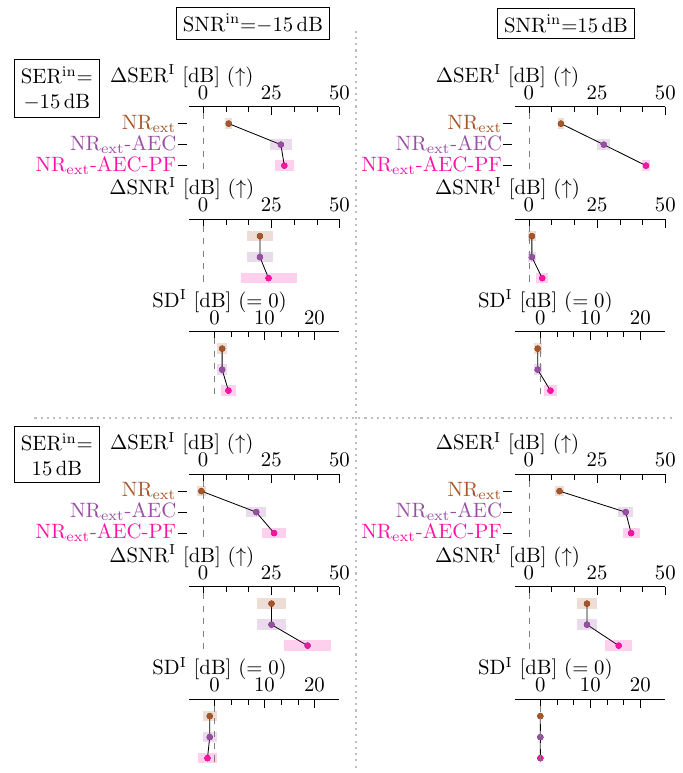}
	\caption{Performance attained by each filter in the NR\textsubscript{ext}-AEC-PF by means of the intelligibility-weighted signal-to-echo ratio improvement $\left(\Delta\text{SER\textsuperscript{I}}\right)$, intelligibility-weighted signal-to-noise ratio improvement $\left(\Delta\text{SNR\textsuperscript{I}}\right)$, and intelligibility weighted speech distortion. The mean performance is represented by the solid dots and the standard deviation by the shading. The performance generally increases with each filter in the cascade.}
	\label{fig:NRext}
\end{figure}

Fig. \ref{fig:NRext} shows the performance attained by each filter in the NR\textsubscript{ext}-AEC-PF cascade algorithm implementation to investigate the improvement offered by each filter in Setup-1. 

The NR\textsubscript{ext} already improves the SER\textsuperscript{I} and SNR\textsuperscript{I} as the NR\textsubscript{ext} aims at jointly suppressing the far-end room noise component in the echo and the near-end room noise. This SER\textsuperscript{I} improvement is larger for echo-dominant than for noise-dominant settings, and vice versa for the SNR\textsuperscript{I} improvement, since the NR\textsubscript{ext} attributes more degrees of freedom towards suppressing the power-dominant interferer. 

The AEC further improves the SER\textsuperscript{I} as the AEC aims at suppressing the echo, and as the excess error in the AEC due to noise perturbations is diminished by prior application of the NR\textsubscript{ext} filter. The SNR\textsuperscript{I} and SD\textsuperscript{I} remain unaffected by the AEC, as the AEC does not filter the microphone signals, and so does not alter the near-end room noise or the desired speech.

The PF aims at further suppressing the near-end room noise and the echo, such that the PF improves both the SNR\textsuperscript{I} and SER\textsuperscript{I} with respect to the NR\textsubscript{ext}-AEC. This SNR\textsuperscript{I} and SER\textsuperscript{I} improvement comes at the expense of a slightly increased SD\textsuperscript{I} as both the NR\textsubscript{ext} and the PF distort the desired speech.

\subsection{Comparison} \label{section:results_and_discussion-comparison}
\subsubsection{Setup-1}
\begin{figure*}
	\centering
	\subfloat[Intelligibility-weighted signal-to-echo ratio improvement $\left(\Delta\text{SER\textsuperscript{I}}\right)$, intelligibility-weighted signal-to-noise ratio improvement $\left(\Delta\text{SNR\textsuperscript{I}}\right)$, and intelligibility weighted speech distortion $\left(\text{SD\textsuperscript{I}}\right)$]
	{\includegraphics[width=0.9\linewidth]{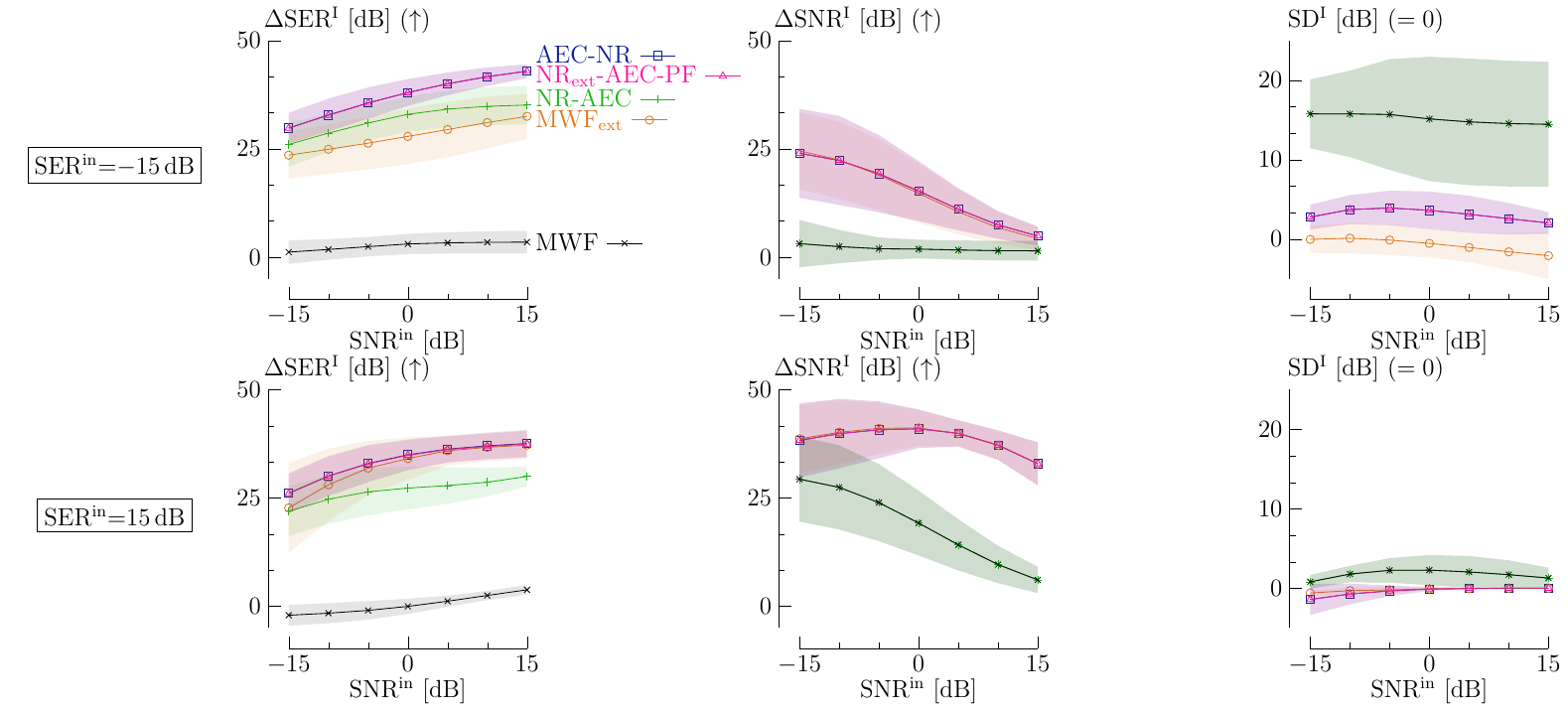}\label{fig:1noise_2loudspeaker_comparison_SNRI_SERI_SDI}
	}\vspace{-0.3em}
	\subfloat[Extended short-time objective intelligibility improvement $\left(\Delta\text{ESTOI}\right)$, hearing-aid speech perception index version 2 improvement $\left(\Delta\text{HASPI}\right)$, hearing-aid speech quality index version 2 improvement $\left(\Delta\text{HASQI}\right)$, and perceptual evaluation of speech quality improvement $\left(\Delta\text{PESQ}\right)$]{
		\includegraphics[width=0.9\linewidth]{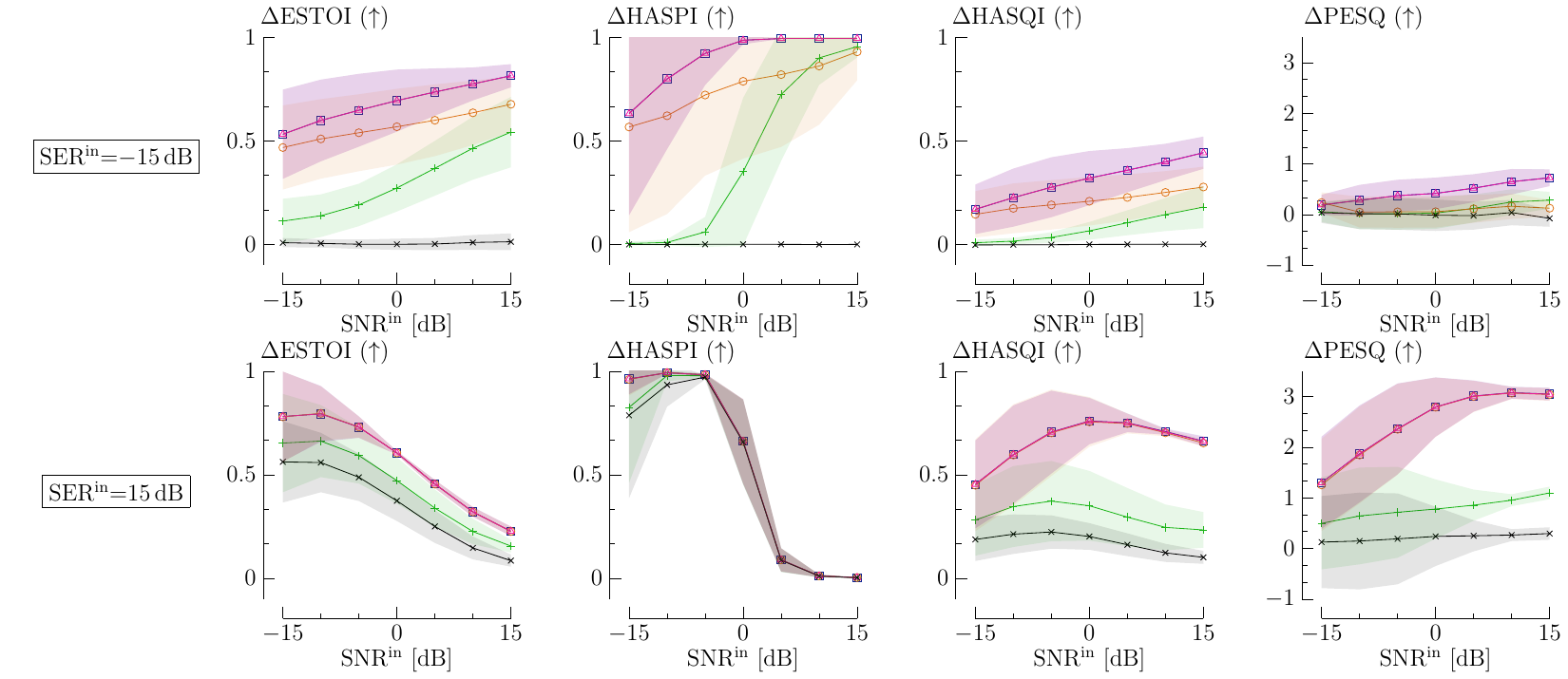}
		\label{fig:1noise_2loudspeaker_comparison_ESTOI_HASPI_HASQI}
	}
	\caption{Performance comparison between the integrated algorithms for Setup-1. The mean performance is represented by the solid line and the standard deviation by the shading. Due to non-stationarities and imperfect correlation matrix estimation the theoretically equivalent integrated algorithms differ in their practical performances. To this end, the AEC-NR and NR\textsubscript{ext}-AEC-PF generally attain the best performance.}
	\label{fig:1noise_2loudspeaker_comparison}
\end{figure*}

Fig. \ref{fig:1noise_2loudspeaker_comparison} compares the performance of the integrated algorithms, as according to Section \ref{section:practial_considerations} the theoretically equivalent algorithms of Section \ref{section:integrated_approach} differ practically due to non-stationarities and imperfect correlation matrix estimation. To this end, Fig. \ref{fig:1noise_2loudspeaker_comparison_SNRI_SERI_SDI} shows the echo cancellation, noise reduction and speech distortion performance separately, which are amalgamated into the instrumental measures of Fig. \ref{fig:1noise_2loudspeaker_comparison_ESTOI_HASPI_HASQI}. The AEC-NR and NR\textsubscript{ext}-AEC-PF generally attain top performance.

The MWF attains lower performance than the AEC-NR, as the MWF does not take the loudspeaker information into account. Indeed there are only two microphones, such that the MWF can only completely cancel one interfering source. However, as there are two loudspeakers and one near-end room noise source, there are three interfering sources, such that the MWF lacks the degrees of freedom to cancel these interfering sources. This effect is mostly apparent at $\text{SER\textsuperscript{in}}=\SI{-15}{\decibel}$, as the two echo sources then dominate the one near-end room noise source, leading to low SNR\textsuperscript{I}, SER\textsuperscript{I} and instrumental measure improvements, and high SD\textsuperscript{I}. While the NR in the AEC-NR can also be interpreted as an MWF after preprocessing by an AEC, this NR suffers less from a lack of degrees of freedom as the AEC already partially suppresses the echo, such that the AEC-NR scales with the number of loudspeakers. 

The MWF\textsubscript{ext} also scales with the number of loudspeakers, in this way increasing the SNR\textsuperscript{I}, SER\textsuperscript{I} and instrumental measures and decreasing the SD\textsuperscript{I} with respect to the MWF. Nevertheless, the MWF\textsubscript{ext} SER\textsuperscript{I} improvement is lower than for the AEC-NR. As described in Section \ref{section:practial_considerations}, the MWF\textsubscript{ext} suffers from loudspeaker and echo non-stationarities, while the AEC in the AEC-NR is less affected, thereby attaining a lower SER\textsuperscript{I} improvement than the AEC-NR. This effect is primarily visible at $\text{SER\textsuperscript{in}}=\SI{-15}{\decibel}$ as the echo then dominates the noise. 

The NR-AEC realises the same decreased SNR\textsuperscript{I} improvement and high SD\textsuperscript{I} with respect to the AEC-NR as the MWF. Indeed, the NR of the NR-AEC in its modified form (\ref{eq:NRAEC_mod}) is identical to the MWF. Adding an AEC after the NR increases the SER\textsuperscript{I} compared to the MWF, but as the AEC neither alters the near-end room noise nor the desired speech, the AEC does not alter the SNR\textsuperscript{I} or the SD\textsuperscript{I}. Despite the addition of the AEC, the SER\textsuperscript{I} improvement of the NR-AEC is also inferior to the AEC-NR, as the NR in the NR-AEC already distorts the desired speech, and as the AEC needs to track the NR next to the echo paths. Due to the limited SNR\textsuperscript{I} improvement and high SD\textsuperscript{I}, the instrumental measure improvement of the NR-AEC is also limited, as mostly apparent at $\text{SER\textsuperscript{in}}=\SI{-15}{\decibel}$ as the decreased performance of the NR is then mostly apparent. 

The NR\textsubscript{ext}-AEC-PF also incorporates a filter to suppress the near-end room noise preceding the AEC as for the NR-AEC, namely the NR\textsubscript{ext}. However, as opposed to the MWF, the NR\textsubscript{ext} next to suppressing the near-end room noise also suppresses the far-end room noise component in the echo, and therefore scales with the number of loudspeakers, resulting in less SD\textsuperscript{I} and an increased SNR\textsuperscript{I} with respect to the MWF. The SNR\textsuperscript{I} improvement of the NR\textsubscript{ext}-AEC-PF additionally decreases less quickly with an increasing SNR\textsuperscript{in} than the MWF. In fact, the NR\textsubscript{ext}-AEC-PF attains similar performance to the AEC-NR. While, the AEC operates under reduced noise due to the addition of the NR\textsubscript{ext}, the AEC, although theoretically independent of the NR\textsubscript{ext}, is in practice affected by the NR\textsubscript{ext} due to the GEVD and imperfect correlation matrix estimation. This effect is, nevertheless, much more limited than for the NR-AEC due to this theoretical independence between the NR\textsubscript{ext} and the AEC as also argued in \cite{roebbenCascadedNoiseReduction2024}. Both effects of reduced noise and NR\textsubscript{ext} dependence, however, seem to balance each other out, such that the NR\textsubscript{ext}-AEC-PF and the AEC-NR achieve a similar performance gain. Alternatively, the NR\textsubscript{ext}-AEC-PF can also be interpreted as an AEC-NR preceded by an NR\textsubscript{ext}, such that preceding an AEC-NR by an NR\textsubscript{ext} does not seem to further increase performance.

\subsubsection{Setup-2} 
\begin{figure}
\subfloat[Ideal VADs]
	{\includegraphics[width=0.9\linewidth]{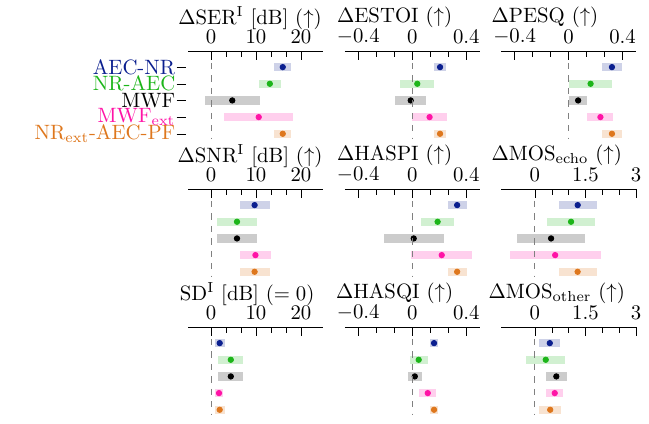}\label{fig:complex_scenario_VAD_power}
	}	\vspace{-0.3em}
\subfloat[Practical (NN-based) VADs]
	{\includegraphics[width=0.9\linewidth]{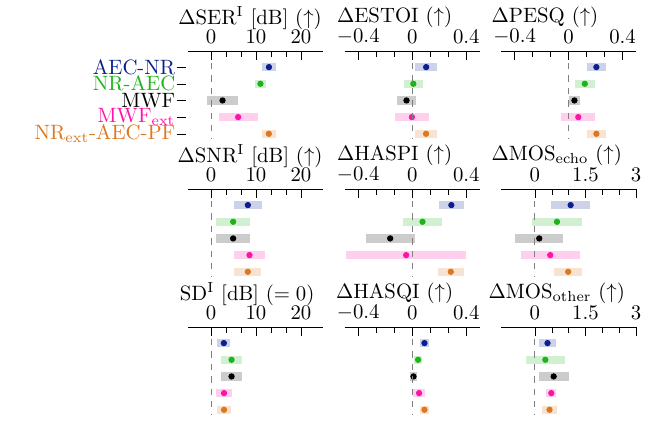}\label{fig:complex_scenario_VAD_DNN}
	}		
	\caption{Performance comparison for Setup-2. The NR\textsubscript{ext}-AEC-PF is still effective despite the additive map assumption no longer holding true due to loudspeaker non-linearities, and performs similar to the AEC-NR. The mean performance only drops slightly when practical VADs are used.}
	\label{fig:complex_scenario}
\end{figure}

Fig. \ref{fig:complex_scenario} compares the performance of the integrated algorithms for measured, rather than simulated, meeting room impulse responses, for (mild) loudspeaker non-linearities, and for office noise from the DEMAND database. The numerical results for each of the scenarios is also available in Supplementary material Section \Romannum{12}. In Fig. \ref{fig:complex_scenario_VAD_power} ideal VADs are still assumed available. The relative differences between the algorithms is similar as in Fig. \ref{fig:1noise_2loudspeaker_comparison}, and again the AEC-NR and NR\textsubscript{ext}-AEC-PF generally attain the highest performance. So although the additive map assumption of (\ref{eq:additive_map}), necessary to prove optimality of the NR\textsubscript{ext}-AEC-PF as detailed in Section \ref{section:integrated_approach-solution_strategy-NRext_AEC}, is no longer satisfied due to the addition of non-linear processing to model loudspeaker non-linearities, the NR\textsubscript{ext}-AEC-PF is still shown to be effective. Compared to Fig. \ref{fig:1noise_2loudspeaker_comparison}, the NR-AEC is also closer to the AEC-NR due to there being only one rather than two noise sources and there being two microphones. Nevertheless, the AEC filters in the NR-AEC should still model the prior NR stage, such that the AEC-NR attains superior performance. Finally, the absolute performance of the algorithms is lower in Setup-2 than in Setup-1 due to the STFT window size being smaller than the length of the impulse responses, which degades performance of the frequency-domain implementation \cite{avargelSystemIdentificationShortTime2007}. Indeed, when the echo path impulse response becomes smaller than the window size, this echo path cannot be modelled exactly by a per-frequency bin estimate \cite{avargelSystemIdentificationShortTime2007}. Further increasing the STFT window size, or switching to a time-domain implementation, as discussed in Section \ref{section:cascaded_approach} and Section \ref{section:practial_considerations}, would improve the performance.
In Fig. \ref{fig:complex_scenario_VAD_DNN}, the same scenarios are considered as in Fig. \ref{fig:complex_scenario_VAD_power} but with practical NN-based VADs. As the performance drops only slightly with respect to the ideal VAD case, the VAD availability seems justifiable in practice.

In conclusion, the AEC-NR and NR\textsubscript{ext}-AEC-PF generally attain top performance as the AEC (theoretically) only needs to model the echo paths, and as the NR operates with reduced echo due to the prior application of the AEC (and NR\textsubscript{ext}). 
 
\section{Conclusion} \label{section:conclusion}
In this paper, an integrated approach is proposed for the combined acoustic echo cancellation (AEC) and noise reduction (NR) problem in the general multi-microphone/multi-loudspeaker setup with possible linear dependencies between the loudspeaker and microphone signals. This integrated approach is achieved by selecting a single signal model of either the microphone signal vector or the extended signal vector by stacking microphone and loudspeaker signals, formulating a single mean squared error cost function, and using a common solution strategy. Employing the microphone signal model, an MWF is derived. Similarly, employing the extended signal, an extended MWF (MWF\textsubscript{ext}) is derived, for which several theoretically equivalent expressions are found that turn out to be interpretable as specific cascade algorithms. More specifically, the MWF\textsubscript{ext} is shown to be theoretically equivalent to the AEC preceding the NR (AEC-NR), the NR preceding the AEC (NR-AEC), and the extended NR (NR\textsubscript{ext}) preceding the AEC and a post-filter (PF) (NR\textsubscript{ext}-AEC-PF). Under rank-deficiency conditions the MWF\textsubscript{ext} is non-unique, such that this theoretical equivalence amounts to the expressions being specific, not necessarily minimum-norm solutions, for this MWF\textsubscript{ext}.

Nevertheless, although the MWF\textsubscript{ext}, AEC-NR, NR-AEC and NR\textsubscript{ext}-AEC-PF are theoretically equivalent, their practical performances differ due to non-stationarities and imperfect correlation matrix estimation, leading to the AEC-NR and NR\textsubscript{ext}-AEC-PF generally attaining best overall performance.

\section*{Acknowledgements}
We thank Prof. J. Kates for providing the MATLAB implementations for the HASPI and HASQI measures.

\bibliographystyle{IEEEtran} 
\bibliography{ref.bib}
\begin{IEEEbiography}[{\includegraphics[width=1in,height=1.25in,clip,keepaspectratio]{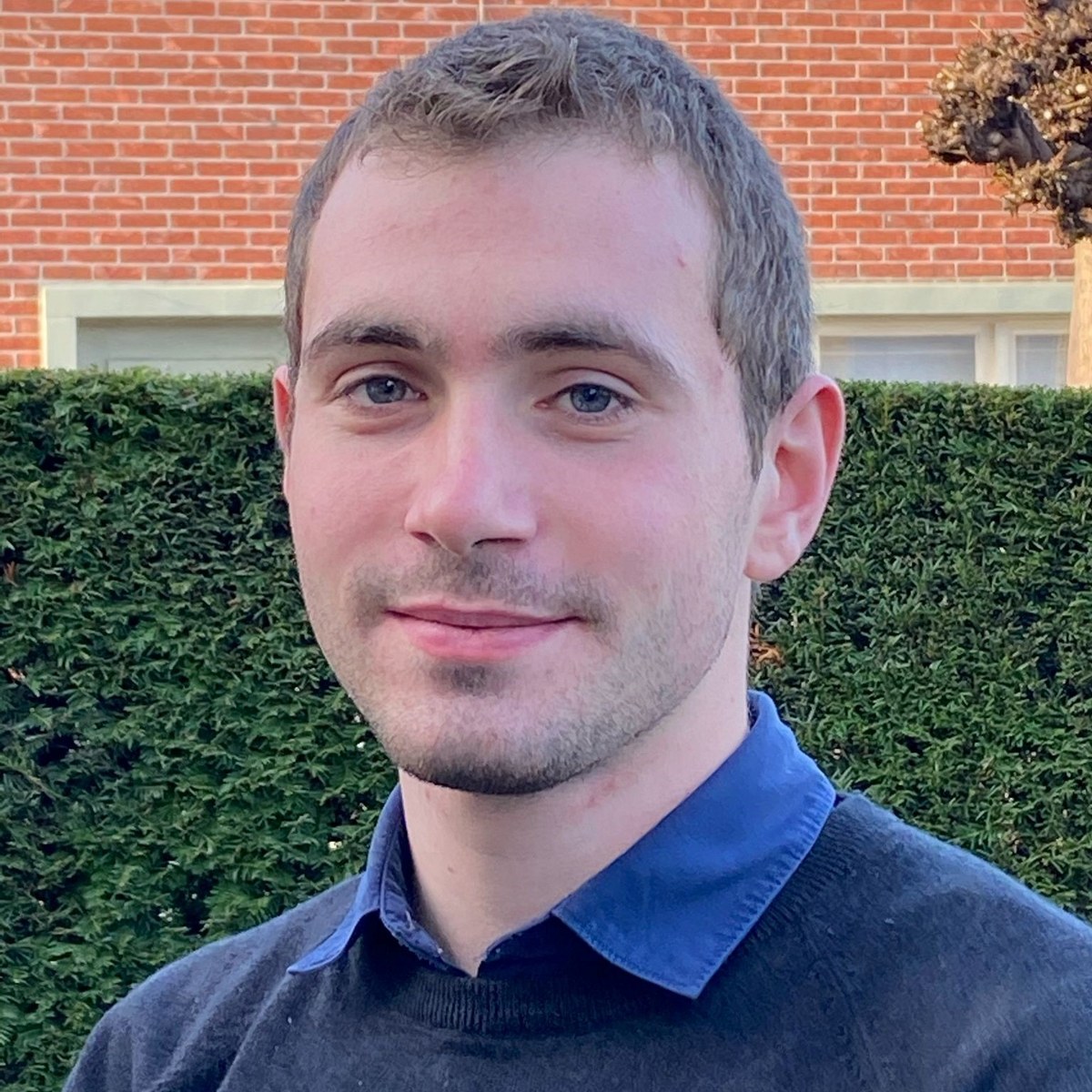}}]
{Arnout Roebben}
obtained a Bachelor of Engineering, and a Master of Electrical Engineering from KU Leuven, Belgium, in 2020 and 2022, respectively. He has been a Ph.D. researcher at KU Leuven since September 2022. His research is a joint venture between the Department of Electrical Engineering and the Department of Neurosciences, and focuses on the design of integrated digital signal processing algorithms. 
\end{IEEEbiography}

\begin{IEEEbiography}[{\includegraphics[width=1in,height=1.25in,clip,keepaspectratio]{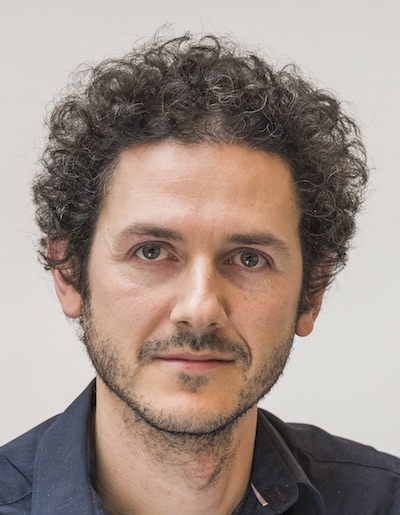}}]
{Toon van Waterschoot} received MSc (2001) and PhD (2009) degrees in Electrical Engineering, both from KU Leuven, Belgium, where he is currently a Professor. His research interests are in signal processing, machine learning, and numerical optimization, applied to acoustic signal enhancement, acoustic modeling, audio analysis, and audio reproduction. He has been the Scientific Coordinator for several major European research projects: the FP7-PEOPLE Marie Curie Initial Training Network "Dereverberation and Reverberation of Audio, Music, and Speech (DREAMS, 2013-2016)", the H2020 ERC Consolidator Grant "The Spatial Dynamics of Room Acoustics (SONORA, 2018-2023)", and the H2020 MSCA European Training Network "Service-Oriented Ubiquitous Network-Driven Sound (SOUNDS, 2021-2025)".

He has been serving as an Associate Editor for the Journal of the Audio Engineering Society and for the EURASIP Journal on Audio, Music, and Speech Processing. He is Executive Director of the European Association for Signal Processing (EURASIP) and Founding Member of the EAA Technical Committee in Audio Signal Processing. He was the General Chair of the 60th AES International Conference in Leuven, Belgium (2016), and has been serving on the Organizing Committee of the European Conference on Computational Optimization (EUCCO 2016), the IEEE Workshop on Applications of Signal Processing to Audio and Acoustics (WASPAA 2017), and the 28th and 29th European Signal Processing Conferences (EUSIPCO 2020 and 2021). He is a member of EURASIP, IEEE, ASA, and AES.

\end{IEEEbiography}

\begin{IEEEbiography}[{\includegraphics[width=1in,height=1.25in,clip,keepaspectratio]{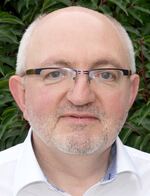}}]
{Jan Wouters}
is a Full Professor at the Dept. of Neurosciences, KU Leuven, Leuven, Belgium, since 2005 and received the Master’s and Ph.D. degrees in physics from the University of Leuven, KU Leuven, in 1982 and 1989, respectively. From 1989 to 1992, he was a Postdoctoral Research Fellow with the Belgian National Fund for Scientific Research, Institute of Nuclear Physics, Université de Louvain UCL and at NASA Goddard Space Flight Center, USA. His research interests include hearing sciences, audiology and auditory neural processing, signal processing for cochlear implants and hearing aids. He is on the Editorial Board of the International Journal of Audiology.

Jan Wouters served as President and Secretary-General of the European Federation of Audiological Societies (EFAS), as President of the Belgian Audiological Society (B-Audio), and as Executive Board member of the International Collegium for Rehabilitative Audiology (ICRA) and of the Dutch Acoustical Society (NAG). He is an elected member of the International Collegium for ORL (CORLAS). He is a Honorary Member of the Deutsche Gesellschaft für Audiologie DGA (2020) and received the EFAS Lifetime Achievement Award (2021).
\end{IEEEbiography}

\begin{IEEEbiography}[{\includegraphics[width=1in,height=1.25in,clip,keepaspectratio]{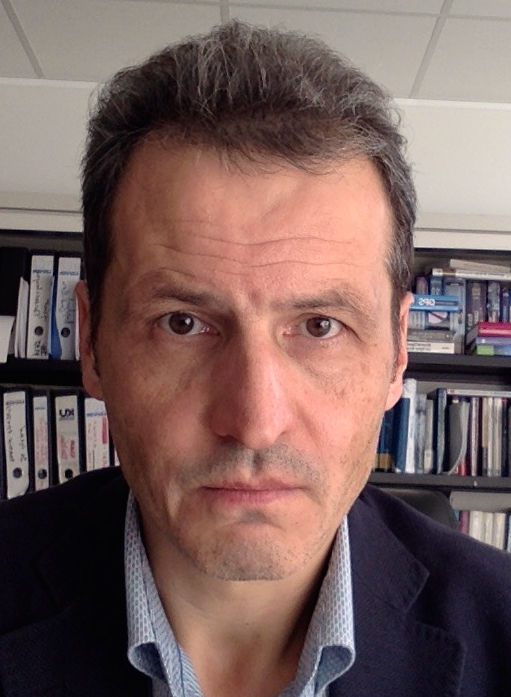}}]
{Marc Moonen}
is a Full Professor with the Electrical Engineering Department, KU Leuven, Belgium. He is a Fellow of the IEEE (2007), a Fellow of EURASIP (European Association for Signal Processing, 2018) and the 2024 recipient of the IEEE Signal Processing Society Claude Shannon-Harry Nyquist Technical Achievement Award. He was President of EURASIP (2007-2008 and 2011-2012), Vice-President for Publications of the IEEE Signal Processing Society (2021-2023) and is currently President-Elect of the IEEE Signal Processing Society (2026-2027). He has served as Editor-in-Chief for the EURASIP Journal on Applied Signal Processing (2003-2005), Area Editor for Feature Articles in the IEEE Signal Processing Magazine (2012-2014), and has been a member of the Editorial Board of Signal Processing, IEEE Transactions on Circuits and Systems-II, IEEE Signal Processing Magazine, EURASIP Journal on Wireless Communications and Networking, and EURASIP Journal on Advances in Signal Processing. 
\end{IEEEbiography}

\clearpage
\newpage
	
\section*{Integrated Minimum Mean Squared Error Algorithms for Combined Acoustic Echo Cancellation and Noise Reduction: Supplementary Material} 
\setcounter{page}{18}

This supplementary material relates to the paper \textit{Integrated minimum mean squared error algorithms for combined acoustic echo cancellation and noise reduction}. For consistency with the main text, all equation, citation, section and page numbers in this supplementary material are compatible with the corresponding numbers in the main text.

In Section \ref{section:supp_material_AECNR_Rxxinv}, it will be shown that (23) corresponds to a valid generalised inverse in general, and to a pseudo-inverse when $\text{rank}\!\left(\Sigma_{mm}\right)=M$. In Section \ref{section:supp_material_NRextAEC}, it will subsequently be shown that (31) is equivalent to (22) if $\text{rank}\!\left(\Sigma_{mm}\right)=M$ as the MWF\textsubscript{ext} is then uniquely defined and $R_{\tilde{m}\tilde{m}}^{g,g}$ corresponds to the uniquely-defined pseudoinverse. If $\text{rank}\!\left(\Sigma_{mm}\right)<M$, the MWF\textsubscript{ext} (22) is not unique, (31) then still corresponds to one of the solutions of MWF\textsubscript{ext}, be it not necessarily the minimum-norm solution. In Section \ref{section:supp_material_NRextAEC-lin}, it will be shown that (31) simplifies to (33) when the echo path is linear and the AEC filters are chosen sufficiently long to model this echo path, i.e., $\Vec{e}=F_{\text{lin}}\Vec{l}$ and $R_{ee}=R_{ll}F_{\text{lin}}^H$. Finally, in Section \ref{section:supp-numerical_results}, numerical results are provided for Fig. 5.

\setcounter{section}{8}
\setcounter{equation}{40}
\section{Proof of (23)} \label{section:supp_material_AECNR_Rxxinv}
In this section, it is shown that $R_{\tilde{m}\tilde{m}}^{g,g}$ (23) corresponds to a valid generalised inverse of $R_{\tilde{m}\tilde{m}}$ \cite{jamesGeneralisedInverse1978} in general, and to a pseudo-inverse of $R_{\tilde{m}\tilde{m}}$ \cite{penroseGeneralizedInverseMatrices1955} if $\text{rank}\!\left(\Sigma_{mm}\right)=M$.

$R_{\tilde{m}\tilde{m}}$ can be expanded as
\begin{equation} \label{eq:Rmm_expanded}
	\begin{IEEEeqnarraybox}{rll}
		R_{\tilde{m}\tilde{m}} = &\begin{bmatrix}
			\mathbb{I}_{M\times M} & R_{el}R_{ll}^\dagger\\ 0_{L\times M} & \mathbb{I}_{L\times L}
		\end{bmatrix}\cdot\\ &\begin{bmatrix}
			\Sigma_{mm} & 0_{M\times L}\\ 0_{L\times M} & R_{ll}
		\end{bmatrix}\cdot\\ &\begin{bmatrix}
			\mathbb{I}_{M\times M} & 0_{M\times L}\\ R_{ll}^\dagger R_{le} & \mathbb{I}_{L\times L}\end{bmatrix}_{\textstyle \raisebox{2pt}{.}}
	\end{IEEEeqnarraybox}
\end{equation}
Similarly, $R_{\tilde{m}\tilde{m}}^{g,g}$ can be expanded as
\begin{equation}\label{eq:Rmmg_expanded}
	\begin{IEEEeqnarraybox}{rll}
		R_{\tilde{m}\tilde{m}}^{g,g} = &\begin{bmatrix} \mathbb{I}_{M\times M} & 0_{M\times L} \\ -R_{ll}^\dagger R_{le} & \mathbb{I}_{L\times L}\end{bmatrix}\cdot \\ 
		&\begin{bmatrix}
			\Sigma_{mm}^\dagger & 0_{M\times L}\\ 0_{L\times M} & R_{ll}^\dagger
		\end{bmatrix}\cdot \\ 
		&\begin{bmatrix}
			\mathbb{I}_{M\times M} & -R_{el}R_{ll}^\dagger\\ 0_{L\times M} & \mathbb{I}_{L\times L}
		\end{bmatrix}_{\textstyle \raisebox{2pt}{.}}
	\end{IEEEeqnarraybox}
\end{equation}
The condition \ref{section:RmmRmmgRmm} has to be satisfied for $R_{\tilde{m}\tilde{m}}^{g,g}$ to be a generalised inverse and the conditions \ref{section:RmmRmmgRmm}-\ref{section:RmmgRmm} have to be satisfied for $R_{\tilde{m}\tilde{m}}^{g,g}$ to be a pseudo-inverse.

\begin{definition}{$R_{\tilde{m}\tilde{m}}R_{\tilde{m}\tilde{m}}^{g,g}R_{\tilde{m}\tilde{m}}=R_{\tilde{m}\tilde{m}}$} \label{section:RmmRmmgRmm}
	
	This condition, which is necessary and sufficient for generalised inverses \cite{jamesGeneralisedInverse1978}, and necessary for pseudo-inverses \cite{penroseGeneralizedInverseMatrices1955}, is satisfied for (23). 
\end{definition} 
\begin{proof}
	Using (\ref{eq:Rmm_expanded}) and (\ref{eq:Rmmg_expanded}), $R_{\tilde{m}\tilde{m}}R_{\tilde{m}\tilde{m}}^{g,g}R_{\tilde{m}\tilde{m}}$ can be expanded as
	\begin{equation} \label{eq:supp1_1}
		\begin{IEEEeqnarraybox}{rll}
			R_{\tilde{m}\tilde{m}}R_{\tilde{m}\tilde{m}}^{g,g}R_{\tilde{m}\tilde{m}} = &\begin{bmatrix}
				\mathbb{I}_{M\times M} & R_{el}R_{ll}^\dagger\\ 0_{L\times M} & \mathbb{I}_{L\times L}
			\end{bmatrix}\begin{bmatrix}
				\Sigma_{mm} & 0_{M\times L}\\ 0_{L\times M} & R_{ll}
			\end{bmatrix}\cdot\\ &\begin{bmatrix}
				\mathbb{I}_{M\times M} & 0_{M\times L}\\ R_{ll}^\dagger R_{le} & \mathbb{I}_{L\times L}\end{bmatrix}\begin{bmatrix} \mathbb{I}_{M\times M} & 0_{M\times L} \\ -R_{ll}^\dagger R_{le} & \mathbb{I}_{L\times L}\end{bmatrix} \cdot\\
			& \begin{bmatrix}
				\Sigma_{mm}^\dagger & 0_{M\times L}\\ 0_{L\times M} & R_{ll}^\dagger
			\end{bmatrix}\begin{bmatrix}
				\mathbb{I}_{M\times M} & -R_{el}R_{ll}^\dagger\\ 0_{L\times M} & \mathbb{I}_{L\times L}
			\end{bmatrix} \cdot\\
			&\begin{bmatrix}
				\mathbb{I}_{M\times M} & R_{el}R_{ll}^\dagger\\ 0_{L\times M} & \mathbb{I}_{L\times L}
			\end{bmatrix}\begin{bmatrix}
				\Sigma_{mm} & 0_{M\times L}\\ 0_{L\times M} & R_{ll}
			\end{bmatrix}\cdot \\ &\begin{bmatrix}
				\mathbb{I}_{M\times M} & 0_{M\times L}\\ R_{ll}^\dagger R_{le} & \mathbb{I}_{L\times L}\end{bmatrix}_{\textstyle \raisebox{2pt}{,}}
		\end{IEEEeqnarraybox}
	\end{equation}
	which can be simplified using $\begin{bmatrix}\mathbb{I}_{M\times M} & 0_{M\times L}\\ R_{ll}^\dagger R_{le} & \mathbb{I}_{L\times L}\end{bmatrix}\begin{bmatrix} \mathbb{I}_{M\times M} & 0_{M\times L} \\ -R_{ll}^\dagger R_{le} & \mathbb{I}_{L\times L}\end{bmatrix}=\mathbb{I}_{\left(M+L\right)\times \left(M+L\right)}$ and $\begin{bmatrix} \mathbb{I}_{M\times M} & -R_{el}R_{ll}^\dagger \\ 0_{L\times M} & \mathbb{I}_{L\times L}\end{bmatrix}\begin{bmatrix}\mathbb{I}_{M\times M} & R_{el}R_{ll}^\dagger\\ 0_{L\times M} & \mathbb{I}_{L\times L}\end{bmatrix}=\mathbb{I}_{\left(M+L\right)\times\left(M+L\right)}$. (\ref{eq:supp1_1}) can then indeed be simplified to
	\begin{equation} \label{eq:sup1_2}
		\begin{IEEEeqnarraybox}{rll}
			R_{\tilde{m}\tilde{m}}R_{\tilde{m}\tilde{m}}^{g,g}R_{\tilde{m}\tilde{m}} = &\begin{bmatrix}
				\mathbb{I}_{M\times M} & R_{el}R_{ll}^\dagger\\ 0_{L\times M} & \mathbb{I}_{L\times L}
			\end{bmatrix}\cdot \\ &\begin{bmatrix}
				\Sigma_{mm}\Sigma_{mm}^\dagger\Sigma_{mm} & 0_{M\times L}\\ 0_{L\times M} & R_{ll}R_{ll}^\dagger R_{ll}
			\end{bmatrix}\cdot \\ &\begin{bmatrix}
				\mathbb{I}_{M\times M} & 0_{M\times L}\\ R_{ll}^\dagger R_{le} & \mathbb{I}_{L\times L}\end{bmatrix}_{\textstyle \raisebox{2pt}{.}}
		\end{IEEEeqnarraybox}
	\end{equation}
	Finally, $\Sigma_{mm}\Sigma_{mm}^\dagger\Sigma_{mm}=\Sigma_{mm}$ and $R_{ll}R_{ll}^\dagger R_{ll}=R_{ll}$ as per definition of the pseudo-inverse, such that (\ref{eq:sup1_2}) corresponds to
	\begin{equation}
		R_{\tilde{m}\tilde{m}}R_{\tilde{m}\tilde{m}}^{g,g}R_{\tilde{m}\tilde{m}}=R_{{\tilde{m}\tilde{m}}_{\textstyle \raisebox{2pt}{.}}}
	\end{equation}
\end{proof}

\begin{definition}{$R_{\tilde{m}\tilde{m}}^{g,g}R_{\tilde{m}\tilde{m}}R_{\tilde{m}\tilde{m}}^{g,g}=R_{\tilde{m}\tilde{m}}^{g,g}$}
	
This condition, which is necessary for pseudo-inverses \cite{penroseGeneralizedInverseMatrices1955}, is satisfied for (23). 
\end{definition}
\begin{proof}
Using (\ref{eq:Rmm_expanded}) and (\ref{eq:Rmmg_expanded}), $R_{\tilde{m}\tilde{m}}^{g,g}R_{\tilde{m}\tilde{m}}R_{\tilde{m}\tilde{m}}^{g,g}$ can be expanded as
\begin{equation} \label{eq:supp2_1}
	\begin{IEEEeqnarraybox}{rll}
		R_{\tilde{m}\tilde{m}}^{g,g}R_{\tilde{m}\tilde{m}}R_{\tilde{m}\tilde{m}}^{g,g} = &\begin{bmatrix} \mathbb{I}_{M\times M} & 0_{M\times L} \\ -R_{ll}^\dagger R_{le} & \mathbb{I}_{L\times L}\end{bmatrix}\begin{bmatrix}
			\Sigma_{mm}^\dagger & 0_{M\times L}\\ 0_{L\times M} & R_{ll}^\dagger
		\end{bmatrix}\cdot \\ 
		&\begin{bmatrix}
			\mathbb{I}_{M\times M} & -R_{el}R_{ll}^\dagger\\ 0_{L\times M} & \mathbb{I}_{L\times L}
		\end{bmatrix}\begin{bmatrix}
			\mathbb{I}_{M\times M} & R_{el}R_{ll}^\dagger\\ 0_{L\times M} & \mathbb{I}_{L\times L}
		\end{bmatrix}\cdot\\
		& \begin{bmatrix}
			\Sigma_{mm} & 0_{M\times L}\\ 0_{L\times M} & R_{ll}
		\end{bmatrix}\begin{bmatrix}
			\mathbb{I}_{M\times M} & 0_{M\times L}\\ R_{ll}^\dagger R_{le} & \mathbb{I}_{L\times L}\end{bmatrix}\cdot\\&\begin{bmatrix} \mathbb{I}_{M\times M} & 0_{M\times L} \\ -R_{ll}^\dagger R_{le} & \mathbb{I}_{L\times L}\end{bmatrix}\begin{bmatrix}
			\Sigma_{mm}^\dagger & 0_{M\times L}\\ 0_{L\times M} & R_{ll}^\dagger
		\end{bmatrix}\cdot \\ 
		&\begin{bmatrix}
			\mathbb{I}_{M\times M} & -R_{el}R_{ll}^\dagger\\ 0_{L\times M} & \mathbb{I}_{L\times L}
		\end{bmatrix}_{\textstyle \raisebox{2pt}{,}}
	\end{IEEEeqnarraybox}
\end{equation}
which can be simplified using $\begin{bmatrix} \mathbb{I}_{M\times M} & -R_{el}R_{ll}^\dagger \\ 0_{L\times M}& \mathbb{I}_{L\times L}\end{bmatrix}\begin{bmatrix}\mathbb{I}_{M\times M} & R_{el}R_{ll}^\dagger\\ 0_{L\times M} & \mathbb{I}_{L\times L}\end{bmatrix}=\mathbb{I}_{\left(M+L\right)\times \left(M+L\right)}$ and $\begin{bmatrix}\mathbb{I}_{M\times M} & 0_{M\times L}\\ R_{ll}^\dagger R_{le} & \mathbb{I}_{L\times L}\end{bmatrix}\begin{bmatrix} \mathbb{I}_{M\times M} & 0_{M\times L} \\ -R_{ll}^\dagger R_{le} & \mathbb{I}_{L\times L}\end{bmatrix}=\mathbb{I}_{\left(M+L\right)\times \left(M+L\right)}$. (\ref{eq:supp2_1}) can then indeed be simplified to
\begin{equation} \label{eq:sup2_2}
	\begin{IEEEeqnarraybox}{rll}
		R_{\tilde{m}\tilde{m}}^{g,g}R_{\tilde{m}\tilde{m}}R_{\tilde{m}\tilde{m}}^{g,g} = &\begin{bmatrix} \mathbb{I}_{M\times M} & 0_{M\times L} \\ -R_{ll}^\dagger R_{le} & \mathbb{I}_{L\times L}\end{bmatrix}\cdot \\ &\begin{bmatrix}
			\Sigma_{mm}^\dagger\Sigma_{mm}\Sigma_{mm}^\dagger & 0_{M\times L}\\ 0_{L\times M} & R_{ll}^\dagger R_{ll} R_{ll}^\dagger
		\end{bmatrix}\cdot \\ 
		&\begin{bmatrix}
			\mathbb{I}_{M\times M} & -R_{el}R_{ll}^\dagger\\ 0_{L\times M} & \mathbb{I}_{L\times L}
		\end{bmatrix}_{\textstyle \raisebox{2pt}{.}}
	\end{IEEEeqnarraybox}
\end{equation}
Finally, $\Sigma_{mm}^\dagger\Sigma_{mm}\Sigma_{mm}^\dagger=\Sigma_{mm}^\dagger$ and $R_{ll}^\dagger R_{ll} R_{ll}^\dagger=R_{ll}^\dagger$ as per definition of the pseudo-inverse, such that (\ref{eq:sup2_2}) corresponds to
\begin{equation}
	R_{\tilde{m}\tilde{m}}^{g,g}R_{\tilde{m}\tilde{m}}R_{\tilde{m}\tilde{m}}^{g,g}=R_{{\tilde{m}\tilde{m}}_{\textstyle \raisebox{2pt}{.}}}^{g,g}
\end{equation}
\end{proof}

\begin{definition}{$\left(R_{\tilde{m}\tilde{m}}R_{\tilde{m}\tilde{m}}^{g,g}\right)^H=R_{\tilde{m}\tilde{m}}R_{\tilde{m}\tilde{m}}^{g,g}$}
	
This condition, which is necessary for pseudo-inverses \cite{penroseGeneralizedInverseMatrices1955}, is only satisfied for (23) if $\text{rank}\!\left(\Sigma_{mm}\right)=M$.
\end{definition}
\begin{proof}
Using (\ref{eq:Rmm_expanded}) and (\ref{eq:Rmmg_expanded}), $R_{\tilde{m}\tilde{m}}R_{\tilde{m}\tilde{m}}^{g,g}$ can be expanded as
\begin{equation} \label{eq:sup3_1}
	\begin{IEEEeqnarraybox}{rll}
		R_{\tilde{m}\tilde{m}}R_{\tilde{m}\tilde{m}}^{g,g} =&\begin{bmatrix}
			\mathbb{I}_{M\times M} & R_{el}R_{ll}^\dagger\\ 0_{L\times M} & \mathbb{I}_{L\times L}
		\end{bmatrix}\begin{bmatrix}
			\Sigma_{mm} & 0_{M\times L}\\ 0_{L\times M} & R_{ll}
		\end{bmatrix}\cdot \\ &\begin{bmatrix}
			\mathbb{I}_{M\times M} & 0_{M\times L}\\ R_{ll}^\dagger R_{le} & \mathbb{I}_{L\times L}\end{bmatrix} \begin{bmatrix} \mathbb{I}_{M\times M} & 0_{M\times L} \\ -R_{ll}^\dagger R_{le} & \mathbb{I}_{L\times L}\end{bmatrix}\cdot\\ &\begin{bmatrix}
			\Sigma_{mm}^\dagger & 0_{M\times L}\\ 0_{L\times M} & R_{ll}^\dagger
		\end{bmatrix}\begin{bmatrix}
			\mathbb{I}_{M\times M} & -R_{el}R_{ll}^\dagger\\ 0_{L\times M} & \mathbb{I}_{L\times L}
		\end{bmatrix}_{\textstyle \raisebox{2pt}{,}}
	\end{IEEEeqnarraybox}
\end{equation}
which can be simplified using $\begin{bmatrix}\mathbb{I}_{M\times M} & 0_{M\times L}\\ R_{ll}^\dagger R_{le} & \mathbb{I}_{L\times L}\end{bmatrix} \begin{bmatrix} \mathbb{I}_{M\times M} & 0_{M\times L} \\ -R_{ll}^\dagger R_{le} & \mathbb{I}_{L\times L}\end{bmatrix}=\mathbb{I}_{\left(M+L\right)\times\left(M+L\right)}$. (\ref{eq:sup3_1}) can then be simplified to
\begin{equation}
	R_{\tilde{m}\tilde{m}}R_{\tilde{m}\tilde{m}}^{g,g} = \begin{bmatrix}
		\Sigma_{mm}\Sigma_{mm}^\dagger & -\Sigma_{mm}\Sigma_{mm}^\dagger R_{el}R_{ll}^\dagger +R_{el}R_{ll}^\dagger\\ 0_{L\times M} & R_{ll}R_{ll}^\dagger
	\end{bmatrix}_{\textstyle \raisebox{2pt}{.}}
\end{equation} 
It is seen that $R_{\tilde{m}\tilde{m}}R_{\tilde{m}\tilde{m}}^{g,g}$ is not Hermitian unless $\text{rank}\!\left(\Sigma_{mm}\right)=M$. Indeed, if $\text{rank}\!\left(\Sigma_{mm}\right)=M$, the pseudo-inverse corresponds to a regular inverse, such that $\Sigma_{mm}\Sigma_{mm}^\dagger=\Sigma_{mm}\Sigma_{mm}^{-1}=\mathbb{I}_{M\times M}$ and
\begin{equation}
	R_{\tilde{m}\tilde{m}}R_{\tilde{m}\tilde{m}}^{g,g}=\begin{bmatrix}
		\mathbb{I}_{M\times M} & 0_{M\times L}\\ 0_{L\times M} & R_{ll}R_{ll}^\dagger
	\end{bmatrix}_{\textstyle \raisebox{2pt}{,}}
\end{equation}
which is Hermitian.
\end{proof}

\begin{definition}{$\left(R_{\tilde{m}\tilde{m}}^{g,g}R_{\tilde{m}\tilde{m}}\right)^H=R_{\tilde{m}\tilde{m}}^{g,g}R_{\tilde{m}\tilde{m}}$} \label{section:RmmgRmm}
	
This condition, which is necessary for pseudo-inverses \cite{penroseGeneralizedInverseMatrices1955}, is only satisfied for (23) if $\text{rank}\!\left(\Sigma_{mm}\right)=M$.
\end{definition} 

\begin{proof}
Using (\ref{eq:Rmm_expanded}) and (\ref{eq:Rmmg_expanded}), $R_{\tilde{m}\tilde{m}}^{g,g}R_{\tilde{m}\tilde{m}}$ can be expanded as
\begin{equation} \label{eq:sup4_1}
	\begin{IEEEeqnarraybox}{rll}
		R_{\tilde{m}\tilde{m}}^{g,g}R_{\tilde{m}\tilde{m}} =&\begin{bmatrix} \mathbb{I}_{M\times M} & 0_{M\times L} \\ -R_{ll}^\dagger R_{le} & \mathbb{I}_{L\times L}\end{bmatrix}\begin{bmatrix}
			\Sigma_{mm}^\dagger & 0_{M\times L}\\ 0_{L\times M} & R_{ll}^\dagger
		\end{bmatrix}\cdot\\ &\begin{bmatrix}
			\mathbb{I}_{M\times M} & -R_{el}R_{ll}^\dagger\\ 0_{L\times M} & \mathbb{I}_{L\times L}
		\end{bmatrix}\begin{bmatrix}
			\mathbb{I}_{M\times M} & R_{el}R_{ll}^\dagger\\ 0_{L\times M} & \mathbb{I}_{L\times L}
		\end{bmatrix}\cdot\\&\begin{bmatrix}
			\Sigma_{mm} & 0_{M\times L}\\ 0_{L\times M} & R_{ll}
		\end{bmatrix}\begin{bmatrix}
			\mathbb{I}_{M\times M} & 0_{M\times L}\\ R_{ll}^\dagger R_{le} & \mathbb{I}_{L\times L}\end{bmatrix}	_{\textstyle \raisebox{2pt}{,}}
	\end{IEEEeqnarraybox}
\end{equation}
which can be simplified using $\begin{bmatrix}
	\mathbb{I}_{M\times M} & -R_{el}R_{ll}^\dagger\\ 0_{L\times M} & \mathbb{I}_{L\times L}
\end{bmatrix}\begin{bmatrix}
	\mathbb{I}_{M\times M} & R_{el}R_{ll}^\dagger\\ 0_{L\times M} & \mathbb{I}_{L\times L}
\end{bmatrix}=\mathbb{I}_{\left(M+L\right)\times\left(M+L\right)}$. (\ref{eq:sup4_1}) can then indeed be simplified to
\begin{equation}
	R_{\tilde{m}\tilde{m}}^{g,g}R_{\tilde{m}\tilde{m}} = \begin{bmatrix}
		\Sigma_{mm}^\dagger \Sigma_{mm} & 0_{M\times L}\\ -R_{ll}^\dagger R_{le} \Sigma_{mm}^\dagger \Sigma_{mm} + R_{ll}^\dagger R_{le} & R_{ll}^\dagger R_{ll}
	\end{bmatrix}_{\textstyle \raisebox{2pt}{.}}
\end{equation} 
It is seen that $R_{\tilde{m}\tilde{m}}^{g,g}R_{\tilde{m}\tilde{m}}$ is not Hermitian unless $\text{rank}\!\left(\Sigma_{mm}\right)=M$. Indeed, if $\text{rank}\!\left(\Sigma_{mm}\right)=M$, the pseudo-inverse corresponds to a regular inverse, such that $\Sigma_{mm}^\dagger\Sigma_{mm}=\Sigma_{mm}^{-1}\Sigma_{mm}=\mathbb{I}_{M\times M}$ and
\begin{equation}
	R_{\tilde{m}\tilde{m}}^{g,g}R_{\tilde{m}\tilde{m}}=\begin{bmatrix}
		\mathbb{I}_{M\times M} & 0_{M\times L}\\ 0_{L\times M} & R_{ll}^\dagger R_{ll}
	\end{bmatrix}_{\textstyle \raisebox{2pt}{,}}
\end{equation}
which is Hermitian.
\end{proof}

\section{Proof of (31)}\label{section:supp_material_NRextAEC}
This section proves that the NR\textsubscript{ext}-AEC-PF (31) and MWF\textsubscript{ext} (22) are equivalent when choosing $R^{g,g}_{\tilde{m}\tilde{m}}$ as $R^g_{\tilde{m}\tilde{m}}$. 

The following theorem is used:
\begin{theorem}\label{theorem:theorem1}
	Define the Hermitian positive-semidefinite matrices $A\in\mathbb{C}^{M\times M}$ and $B\in\mathbb{C}^{M\times M}$, where the column space of $A$ is contained within the column space of $B$. Further define $C\in\mathbb{C}^{M\times L}$, whose column space is contained within the column space of $A$. Then, 
	\begin{equation}
		(ABA)^\dagger ABC=A^\dagger C_{\textstyle \raisebox{2pt}{.}}
	\end{equation}
\end{theorem}
\begin{proof}
	Since the column space of $C$ is contained within $A$, $C$ can be rewritten as $C=AP$ with $P=A^\dagger C$, such that $(ABA)^\dagger ABC = (ABA)^\dagger (ABA)P$. Herein, $(ABA)^\dagger (ABA)$ is a projection matrix on the row space of $ABA$. As $A$ and $B$ are Hermitian positive-semidefinite matrices, where the column space of $A$ is contained within the column space of $B$, the row space of $A$ equals the row space of $ABA$. Indeed, joint diagonalisation of $A$ and $B$ using a generalised eigenvalue decomposition (GEVD) with $Q\in\mathbb{C}^{M\times M}$ the generalised eigenvectors, $\Sigma_A\in\mathbb{C}^{R_A\times R_A}$ containing the $R_A$ non-zero generalised eigenvalues corresponding to $A$, and $\Sigma_B=\begin{bmatrix}
		\Sigma_{B_1} & 0_{R_A\times (R_B-R_A)}\\
		0_{(R_B-R_A)\times R_A} & \Sigma_{B_2}
	\end{bmatrix}\in\mathbb{C}^{R_B\times R_B}$ containing the $R_B$ non-zero generalised eigenvalues corresponding to $B$, yields
	\begin{subequations}
		\begin{align}
			A &= Q\begin{bmatrix}
				\Sigma_A & 0_{R_A\times (M-R_A)}\\
				0_{(M-R_A)\times R_A} & 0_{(M-R_A)\times(M-R_A)}\end{bmatrix} Q^H\\
			B &= Q\begin{bmatrix}
				\Sigma_B & 0_{R_B\times (M-R_B)}\\
				0_{(M-R_B)\times R_B} & 0_{(M-R_B)\times(M-R_B)}\end{bmatrix} Q^H_{\textstyle \raisebox{2pt}{.}}			
		\end{align}
	\end{subequations}
With $Q=\begin{bmatrix}
	\smash[b]{\underbrace{Q_{1}}_{R_A}} & \smash[b]{\underbrace{Q_{2}}_{(R_B-R_A)}} & \smash[b]{\underbrace{Q_{3}}_{(M-R_B)}}\end{bmatrix}$, $ABA$ can\\ \\ subsequently be written as
\begin{equation}
	\begin{aligned}
	&ABA =\\& Q_1 \left(\begin{bmatrix}
		\Sigma_A & 0_{R_A\times(R_B-R_A)}
	\end{bmatrix}\begin{bmatrix}Q_1 & Q_2\end{bmatrix}^H\begin{bmatrix}Q_1 & Q_2\end{bmatrix}\Sigma_{B}^{1/2}\right)\cdot\\ &\left(\Sigma_{B}^{1/2}\begin{bmatrix}Q_1 & Q_2\end{bmatrix}^H\begin{bmatrix}Q_1 & Q_2\end{bmatrix}\begin{bmatrix}
	\Sigma_A\\ 0_{(R_B-R_A)\times R_A}
\end{bmatrix}\right)Q^H_{1_{\textstyle \raisebox{2pt}{,}}}
\end{aligned}
\end{equation}
where both $Q_1$ and\\ $\left(\begin{bmatrix}
	\Sigma_A & 0_{R_A\times(R_B-R_A)}
\end{bmatrix}\begin{bmatrix}Q_1 & Q_2\end{bmatrix}^H\begin{bmatrix}Q_1 & Q_2\end{bmatrix}\Sigma_{B}^{1/2}\right)\cdot\\ \left(\Sigma_{B}^{1/2}\begin{bmatrix}Q_1 & Q_2\end{bmatrix}^H\begin{bmatrix}Q_1 & Q_2\end{bmatrix}\begin{bmatrix}
	\Sigma_A\\ 0_{(R_B-R_A)\times R_A}
\end{bmatrix}\right)$ are of rank $R_A$, such that the row space of $A$ equals the row space of $ABA$. Consequently, the projection matrix on the row space of $ABA$ corresponds to the projection matrix on the row space of $A$, such that, finally, $(ABA)^\dagger (ABA)=A^\dagger A$ and $(ABA)^\dagger (ABA)P=A^\dagger AP=A^\dagger C$.
\end{proof}

Using the assumption described in Section \Romannum{3}-C5 of the main text that $R_{ll}^\dagger R_{le}$ is a solution to the Wiener-Hopf equations $R_{l^sl^s}W=R_{l^se^s}$, i.e., $R_{l^sl^s}R_{ll}^\dagger R_{le}=R_{l^se^s}$, $W_{\text{NR\textsubscript{ext}}}=R_{\tilde{m}\tilde{m}}^{g,g}\left(R_{\tilde{s}\tilde{s}}+R_{\tilde{e}^s\tilde{e}^s}\right)$ corresponds to
\begin{equation} \label{eq:NRextAEC-WNRext2}
	W_{\text{NR\textsubscript{ext}}}=\begin{bmatrix}
		W_{\text{NR\textsubscript{ext}}}^{11} & W_{\text{NR\textsubscript{ext}}}^{12}\\
		W_{\text{NR\textsubscript{ext}}}^{21} & W_{\text{NR\textsubscript{ext}}}^{22}
	\end{bmatrix}_{\textstyle \raisebox{2pt}{,}}
\end{equation}
with
\begin{subequations}
	\begin{align} 
		W_{\text{NR\textsubscript{ext}}}^{11} &= \Sigma_{mm}^\dagger\left(R_{ss}+R_{e^se^s}-R_{el}R_{ll}^\dagger R_{l^se^s}\right)\\
		W_{\text{NR\textsubscript{ext}}}^{12} &= 0_{M\times L}\\
		W_{\text{NR\textsubscript{ext}}}^{21} &= -R_{ll}^\dagger R_{le}W_{\text{NR\textsubscript{ext}}}^{11}+R_{ll}^\dagger R_{l^se^s}\\
		W_{\text{NR\textsubscript{ext}}}^{22} &= R_{ll}^\dagger R_{{l^sl^s}_{\textstyle \raisebox{2pt}{.}}}
	\end{align}
\end{subequations}
Using (\ref{eq:NRextAEC-WNRext2}) and $R_{ll}^\dagger R_{ll} R_{ll}^\dagger=R_{ll}^\dagger$ as per definition of the pseudo-inverse, the correlation matrices $R_{m^`m^`}$, $R_{l^`l^`}$, $R_{m^`l^`}$, $R_{m^`l^`}R_{l^`l^`}^\dagger R_{l^`m^`}$, and $R_{l^`l^`}^\dagger R_{l^`m^`}$ in the NR\textsubscript{ext}-AEC-PF (31) can be rewritten in terms of the correlation matrices before applying the NR\textsubscript{ext}. Indeed, $R_{m^`m^`}$ can be rewritten as
\begin{subequations}
	\begin{align}
		R_{m^`m^`} &= W_{\text{NR\textsubscript{ext}}}^{11^H}R_{mm}W_{\text{NR\textsubscript{ext}}}^{11}+W_{\text{NR\textsubscript{ext}}}^{11^H}R_{el}W_{\text{NR\textsubscript{ext}}}^{21}\nonumber\\&+W_{\text{NR\textsubscript{ext}}}^{21^H}R_{le}W_{\text{NR\textsubscript{ext}}}^{11}+W_{\text{NR\textsubscript{ext}}}^{21^H}R_{ll}W_{\text{NR\textsubscript{ext}}}^{21}\\
		&= W_{\text{NR\textsubscript{ext}}}^{11^H}\Sigma_{mm}W_{\text{NR\textsubscript{ext}}}^{11}+R_{e^sl^s}R_{ll}^\dagger R_{{l^se^s}_{\textstyle \raisebox{2pt}{.}}} \label{eq:NRextAEC-Rxx}
	\end{align}
\end{subequations}
Similarly, $R_{l^`l^`}$ can be rewritten as
\begin{subequations}
	\begin{align}
		R_{l^`l^`} &= W_{\text{NR\textsubscript{ext}}}^{22^H}R_{ll}W_{\text{NR\textsubscript{ext}}}^{22}\\
		&= R_{l^sl^s}R_{ll}^\dagger R_{{l^sl^s}_{\textstyle \raisebox{2pt}{,}}} \label{eq:NRextAEC-Rll}
	\end{align}
\end{subequations}
and $R_{l^`m^`}$ can be rewritten as
\begin{subequations}
	\begin{align}
		R_{l^`m^`} &= W_{\text{NR\textsubscript{ext}}}^{22^H}R_{le}W_{\text{NR\textsubscript{ext}}}^{11}+W_{\text{NR\textsubscript{ext}}}^{22^H}R_{ll}W_{\text{NR\textsubscript{ext}}}^{21}\\
		&= R_{l^sl^s}R_{ll}^\dagger R_{{l^se^s}_{\textstyle \raisebox{2pt}{.}}} \label{eq:NRextAEC-Rlx}
	\end{align}
\end{subequations}
Combining (\ref{eq:NRextAEC-Rll}) and (\ref{eq:NRextAEC-Rlx}), $R_{m^`l^`}R_{l^`l^`}^\dagger R_{l^`m^`}$ then corresponds to
\begin{equation} \label{eq:NRextAEC-RxlRllRlx}
	\begin{IEEEeqnarraybox}{rll}
		R_{m^`l^`}R_{l^`l^`}^\dagger R_{l^`m^`} =& R_{e^sl^s}R_{ll}^\dagger R_{l^sl^s}\left(R_{l^sl^s}R_{ll}^\dagger R_{l^sl^s}\right)^\dagger\cdot\\& R_{l^sl^s}R_{ll}^\dagger R_{{l^se^s}_{\textstyle \raisebox{2pt}{.}}}
	\end{IEEEeqnarraybox}
\end{equation}
Using the assumption $R_{l^sl^s}R_{ll}^\dagger R_{le}=R_{l^se^s}$, (\ref{eq:NRextAEC-RxlRllRlx}) can further be expanded as
\begin{equation} \label{eq:NRextAEC-RxlRllRlx2}
	\begin{IEEEeqnarraybox}{rll}
		&R_{m^`l^`}R_{l^`l^`}^\dagger R_{l^`m^`} = R_{el}R_{ll}^\dagger \left(R_{l^sl^s}R_{ll}^\dagger R_{l^sl^s}\right)\cdot\\&\qquad \left(R_{l^sl^s}R_{ll}^\dagger R_{l^sl^s}\right)^\dagger \left( R_{l^sl^s}R_{ll}^\dagger R_{l^sl^s}\right)R_{ll}^\dagger R_{{le}_{\textstyle \raisebox{2pt}{,}}}
	\end{IEEEeqnarraybox}
\end{equation}
where $\left(R_{l^sl^s}R_{ll}^\dagger R_{l^sl^s}\right)\left(R_{l^sl^s}R_{ll}^\dagger R_{l^sl^s}\right)^\dagger \left( R_{l^sl^s}R_{ll}^\dagger R_{l^sl^s}\right)=\left(R_{l^sl^s}R_{ll}^\dagger R_{l^sl^s}\right)$ as per definition of the pseudo-inverse. Again using the assumption $R_{l^sl^s}R_{ll}^\dagger R_{le}=R_{l^se^s}$, (\ref{eq:NRextAEC-RxlRllRlx2}) can subsequently be rewritten as
\begin{equation}\label{eq:NRextAEC-RxlRllRlx3}
	R_{m^`l^`}R_{l^`l^`}^\dagger R_{l^`m^`} = R_{e^sl^s}R_{ll}^\dagger R_{{l^se^s}_{\textstyle \raisebox{2pt}{.}}}
\end{equation}
$R_{l^`l^`}^\dagger R_{l^`m^`}$ similarly corresponds to
\begin{equation} \label{eq:NRextAEC-RllRlx}
	R_{l^`l^`}^\dagger R_{l^`m^`} = \left(R_{l^sl^s}R_{ll}^\dagger R_{l^sl^s}\right)^\dagger R_{l^sl^s}R_{ll}^\dagger R_{{l^se^s}_{\textstyle \raisebox{2pt}{.}}}
\end{equation} 
As $R_{l^sl^s}$ and $R_{ll}$ are Hermitian positive-semidefinite matrices, the column space of $R_{l^sl^s}$ is contained within the column space of $R_{ll}$, and the column space of $R_{l^se^s}$ is contained within the column space of $R_{l^sl^s}$, Theorem \ref{theorem:theorem1} can thus be applied to (\ref{eq:NRextAEC-RllRlx}), resulting in
\begin{equation} \label{eq:NRextAEC-RllRlm}
	R_{l^`l^`}^\dagger R_{l^`m^`} = R_{l^sl^s}^\dagger R_{{l^se^s}_{\textstyle \raisebox{2pt}{.}}}
\end{equation}

Using these rewritten correlation matrices, it can be shown that $\left(R_{\tilde{s}\tilde{s}}+R_{\tilde{e}^s\tilde{e}^s}\right)\begin{bmatrix}
	\Sigma_{m^`m^`}^\dagger R_{s^`s} & 0_{M\times L}\\ -R_{l^`l^`}^\dagger R_{l^`m^`}\Sigma_{m^`m^`}^\dagger R_{s^`s} & 0_{L\times L}
\end{bmatrix}=R_{\tilde{s}\tilde{s}}$, as this expression can be expanded using (\ref{eq:NRextAEC-RllRlm}) as
\begin{equation}
	\begin{IEEEeqnarraybox}{rll}
		&\left(R_{\tilde{s}\tilde{s}}+R_{\tilde{e}^s\tilde{e}^s}\right)\begin{bmatrix}
			\Sigma_{m^`m^`}^\dagger R_{s^`s} & 0_{M\times L}\\ -R_{l^`l^`}^\dagger R_{l^`m^`}\Sigma_{m^`m^`}^\dagger R_{s^`s} & 0_{L\times L}
		\end{bmatrix} =\\ &\begin{bmatrix}
			\left(R_{ss}+R_{e^se^s}-R_{e^sl^s}R_{l^sl^s}^\dagger R_{l^se^s}\right)\Sigma_{m^`m^`}^\dagger R_{s^`s} & 0_{M\times L}\\ \left(R_{l^se^s}-R_{l^sl^s}R_{l^sl^s}^\dagger R_{l^se^s}\right)\Sigma_{m^`m^`}^\dagger R_{s^`s} & 0_{L\times L}
		\end{bmatrix}_{\textstyle \raisebox{2pt}{.}}
	\end{IEEEeqnarraybox}
\end{equation}
Here, $R_{l^se^s}-R_{l^sl^s}R_{l^sl^s}^\dagger R_{l^se^s}=0_{L\times M}$ as $R_{l^sl^s}R_{l^sl^s}^\dagger$ is a projection on the column space of $R_{l^sl^s}$ and the column space of $R_{l^se^s}$ is contained within the column space of $R_{l^sl^s}$. Furthermore, using (\ref{eq:NRextAEC-Rxx}) and (\ref{eq:NRextAEC-RxlRllRlx3}) $\left(R_{ss}+R_{e^se^s}-R_{e^sl^s}R_{l^sl^s}^\dagger R_{l^se^s}\right)\Sigma_{m^`m^`}^\dagger R_{s^`s}$ corresponds to
\begin{equation}
	\begin{IEEEeqnarraybox}{rll}
		&\left(R_{ss}+R_{e^se^s}-R_{e^sl^s}R_{l^sl^s}^\dagger R_{l^se^s}\right)\Sigma_{m^`m^`}^\dagger R_{s^`s} = \\& \left(R_{ss}+R_{e^se^s}-R_{e^sl^s}R_{l^sl^s}^\dagger R_{l^se^s}\right)\left(W^{11^H}_{\text{NR\textsubscript{ext}}}\Sigma_{mm}W^{11}_{\text{NR\textsubscript{ext}}}\right)^\dagger\cdot\\&W^{11^H}_{\text{NR\textsubscript{ext}}}R_{{ss}_{\textstyle \raisebox{2pt}{,}}}
	\end{IEEEeqnarraybox}
\end{equation}
which can be further expanded using (\ref{eq:NRextAEC-WNRext2}) and $\Sigma_{mm}^\dagger\Sigma_{mm}\Sigma_{mm}^\dagger=\Sigma_{mm}^\dagger$ as per definition of the pseudo-inverse as
\begin{equation} \label{eq:NRextAEC-Rsseq1}
	\begin{IEEEeqnarraybox}{rll}
		&\left(R_{ss}+R_{e^se^s}-R_{e^sl^s}R_{l^sl^s}^\dagger R_{l^se^s}\right)\Sigma_{m^`m^`}^\dagger R_{s^`s} = \\
		&\left(R_{ss}+R_{e^se^s}-R_{e^sl^s}R_{l^sl^s}^\dagger R_{l^se^s}\right)\cdot\\ &\biggl[\left(R_{ss}+R_{e^se^s}-R_{el}R_{ll}^\dagger R_{l^se^s}\right)\Sigma_{mm}^\dagger\cdot\\&\left(R_{ss}+R_{e^se^s}-R_{el}R_{ll}^\dagger R_{l^se^s}\right)\biggr]^\dagger\cdot\\&\left(R_{ss}+R_{e^se^s}-R_{el}R_{ll}^\dagger R_{l^se^s}\right)\Sigma_{mm}^\dagger R_{ss_{\textstyle \raisebox{2pt}{.}}}
	\end{IEEEeqnarraybox}
\end{equation}
Further, $R_{el}R_{ll}^\dagger R_{l^se^s}=R_{e^sl^s}R_{l^sl^s}^\dagger R_{l^se^s}$ due to the assumption $R_{l^sl^s}R_{ll}^\dagger R_{le}=R_{l^se^s}$ and due to $R_{l^sl^s}R_{l^sl^s}^\dagger R_{l^se^s}=R_{l^se^s}$, such that $R_{ss}$ + $R_{e^se^s}-R_{el}R_{ll}^\dagger R_{l^se^s}=R_{ss}$ + $R_{e^se^s}-R_{e^sl^s}R_{l^sl^s}^\dagger R_{l^se^s}$ can be seen to be a Hermitian matrix. Thus, since $R_{ss}+R_{e^se^s}-R_{e^sl^s}R_{l^sl^s}^\dagger R_{l^se^s}$ and $\Sigma_{mm}$ are Hermitian positive-semidefinite matrices, and since the column space of $R_{ss}$ + $R_{e^se^s}-R_{e^sl^s}R_{l^sl^s}^\dagger R_{l^se^s}$ is contained within the column space of $\Sigma_{mm}$, as per Theorem \ref{theorem:theorem1}, (\ref{eq:NRextAEC-Rsseq1}) corresponds to 
\begin{equation} \label{eq:NRextAEC-Rsseq3}
	\begin{IEEEeqnarraybox}{rll}
		&\left(R_{ss}+R_{e^se^s}-R_{e^sl^s}R_{l^sl^s}^\dagger R_{l^se^s}\right)\Sigma_{m^`m^`}^\dagger R_{s^`s} = \\
		&\left(R_{ss}+R_{e^se^s}-R_{e^sl^s}R_{l^sl^s}^\dagger R_{l^se^s}\right)\cdot\\ &\left(R_{ss}+R_{e^se^s}-R_{e^sl^s}R_{l^sl^s}^\dagger R_{l^se^s}\right)^\dagger R_{ss_{\textstyle \raisebox{2pt}{.}}}
	\end{IEEEeqnarraybox}
\end{equation}
Additionally, since the column space of $R_{ss}$ is contained within the column space of $R_{ss}+R_{e^se^s}-R_{e^sl^s}R_{l^sl^s}^\dagger R_{l^se^s}$, (\ref{eq:NRextAEC-Rsseq3}) corresponds to
\begin{equation} \label{eq:NRextAEC-Rss}
	\begin{IEEEeqnarraybox}{rll}
		&\left(R_{ss}+R_{e^se^s}-R_{e^sl^s}R_{l^sl^s}^\dagger R_{l^se^s}\right)\Sigma_{m^`m^`}^\dagger R_{s^`s} = R_{ss_{\textstyle \raisebox{2pt}{.}}}
	\end{IEEEeqnarraybox}
\end{equation}
Finally, using (\ref{eq:NRextAEC-Rss}), the MWF\textsubscript{ext} (22) is obtained, such that the NR\textsubscript{ext}-AEC-PF (31) is shown to correspond to the MWF\textsubscript{ext} (22) when $R_{\tilde{m}\tilde{m}}^{g,g}$ is selected for $R_{\tilde{m}\tilde{m}}^g$, i.e.,
\begin{equation}
	\begin{IEEEeqnarraybox}{rll}
		&R_{\tilde{m}\tilde{m}}^{g,g}\left(R_{\tilde{s}\tilde{s}}+R_{\tilde{e}^s\tilde{e}^s}\right)\begin{bmatrix}
			\mathbb{I}_{M\times M} & 0_{M\times L}\\ -R_{l^`l^`}^{\dagger}R_{l^`m^`} & 0_{L\times L}
		\end{bmatrix}\begin{bmatrix}
			\Sigma_{m^`m^`}^\dagger R_{s^`s}\\ {0}_{L\times M}
		\end{bmatrix}{\Vec{t}}_{r}\\
		&= R_{\tilde{m}\tilde{m}}^{g,g}R_{\tilde{s}\tilde{s}}\tilde{\Vec{t}}_{r_{\textstyle \raisebox{2pt}{.}}} 
	\end{IEEEeqnarraybox}
\end{equation}
\begin{equation}\nonumber
\IEEEQEDhereeqn
\end{equation}

\section{Proof of (33)} \label{section:supp_material_NRextAEC-lin}
This section proves that (33) is obtained from the general NR\textsubscript{ext}-AEC-PF (31) expression if the echo path is linear and the filters are sufficiently long to model this echo path, i.e., $\Vec{e}=F_{\text{lin}}\Vec{l}$ and $R_{le}=R_{ll}F_{\text{lin}}^H$.

With $\Vec{e}=F_{\text{lin}}\Vec{l}$, $R_{le}=R_{ll}F_{\text{lin}}^H$ and $R_{l^se^s}=R_{l^sl^s}F_{\text{lin}}^H$, the $W_{\text{NR\textsubscript{ext}}}$ filter (\ref{eq:NRextAEC-WNRext2}) simplifies to
\begin{equation}
	\begin{IEEEeqnarraybox}{rll}
		&W_{\text{NR\textsubscript{ext}}} =\\ &\qquad\begin{bmatrix}
			\left(R_{ss}+R_{nn}\right)^\dagger R_{ss} & 0_{M\times L}\\ -R_{ll}^\dagger R_{le}\left(R_{ss}+R_{nn}\right)^\dagger R_{ss} +R_{ll}^\dagger R_{l^se^s} & R_{ll}^\dagger R_{l^sl^s}
		\end{bmatrix}_{\textstyle \raisebox{2pt}{.}}
	\end{IEEEeqnarraybox}
\end{equation}
In the AEC, $R_{l^`l^`}^\dagger R_{l^`m^`}$ corresponds to $R_{l^sl^s}^\dagger R_{l^se^s}$ (cfr. Supplementary material Section \ref{section:supp_material_NRextAEC}). Finally, in the PF, $\Sigma_{m^`m^`}^\dagger R_{s^`s}$ corresponds to
\begin{equation}
	\begin{IEEEeqnarraybox}{rll}
		&\Sigma_{m^`m^`}^\dagger R_{s^`s} =\\&\qquad \left(R_{ss}\left(R_{ss}+R_{nn}\right)^\dagger R_{ss}\right)^\dagger R_{ss}\left(R_{ss}+R_{nn}\right)^\dagger R_{{ss}_{\textstyle \raisebox{2pt}{,}}}
	\end{IEEEeqnarraybox}
\end{equation} 
which can be further simplified using Theorem \ref{theorem:theorem1} as $R_{ss}+R_{nn}$ and $R_{ss}$ are Hermitian positive-semidefinite matrices, and the column space of $R_{ss}$ is contained within the column space of $R_{ss}+R_{nn}$. Thus, $\Sigma_{m^`m^`}^\dagger R_{s^`s}$ corresponds to
\begin{equation}
	\Sigma_{m^`m^`}^\dagger R_{s^`s} = R_{ss}^\dagger R_{{ss}_{\textstyle \raisebox{2pt}{.}}}
\end{equation}
Consequently, the general NR\textsubscript{ext}-AEC-PF (31) expression can be simplified to $\tilde{\Vec{w}}_{\text{int,lin}}$ with
\begin{equation} \label{eq:NRextAEC-lin2}
	\begin{IEEEeqnarraybox}{rll}
		\tilde{\Vec{w}}_{\text{int,lin}} &{=}\underbrace{\begin{bmatrix}
				\left(R_{ss}+R_{nn}\right)^\dagger R_{ss} & 0_{M\times L}\\ -R_{ll}^\dagger R_{le}\left(R_{ss}+R_{nn}\right)^\dagger R_{ss} + R_{ll}^\dagger R_{l^se^s} & R_{ll}^\dagger R_{l^sl^s}
		\end{bmatrix}}_{\text{NR\textsubscript{ext}}}\cdot\\
		&\underbrace{\begin{bmatrix}
				\mathbb{I}_{M\times M} & 0_{M\times L}\\ -R_{l^sl^s}^{\dagger}R_{l^se^s} & 0_{L\times L}
		\end{bmatrix}}_{\text{AEC}}\underbrace{\begin{bmatrix}
				R_{ss}^\dagger R_{ss}\\ {0}_{L\times M}
		\end{bmatrix}}_{\text{PF}}{\Vec{t}_{r_{\textstyle \raisebox{2pt}{.}}}}
	\end{IEEEeqnarraybox}
\end{equation}
Finally, as $R_{l^se^s}=R_{l^sl^s}R_{l^sl^s}^\dagger R_{l^se^s}$ (cfr. Supplementary material Section \ref{section:supp_material_NRextAEC}) and $R_{ss}R_{ss}^\dagger R_{ss}=R_{ss}$ as per definition of the pseudo-inverse, (33) is obtained
\begin{equation}
		\tilde{\Vec{w}}_{\text{int}} {=}{\underbrace{R_{\tilde{m}\tilde{m}}^{g,g}\left(R_{\tilde{s}\tilde{s}}+R_{\tilde{e}^s\tilde{e}^s}\right)}_{\text{NR\textsubscript{ext}}}\underbrace{\begin{bmatrix}
					\mathbb{I}_{M\times M}\\ -R_{l^sl^s}^{\dagger}R_{l^sl^s}F_{\text{lin}}^H
			\end{bmatrix}}_{\text{AEC}}}{\Vec{t}}_{r_{\textstyle \raisebox{2pt}{.}}}
\end{equation}
\begin{equation}\nonumber
\IEEEQEDhereeqn
\end{equation}

\section{Numerical results} \label{section:supp-numerical_results}
Table \ref{table:scenario2} and Table \ref{table:scenario3} show the numerical results, as summarised in Fig. 5(a) and Fig. 5(b), respectively.

\setcounter{table}{1}
\begin{table}
\subfloat[][$\Delta \text{SER\textsuperscript{I}} \text{ }{[\text{dB}]} \text{ }(\uparrow)$]{
\begin{tabular}{ l l  l  l l l}
\hline
\textbf{Algorithm} & \multicolumn{5}{l}{\textbf{Scenario}} \\
\cmidrule{2-6}
& \textbf{1} & \textbf{2} & \textbf{3} & \textbf{4} & \textbf{5}\\
\hline
MWF & 0.31 &0.59 & 12.92& 0.24& 9.57\\
MWF\textsubscript{ext} &1.33 &3.36 & 17.63&15.29 & 15.43\\
AEC-NR  & 13.35& 15.14&18.44 &16.17 & 16.64\\
NR-AEC  &13.29 & 13.58 & 14.97& 8.99& 14.57\\
NR\textsubscript{ext}-AEC-PF  &13.35 &15.14 &18.44 &16.17 & 16.64
\end{tabular}
\label{table:scenario2_ser}
}
\subfloat[][$\Delta \text{SNR\textsuperscript{I}} \text{ }{[\text{dB}]} \text{ }(\uparrow)$]{
\begin{tabular}{l  l  l l l}
\hline
\multicolumn{5}{l}{\textbf{Scenario}} \\
\cmidrule{1-5}
\textbf{1} & \textbf{2} & \textbf{3} & \textbf{4} & \textbf{5}\\
\hline
10.71& 10.43&1.62& 2.39& 3.75\\
13.72 &13.48 &6.45 & 8.91&6.82 \\
 12.86 &13.60 & 6.62& 8.76&6.63 \\
 10.71 & 10.43 & 1.62 &2.39  & 3.75\\
12.86 & 13.60&6.62 & 8.76& 6.63
\end{tabular}
\label{table:scenario2_snr}
} \\
\subfloat[][$\text{SD\textsuperscript{I}} \text{ }{[\text{dB}]} \text{ }(=0)$]{
\begin{tabular}{ l l  l  l l l}
\hline
\textbf{Algorithm} & \multicolumn{5}{l}{\textbf{Scenario}} \\
\cmidrule{2-6}
& \textbf{1} & \textbf{2} & \textbf{3} & \textbf{4} & \textbf{5}\\
\hline
MWF & 2.59&3.73&3.18&9.37&3.14\\
MWF\textsubscript{ext} & 2.90&2.71&0.94&1.27&1.10\\
AEC-NR & 3.09&3.11&1.03&1.35&1.07\\
NR-AEC & 2.59&3.73&3.18&9.37&3.14\\
NR\textsubscript{ext}-AEC-PF & 3.09&3.11&1.03&1.35&1.07
\end{tabular}
\label{table:scenario2_sd}
} \hfil
\subfloat[$\Delta \text{ESTOI} \text{ }(\uparrow)$]{
\begin{tabular}{ l l  l  l l l}
\hline
\multicolumn{5}{l}{\textbf{Scenario}} \\
\cmidrule{1-5}
\textbf{1} & \textbf{2} & \textbf{3} & \textbf{4} & \textbf{5}\\
\hline
0.01&0.06&0.08&-0.20&0.00\\
-0.03&0.02&0.26&0.20&0.19\\
0.15&0.21&0.27&0.20&0.20\\
0.10&0.15&0.09&-0.17&0.01\\
0.15&0.21&0.27&0.20&0.20
\end{tabular}
\label{table:scenario2_estoi}
} \hfil
\subfloat[$\Delta \text{HASPI}\text{ } (\uparrow)$]{
\begin{tabular}{ l l  l  l l l}
\hline
\textbf{Algorithm} & \multicolumn{5}{l}{\textbf{Scenario}} \\
\cmidrule{2-6}
& \textbf{1} & \textbf{2} & \textbf{3} & \textbf{4} & \textbf{5}\\
\hline
MWF & -0.07&0.24&0.12&-0.33&0.09\\
MWF\textsubscript{ext} & -0.15&0.17&0.31&0.43&0.33\\
AEC-NR & 0.24&0.34&0.32&0.43&0.33\\
NR-AEC & 0.30&0.32&0.08&0.05&0.19\\
NR\textsubscript{ext}-AEC-PF & 0.24&0.34&0.32&0.43&0.33
\end{tabular}
\label{table:scenario2_haspi}
}\hfil
\subfloat[$\Delta \text{HASQI}\text{ } (\uparrow)$]{
\begin{tabular}{ l l  l  l l l}
\hline
\multicolumn{5}{l}{\textbf{Scenario}} \\
\cmidrule{1-5}
\textbf{1} & \textbf{2} & \textbf{3} & \textbf{4} & \textbf{5}\\
\hline
0.04&0.08&0.03&-0.06&0.00\\
0.03&0.07&0.17&0.16&0.15\\
0.12&0.19&0.17&0.16&0.16\\
0.09&0.13&0.03&-0.04&0.02\\
0.12&0.19&0.17&0.16&0.16
\end{tabular}
\label{table:scenario2_hasqi}
}\\
\subfloat[$\Delta \text{PESQ} \text{ }(\uparrow)$]{
\begin{tabular}{ l l  l  l l l}
\hline
\textbf{Algorithm} & \multicolumn{5}{l}{\textbf{Scenario}} \\
\cmidrule{2-6}
& \textbf{1} & \textbf{2} & \textbf{3} & \textbf{4} & \textbf{5}\\
\hline
MWF & 0.13&0.09&0.10&-0.05&0.08\\
MWF\textsubscript{ext} & 0.08&0.21&0.34&0.27&0.27\\
AEC-NR & 0.26&0.43&0.37&0.26&0.28\\
NR-AEC & 0.25&0.38&0.11&-0.04&0.12\\
NR\textsubscript{ext}-AEC-PF & 0.26&0.43&0.37&0.26&0.28
\end{tabular}
\label{table:scenario2_pesq}
}\hfil
\subfloat[$\Delta \text{MOS\textsubscript{echo}} \text{ }(\uparrow)$]{
\begin{tabular}{ l l  l  l l l}
\hline
\multicolumn{5}{l}{\textbf{Scenario}} \\
\cmidrule{1-5}
\textbf{1} & \textbf{2} & \textbf{3} & \textbf{4} & \textbf{5}\\
\hline
-0.39&-0.42&1.91&0.17&1.14\\
-0.84&-0.83&2.08&1.17&1.42\\
0.57&1.09&2.09&1.14&1.47\\
0.35&1.28&2.01&0.39&1.34\\
0.57&1.09&2.09&1.13&1.46
\end{tabular}
\label{table:scenario2_mos_echo}
}\\
\subfloat[$\Delta \text{MOS\textsubscript{other}} \text{ }(\uparrow)$]{
\begin{tabular}{ l l  l  l l l}
\hline
\textbf{Algorithm} & \multicolumn{5}{l}{\textbf{Scenario}} \\
\cmidrule{2-6}
& \textbf{1} & \textbf{2} & \textbf{3} & \textbf{4} & \textbf{5}\\
\hline
MWF & 0.31&0.61&0.34&0.95&0.97\\
MWF\textsubscript{ext} & 0.85&0.55&0.24&0.49&0.82\\
AEC-NR & 0.09&0.55&0.18&0.55&0.85\\
NR-AEC & -0.51&0.15&0.25&0.81&0.92\\
NR\textsubscript{ext}-AEC-PF & 0.09&0.55&0.18&0.56&0.89
\end{tabular}
\label{table:scenario2_mos_other}
}
\caption{Numerical results of Fig. 5(a).}
\label{table:scenario2}
\end{table}

\begin{table}
\subfloat[][$\Delta \text{SER\textsuperscript{I}} \text{ }{[\text{dB}]} \text{ }(\uparrow)$]{
\begin{tabular}{ l l  l  l l l}
\hline
\textbf{Algorithm} & \multicolumn{5}{l}{\textbf{Scenario}} \\
\cmidrule{2-6}
& \textbf{1} & \textbf{2} & \textbf{3} & \textbf{4} & \textbf{5}\\
\hline
MWF & 0.09&-0.05&7.39&0.09&5.29\\
MWF\textsubscript{ext} & 1.43&2.61&12.36&6.36&7.42\\
NR-AEC & 10.46&12.51&14.10&13.99&13.34\\
NR-AEC & 11.56&11.55&10.20&9.19&12.43\\
NR\textsubscript{ext}-AEC-PF & 10.40&12.45&14.10&13.99&13.34
\end{tabular}
\label{table:scenario3_ser}
} \hfil
\subfloat[][$\Delta \text{SNR\textsuperscript{I}} \text{ }{[\text{dB}]} \text{ }(\uparrow)$]{
\begin{tabular}{ l l  l  l l l}
\hline
\multicolumn{5}{l}{\textbf{Scenario}} \\
\cmidrule{1-5}
\textbf{1} & \textbf{2} & \textbf{3} & \textbf{4} & \textbf{5}\\
\hline
9.57&8.41&1.23&2.90&2.50\\
12.13&12.04&6.02&7.79&4.82\\
10.92&11.84&6.26&7.19&4.78\\
9.57&8.41&1.23&2.90&2.50\\
10.87&11.82&6.26&7.19&4.78
\end{tabular}
\label{table:scenario3_snr}
} \hfil
\subfloat[][$\text{SD\textsuperscript{I}} \text{ }{[\text{dB}]} \text{ }(=0)$]{
\begin{tabular}{ l l  l  l l l}
\hline
\textbf{Algorithm} & \multicolumn{5}{l}{\textbf{Scenario}} \\
\cmidrule{2-6}
& \textbf{1} & \textbf{2} & \textbf{3} & \textbf{4} & \textbf{5}\\
\hline
MWF & 2.73&4.82&2.12&8.05&5.14\\
MWF\textsubscript{ext} & 2.48&3.23&0.19&4.38&4.20\\
AEC-NR & 1.88&3.20&0.85&4.14&4.23\\
NR-AEC & 2.73&4.82&2.12&8.05&5.14\\
NR\textsubscript{ext}-AEC-PF & 1.95&3.31&0.85&4.14&4.23
\end{tabular}
\label{table:scenario3_sd}
} \hfil
\subfloat[$\Delta \text{ESTOI} \text{ }(\uparrow)$]{
\begin{tabular}{ l l  l  l l l}
\hline
\multicolumn{5}{l}{\textbf{Scenario}} \\
\cmidrule{1-5}
\textbf{1} & \textbf{2} & \textbf{3} & \textbf{4} & \textbf{5}\\
\hline
-0.04&-0.01&0.05&-0.13&-0.08\\
-0.09&-0.07&0.21&-0.01&-0.06\\
0.07&0.11&0.24&0.06&0.03\\
0.05&0.06&0.06&-0.09&-0.04\\
0.07&0.10&0.24&0.06&0.03
\end{tabular}
\label{table:scenario3_estoi}
} \hfil
\subfloat[$\Delta \text{HASPI}\text{ } (\uparrow)$]{
\begin{tabular}{ l l  l  l l l}
\hline
\textbf{Algorithm} & \multicolumn{5}{l}{\textbf{Scenario}} \\
\cmidrule{2-6}
& \textbf{1} & \textbf{2} & \textbf{3} & \textbf{4} & \textbf{5}\\
\hline
MWF & -0.03&-0.05&-0.11&-0.47&-0.16\\
MWF\textsubscript{ext} & -0.48&-0.57&0.27&0.36&0.20\\
AEC-NR & 0.14&0.31&0.29&0.39&0.31\\
NR-AEC & 0.29&0.13&-0.03&-0.09&0.08\\
NR\textsubscript{ext}-AEC-PF & 0.13&0.31&0.29&0.39&0.31
\end{tabular}
\label{table:scenario3_haspi}
}\hfil
\subfloat[$\Delta \text{HASQI}\text{ } (\uparrow)$]{
\begin{tabular}{ l l  l  l l l}
\hline
\multicolumn{5}{l}{\textbf{Scenario}} \\
\cmidrule{1-5}
 \textbf{1} & \textbf{2} & \textbf{3} & \textbf{4} & \textbf{5}\\
\hline
0.02&0.04&0.02&-0.04&0.00\\
0.02&0.01&0.13&0.05&0.04\\
0.08&0.09&0.15&0.06&0.08\\
0.07&0.08&0.03&0.00&0.04\\
0.08&0.09&0.15&0.06&0.08
\end{tabular}
\label{table:scenario3_hasqi}
}\hfil
\subfloat[$\Delta \text{PESQ} \text{ }(\uparrow)$]{
\begin{tabular}{ l l  l  l l l}
\hline
\textbf{Algorithm} & \multicolumn{5}{l}{\textbf{Scenario}} \\
\cmidrule{2-6}
& \textbf{1} & \textbf{2} & \textbf{3} & \textbf{4} & \textbf{5}\\
\hline
MWF & 0.08&0.06&0.07&0.01&0.00\\
MWF\textsubscript{ext} & -0.04&-0.01&0.27&0.11&0.03\\
AEC-NR & 0.09&0.22&0.29&0.22&0.20\\
NR-AEC & 0.16&0.22&0.09&0.04&0.09\\
NR\textsubscript{ext}-AEC-PF & 0.09&0.22&0.29&0.22&0.20
\end{tabular}
\label{table:scenario3_pesq}
}\hfil
\subfloat[$\Delta \text{MOS\textsubscript{echo}} \text{ }(\uparrow)$]{
\begin{tabular}{ l l  l  l l l}
\hline
\multicolumn{5}{l}{\textbf{Scenario}} \\
\cmidrule{1-5}
\textbf{1} & \textbf{2} & \textbf{3} & \textbf{4} & \textbf{5}\\
\hline
-0.65&-0.58&0.94&0.31&0.63\\
-0.56&0.02&1.61&0.11&1.11\\
0.69&0.92&1.93&0.47&1.29\\
-0.24&0.75&1.65&0.12&1.01\\
0.78&0.92&1.59&0.48&1.16
\end{tabular}
\label{table:scenario3_mos_echo}
}\hfil
\subfloat[$\Delta \text{MOS\textsubscript{other}} \text{ }(\uparrow)$]{
\begin{tabular}{ l l  l  l l l}
\hline
\textbf{Algorithm} & \multicolumn{5}{l}{\textbf{Scenario}} \\
\cmidrule{2-6}
& \textbf{1} & \textbf{2} & \textbf{3} & \textbf{4} & \textbf{5}\\
\hline
MWF & 0.34&0.25&0.15&0.88&1.17\\
MWF\textsubscript{ext} & 0.54&0.41&0.40&0.36&0.72\\
AEC-NR & 0.00&0.34&0.30&0.52&0.72\\
NR-AEC & -0.52&0.10&0.25&0.75&0.98\\
NR\textsubscript{ext}-AEC-PF & 0.10&0.36&0.48&0.51&0.72
\end{tabular}
\label{table:scenario3_mos_other}
}
\caption{Numerical results of Fig. 5(b).}
\label{table:scenario3}
\end{table}

\end{document}